Preliminary Draft – Comments and Suggestions Welcomed
Version 1.5[1]

# Revolutionary Ideas for Radio Regulation

Douglas A. Galbi
Senior Economist
Federal Communications Commission[2]

June 12, 2002

*Abstract*

Radio technology seems destined to become part of the standard micro-processor input/output system. But unlike for memory or display systems, for radio systems government regulation matters a lot. Much discussion of radio regulation has focused on narrow spectrum management and interference issues. Reflecting on historical experience and centuries of conversation about fundamental political choices, persons working with radio technology should also ponder three questions. First, what is a good separation and balance of powers in radio regulation? Second, how should radio regulation be geographically configured? Third, how should radio regulation understand and respect personal freedom and equality? Working out answering to these questions involves a general process of shaping good government. This process will be hugely important for radio regulation, technology, and applications.

---

[1] The most current version is available from http://www.galbithink.org and http://users.erols.com/dgalbi/telpol/think.htm .

[2] The opinions and conclusions expressed in this paper are those of the author. They do not necessarily reflect the views of the Federal Communications Commission, its Commissioners, or any staff other than the author. I am grateful for numerous FCC colleagues who have helped me and encouraged me over the past seven years of my career at the FCC. Author's address: dgalbi@fcc.gov; FCC, 445 12'th St. SW, Washington, DC 20554, USA.

## Contents

**Contents (cont'd)**

Local telephony is not just about wires, and neither are the Internet and digital devices. Wireless communications revenue has risen from 5% of world telecommunications revenue in 1991 to 28% in the year 2000.[1] The number of wireless subscribers exceeds the number of wireline connections in countries as different as Cambodia, Cote d'Ivoire, and the United Kingdom.[2] Radio technology is growing in importance across information and communications technologies generally, and these areas of the economy drive job-creation. Economic growth and job creation thus depend significantly on radio regulation. But fruitful radio regulation does not come from examining implications for local telephony competition, economic growth, and job creation.

The evolution of the wireless industry may create concerns about market power. Thus far, wireless communications revenue has consisted mainly of undifferentiated voice services. Particularly in Europe and North America, wireless voice services are subject to growth constraints and increasingly intense price competition. In 2001 wireless data accounted for only about 10% of wireless revenue in Europe and about half that in the U.S.[3] The revenue challenge in wireless communications points toward either rapid service differentiation or new ability to raise prices on undifferentiated services, i.e. industry consolidation.[4] But the public interest in radio regulation will not be found in considering implications for service differentiation or industry consolidation.

Wireless technology can bring new communications services to rural and under-served areas. In only five years of growth, mobile telephony has spurred a five-fold increase in the number of telephone subscribers in Uganda and expanded telephone service into the more rural "up-country."[5] In the US, new Internet service providers, small local telephone companies, public utilities, and local governments are using wireless to provide small towns and rural areas with communications services as advanced as those available in much larger cities.[6] Ensuring that rural and under-served areas are not left behind as communications technology advances requires attention to geographic diversity and to local community capabilities. But efficacious radio regulation will not arise from analysis of implications for rural and underserved areas.

Economic growth, job creation, the concentration of economic power, the geographic distribution of communications services, and even efficient use of radio spectrum, all depend significantly on radio regulation. But historically, these issues have not shaped radio regulation. They are unlikely to shape radio regulation in the future.
In recent years the telecom industry has experienced an investment debacle of unprecedented magnitude.[1] How can anyone have faith that industry expertise is the key to reforming radio regulation? Faith is better placed in more promising sources of wisdom.[2]

---

[1] ITU (2002a).

[2] ITU (2001), Figure 2.

[3] In the UK from July to September, 2001, short message service (SMS) revenue amounted to 12% of total cellular service retail revenue. See Oftel (2002), Table 1. In the U.S. in 2000 services other than wireless telephony amounted to about 5% of total wireless revenue. See FCC (2002).

[4] The head of Merrill Lynch's Global Wireless Research stated that, to improve the industry's investment perspective, there needs to be "stabilization of prices," reduction in subscriber churn, and, most importantly, industry consolidation. See Mutschler (2002).. McKinsey consultants say the same. See Isern and Rios (2002), p. 86-9. Pitofsky (1979) explains the significance of economic power to competition policy..

[5] ITU (2000) p. 4 and ITU (2002b).

[6] Beyer, Vestrich, and Garcia-Luna-Aceves (1996) describes non-commercial community wireless networking. See also Prairie inet (www.prairieinet.net), SFLan (http://www.sflan.com) and Technology Review (2001).

Articulating revolutionary ideas about government, persons, and freedom offers the best hope for improving radio regulation. This paper focuses on non-content-related radio licensing and service rules – issues that are generally discussed in terms of regulating radio signal interference.[3] This paper draws on ideas from real, significant, historical revolutions over the past five centuries to create deliberatively productive tension around three important but neglected questions.[4] First, what is a good separation and balance of powers in radio regulation? Second, how should radio regulation be geographically configured? Third, how should radio regulation understand and respect personal freedom?

Even in historically propitious circumstances, these questions have hardly been discussed. Knowledge of centuries of conversation has been absent from consciousness. In some cases the result has been a totalizing system of radio regulation thinly disguised as necessity. Open, free, and vigorous deliberation, among persons with no more expertise than a sense for who they are and where they came from, can produce better ways of governing. Getting right the details of regulation is crucial and requires considerable field-specific knowledge and experience. But persons who have learned only the most revolutionary knowledge about human beings and the world have much unused but useful knowledge for radio regulation.  That knowledge should guide current policy discussions of radio regulation.

---

[1] A Nov. 2001 newspaper article declared: "Bigger than the South Sea bubble. Bigger than tulipmania. Bigger than the dot-bomb. The flameout of the [US] telecommunications sector, when it is over, will wind up costing investors hundreds of billions of dollars." See Morgenson (2001). Since then the situation has worsened considerably.

[2] The U.S. has highly efficient markets for studies and analysis of the communications industry and communications policy. While such markets should be accepted as a feature of open, free deliberation, they do not seem to have promoted much awareness of industry truth, at least over the past few years.

[3] For those interested in programming regulation, Krattenmaker and Powe (1994) provide an outstanding analysis. Galbi (2001a) provides some related data and analysis.

[4] Productive tension depends on boundaries, opposition, and difference. It does not imply animosity or rigid defense of convention. See King (1963) esp. p. 79.

## I. Revolutionary Ideas

Most persons know some revolutionary knowledge about government, persons, and freedom. Profound insights, revelations, and enlightenment come to different persons in different ways, times, places, and forms. Yet history includes periods that have been distinctively and truly educational on a world-wide scale. Such history includes the English Revolution of 1640-1689, the American Revolution of 1776-1789, the French Revolution of 1789-1799, and the revolutions in Central and Eastern Europe, 1989-1990. Many aspects of these revolutions, including their nomenclature, are hotly contested. As one historian recently noted with respect to seventeenth century England:

> *It is customary to point to these historiographical battles as evidence of the continued vitality of our subject, and so they would be if they were leading us anywhere. In fact, however, under questionable generalship, and through the heat, noise, and smoke, the battlefields of the seventeenth century itself are becoming increasingly hard to see under the great pile of bleached bones left by historians murdered by their colleagues….Perhaps we should not, however, confuse this struggle for power with the struggle for knowledge in which we would rather be engaged.*[1]

Invoking the privilege of the popular historian, this section will ignore all academic historiographical battles. It will simply express common knowledge so as to bring it into consciousness within the field of radio regulation.

The English Revolution of 1640-1689 taught dramatically about the separation and balance of powers in government. Battles between King and Parliament, the people beheading the sovereign, a new commonwealth headed by a leader who at one point cleared Parliament at sword-point, restoration of monarchy, secret conversion to Catholicism, family intrigue, continental wars, exile of the monarch: this was not government history easily ignored. In 1642, King Charles I asserted that all was already understood:

> *There being three kinds of government among men, absolute monarchy, aristocracy and democracy, and all these having their particular conveniences and inconveniences, the experience and wisdom of your ancestors hath so moulded this out of a mixture of these as to give this kingdom (as far as human prudence can provide) the conveniences of all three, without the inconveniences of any one, as long as the balance hangs even between the three estates, and they run jointly on in their proper channel (begetting verdure and fertility in the meadows on both sides) and the overflowing of either on either side raise no deluge or inundation.*[2]

---

[1] Scott (2000) p. 22.

[2] Charles I, His Majesties Answer to the XIX Propositions of Both Houses of Parliament (1642), British Library E. 151 (25) (from transcription given in J.P. Kenyon, The Stuart Constitution 1603-1688 at 21 (1965)).

The concept of separation and balance of powers in government has been traced back to ancient Greek thought.[1] Yet these ideas unquestionably were much more truly understood, not at the time of Charles I's portrayal of a well-ordered, green and pleasant England, but after the decades of violent struggle between the King and Parliament. This struggle was at the center of the English Revolution of 1640-1689. Appreciation for this painfully acquired understanding about the separation and balance of powers should inform radio regulation today.

Ideas about interference also developed during the English Revolution. Thomas Hobbes, a leading English intellectual, identified liberty as the absence of interference: "*A* FREE-MAN*, is he, that in those things, which by his strength and wit he is able to do, is not hindered to doe what he has a will to.*"[2] Hobbes also recognized the problem of chaos. He thought the solution to be an absolute monarch. The command of the monarch would eliminate chaos and allow freedom within the boundaries of this law. Hobbes was a profound and fecund thinker. He recognized that one kind of interference (the constraint of the monarch's command) can be beneficial in reducing another kind of interference (chaos). But with respect to the benefits of monarchy, the English Revolution taught that Hobbes was wrong. Monarchy does not promote liberty, as most persons now understand the world.

The American Revolution of 1776-1789 taught important lessons about the geography of governance. Alexis de Tocqueville, a French traveler and thinker early in the nineteenth century, declared:

> *This [US] constitution, which at first sight one is tempted to confuse with the federal constitution that preceded it, in fact rests on an entirely new theory that will be marked as a great discovery in the political science of our day.*[3]

De Tocqueville, however, wasn't quite sure what this great discovery was. He noted that the federal government's powers were enumerated in the constitution, and that plain propositions that persons can easily understand make for a strong and durable government. He observed that the federal government was responsible to citizens, not to state governments. Moreover, the federal government was meant to be able to carry out its responsibilities directly, without requiring the assistance of state governments. This governance structure seemed to combine strengths of both small and large states: "The Union is free and happy like a small nation, glorious and strong like a great one."[4]

De Tocqueville questioned whether these novel principles were applicable elsewhere. This governance structure seemed to depend on extraordinary capacities among citizens:

---

[1] In particular, Polybius' *Histories*. Marshall Davies Lloyd, Polybius and the Founding Fathers: the separation of powers, *available at* http://www.sms.org/mdl-indx/polybius/polybius.htm (Sept. 22, 1998). Montesquieu perceived and disseminated the idea of separation and balance of powers in his influential book, *Spirit of the Laws,* and the idea became a structuring principle for the U.S. Constitution (1789).

[2] Thomas Hobbes, Leviathan 146 (Richard Tuck, ed., Cambridge Univ. Press, 1990) (1651); ch. XXI in alt. online edition at http://www.orst.edu/instruct/phl302/texts/hobbes/leviathan-contents.html.

[3] Alexis de Tocqueville, Democracy in America 147 (Harvey C. Mansfield & Delba Winthrop, trans. & eds., Univ. Chicago Press, 2000) (1835); Bk. 1, ch. 8 in alternative online edition at http://xroads.virginia.edu/~HYPER/DETOC/toc_indx.html..

[4] *Id.* at 154.

*the sovereignty of the Union is so enmeshed in that of the states that it is impossible at first glace to perceive their limits. Everything is conventional and artificial in such a government, and it can be suitable only for a people long habituated to directing its affairs by itself, and in which political science has descended to the last ranks of society. ... I almost never encountered a man of the people of America who did not discern with a surprising facility the obligations arising from laws of Congress and those whose origin is in the laws of his state, and who, after distinguishing the objects placed within the general prerogatives of the Union from those that the local legislature ought to regulate, could not indicate the point at which the competence of the federal courts begins and the limit at which that of the state tribunals stops.*[1]

De Tocqueville worried that the federal government might not be able to carry out policies that state governments opposed. He worried that such a government would be weak and not able to compete militarily with more centralized governments. He argued that considerable uniformity of sentiments was necessary to sustain such government. This sort of Union, he declared, would not do for Europe.

Two hundred years' history enables everyone today to understand more than de Tocqueville's perspicacious mind did. Most persons today do not understand jurisdictional issues and have little interest in them. The U.S. federal government is strong relative to state governments. The U.S. has been militarily successful against more centralized governments. Europe has chosen to adopt a Union for which de Tocqueville's concerns about the U.S. seem equally or more applicable. These facts point to common knowledge flowing from the American Revolution: geographically centralized, unitary government is not necessary. Sub-federal governments give "various and interfering interests" more opportunities to be active in governance, and in doing so they contribute to long-term democratic stability and vitality.[2]

The French Revolution of 1789-1799 proclaimed to the world the "natural, unalienable and sacred rights of man."[3] It produced The Declaration of the Rights of Man and Citizen. This communication set out rights "in a solemn [written] declaration," so that "by being present to all the members of the social body this declaration may always remind them of their rights and duties." Thus the acts of government powers are "liable at every moment to comparison with the aim of any and all political institutions…."[4] The intent was to memorialize that some rights of man do not depend on license from government. The purpose of government is to preserve these natural rights:

*1. Men are born and remain free and equal in rights. Social distinctions may be founded only upon the general good.*

---

[1] *Id.* at 155-6.

[2] James Madison, one of the founders of the U.S. Constitution, recognized this. See Madison (1787), Federalist No. 10.

[3] Declaration of the Rights of Man and Citizen 77 (Lynn Hunt, ed. & trans., St. Martin's Press, 1996) (1789).

[4] *Id.* The declaration, a written document, was a constitutional element of communications policy. It was intended to serve as an enduring mechanism of true memory.

*2. The aim of all political association is the preservation of the natural and imprescriptible rights of man. These rights are liberty, property, security, and resistance to oppression.*

The rights of man are not limited to freedom from interference, but also require persons to be able to share freely ideas and to participate in making law:[1]

*11. The free communication of ideas and opinions is one of the most precious of the rights of man. Every citizen may, accordingly, speak, write, and print with freedom, but shall be responsible for such abuses of this freedom as shall be defined by law.*

*6. Law is the expression of the general will. Every citizen has a right to participate personally, or through his representative, in its foundation. It must be the same for all, whether it protects or punishes.*

"Liberty, equality, and fraternity" resonate deeply in today's understanding of who are persons.[2] Yet the French Revolution provided vivid scenes of both liberation and terror. That persons, not governments, give ultimate meaning to freedom is now well-understood, and still somewhat scary.[3]

The revolutions in Central and Eastern Europe in 1989-1990 showed that systems ignoring the mundane truths of ordinary, daily existence cannot endure. Claiming to be acting on behalf of the proletariat, in the public interest, is not enough. Seeking the true spirit of a revolutionary idea demands attention to the specificities and complexities of real life. The vigorous discussions of liberty, interference, and the state in seventeenth century England ignored the array of institutions between the rulers and the ruled.[4] Early in the nineteenth century, an important thinker, perhaps reflecting upon his experience of

---

[1] Skinner (1998) emphasizes that living in a state with a particular type of governance is central to the neo-Roman idea of (civil) liberty. In this classical understanding of liberty:

*…if you wish to maintain your liberty, you must ensure that you live under a political system in which there is no element of discretionary power, and hence no possibility that your civil rights will be dependent on the goodwill of a ruler, a ruling group, or any other agent of the state. You must live, in other words, under a system in which the sole power of making laws remains with the people or their accredited representatives, and in which all individual members of the body politic – rulers and citizens alike – remain equally subject to whatever laws they choose to impose upon themselves. If and only if you live under such a self-governing system will your rulers be deprived of any discretionary powers of coercion, and in consequence deprived of any tyrannical capacity to reduce you and your fellow-citizens to a condition of dependence on their goodwill, and hence to the status of slaves.*

Skinner (1998) pp. 74-5, internal footnotes omitted. Some critics argue that this idea "is so utopian as to make it irrelevant to the political world in which we live." Skinner responds in part by noting, "One legitimate aspiration of moral and political theory is surely to show us what lines of action we are committed to undertaking by the values we profess to accept." Skinner (1998) pp. 78-9. In practice, considerable discretionary power is exercised in interpreting and applying radio regulation.

[2] That the Declaration of the Rights of Man and of the Citizen literally referred only to male human beings shows a shocking limit of reason at that time. This is an important aspect of what the written text memorializes.

[3] For example, the Universal Declaration of Human Rights, adopted by the UN General Assembly in 1948, states in Article 29, "These rights and freedoms may in no case be exercised contrary to the purposes and principles of the United Nations." The full, frightening and enduring legacy of the French Revolution is well captured by the inclusion of this statement in the Universal Declaration of Human Rights.

[4] Skinner (1998), p. 17.

the French Revolution, described political liberty as persons of all classes who "emerge from the sphere of their usual labors and private industry," and "take in with a glance the whole of France."[1] In modern democratic states, many contemporary appeals to democratic theory bear no relation to most persons' experiences of their lives together. Friendships, families, unions, businesses, religious, scientific, public service and avocational organizations, along with governments of various degrees of subsidiarity, are real and important ways in which persons make up common life. The revolutions in Central and Eastern Europe in 1989-1990 showed that thought that does not incorporate common life will fail disastrously.

To improve radio regulation, these revolutionary ideas must be brought to that specific field in ordinary time. The sections that follow explore in radio regulation the separation and balance of powers, regulatory geography, and personal freedom and licensing. The presentation assumes no expertise in radio regulation. It presents some facts, gleaned from publicly available information, that are not widely recognized even among experts in industry, academia, and government. Most importantly, it presents analysis that everyone can understand, criticize, discuss with each other, and extend.

[1] Constant (1819). Constant's image of a person's consciousness soaring skyward reflects angelism, the origins and importance of which Maritain (1950), pp. 54-89, traces to Descartes.

## II. Separation and Balance of Powers

Prior to the last decade of the twentieth century, radio regulation was almost wholly a minor administrative task for most national governments. State-owned radio and televisions stations organized radio use, and the general structures and process of government administration governed radio.[1] Making, administering, and enforcing rules concerning radio took place within a common, well-established framework of government administration.

The separation and balance of powers in radio regulation has changed significantly over the past decade. Many countries have set up new agencies to regulate radio use. These agencies' authority relative to other legislative, executive, and judicial bodies is often a matter of considerable uncertainty. In addition, private entities now play a much more important role in radio regulation. Governments recognize that allowing private entities to make decisions about radio benefits the public. Private entities, in turn, participate in public regulatory processes, supply radio equipment, develop new technologies, establish non-legal standards, and work with regulators to implement equipment certification, frequency coordination, and dispute resolution. Making, administering, and enforcing rules concerning radio now depends on complex interactions among legislative bodies, regulatory agencies, and private entities.[2]

Property rights, contract rights, and liability rules are an alternative to government administration as a general framework for governance. Property, contract, and liability have been important aspects of human relations throughout human history.[3] Failure to appreciate the significance of such rights and rules to human societies is a fundamental

---

[1] For example, in the UK in 1981 the Home Secretary, the government official responsible for wide range of domestic issues, approved the establishment of 25 additional independent local radio stations. See Carter (1998) p. 5.

[2] Such changes have also occurred in other areas. A lawyer recently did some unpacking of EC environmental law and found, among other things:

> *The non-linear processes of ecosystems, the unpredictability of human behaviour and the problem of scientific uncertainty all make the process of assessing environmental harm an intricate and often intractable business. Scientific uncertainty, in such circumstances, is not simply a 'data gap' but a whole serious of methodological, epistemological and ontological uncertainties that are inherent in the practice of science. [footnotes omitted].*

Unpacking concluded with recognition of the mess created:

> *Despite the appeal of the regulatory toolbox the process of unpacking it reveals that environmental problems are far messier than it assumes and the legitimacy of public institutions matters far more complex than the concept of 'shared responsibility' suggests. Neither of these facts should discount the important role that private actors play in EC environmental law nor the worth of regulatory diversity. What it does negate however is the other facets of the 'new' approach – the presumptions of consensus, the focus on efficiency and effectiveness, and the hallowed and unquestioned status given to private actors.*

See Fisher (2001) pp. 23-4, 37-8. Freeman (1999), using examples mainly from the US, argues that administrative law has suffered from excessive focus on administrative agencies. She urges that private actors' participation in governance be more fully recognized.

[3] For an insightful review of land institutions and property laws in Mesopotamia, Egypt, and Israel between 3000 BCE and 500 BCE, see Ellickson and Thorland (1995).

anthropological error. Property rights, contract rights, and liability rules are essential for fostering individual freedom, human solidarity, and justice. A good configuration of these rights and rules also seems to be associated with material wealth. Realizing the potential contribution of property rights, contract rights, and liability rules to advancing freedom, solidarity, and justice in radio use is in the public interest.

Some policy makers have recognized the importance of property rights, contracts rights, and liability rules. In a March 2002 directive, the European Parliament and the Council of the European Union noted:

> *Transfer of radio frequencies can be an effective means of increasing efficient use of spectrum, as long as there are sufficient safeguards in place to protect the public interest, in particular the need to ensure transparency and regulatory supervision of such transfers.*[1]

There is no evidence that any spectrum trading has yet occurred in the European Union. Further consultations and directives may elaborate on how to achieve transparency. The risk to the public that speculators, traders, and merchants, particularly foreign ones, pose has been well-recognized historically in Europe, and this danger has been regulated in a variety of ways.

Despite more difficult circumstances, Guatemala has taken a more liberal approach to reforming radio regulation. During the European Dark Ages, Guatemala was the center of a highly developed Mayan civilization. But this civilization did not respond successfully to changing circumstances. In 1996, Guatemala, then one of the poorest countries in Central and South America, finally managed to end thirty-six years of civil war. That same year a new telecommunications law was passed. It established *Títulos de Usufructo de Frecuencias* (TUF), which are radio use rights that can be leased, sold, subdivided, or consolidated. TUFs are recorded in a publicly available database. A regulatory body administers simple rules governing TUFs.[2] As of the first half of 2001, about 5000 TUFs have been issued to more than 1000 different persons. About 26% of these TUFs have been sold or otherwise transferred, and banks have judged them to be secure enough rights to be used as collateral for loans. Private mediation of interference disputes is encouraged, and parties can seek from the judiciary system damages for interference.[3]

The Guatemalan reforms have been highly successful. Fixed line telephone subscribership was only 4.1% at the end of 1997, and many Guatemalans, like many persons around the world, lacked important opportunities to communicate. Access to communications capabilities has increased dramatically through deployment of radio services. Mobile telephony subscribership grew from 0.6% of the total population at the end of 1997 to about 9.5% at the end of 2001. Persons have choices between pre-paid and post-paid options, and the average monthly minutes of use is higher in Guatemala than in other Latin American countries. Poor persons in rural areas and urban slums can be seen using mobile phones. Guatemala's approach to radio regulation has provided significant benefits to many of its citizens.[1] Guatemala has probably become the leading practical model of radio regulation reform for rich and poor countries around the world.

---

[1] Directive 2002/21/EC, preface, para. (19).
[2] The Guatemalan regulator is called SIT, the Superintendencia de Telecomunicaciones. SIT's website is http://www.sit.gob.gt/
[3] For an inspiring account of Guatemala radio reform, written by a key participant, see Ibarguen (2001).

Other regions and countries may not have the institutional capacity to implement the type of reforms that Guatemala has. Consider, for example, land titles, an important aspect of property rights in land. The U.S. has a primitive system:

> *The American system of land title records differs from almost all others in developed nations. Our system does not provide citizens with legally binding information about the ownership of land parcels…. It relies on searchers to analyze what they find in the recorded documents and to reach conclusions about the status of title. Moreover, technological advances have had little effect on the recording system. It continues to preserve and provide information in the same fundamental ways used in the Massachusetts Bay Colony in 1620….*[2]

Land titles in the U.S. are registered at the county level in a variety of different institutions and formats. Most countries use a more centralized titling system in which the legal status of title is established and recorded as part of a transaction. Such a system lowers transaction costs and reduces uncertainty regarding property rights. Despite a clearly inferior regulatory system for property rights in land and some recognition of the problem, the U.S. has lacked the capacity to change. The problem does not appear to be related to the judicial system, but to the legislative and administrative process.[3]

Enforcing property rights, contract rights, and liability rules for radio rights is not necessarily a task best assigned only to courts. The judicial system in many countries is weak and susceptible to extra-legal private power. In such circumstances administrative agencies may be much more propitious institutions for establishing and enforcing property rights, contract rights, and liability rules that challenge the private status quo.[4] Even a highly competent judiciary might not be able to translate general legal standards for property, contract, and liability to radio use as quickly and appropriately as administrative action could. On the other hand, a general function of courts is to protect persons against government action that treats persons as legally unequal or violates law on the basis of informal consensus and expediency. While administrative actions may be able to establish rights and rules quickly, they are less likely to be credible long-term without independent judicial support.

Establishing a good separation and balance of powers in radio regulation requires attention to what particular institutions actually do, how they evolve, and how they interact. The effort must encompass not only government institutions but also private entities and their role in regulation. As scholars have pointed out, undoubtedly there is no completely certain formula for success. Circumstances vary, and institutional details are likely to matter a great deal. Yet knowledge and meaningful discussion about these issues are possible. This section will explore patterns, trends and relationships that provide general insight into achieving a better separation and balance of powers in radio regulation.

---

[1] The above facts, and additional ones, can be found in Ibarquen (2001).

[2] Whitman (1999) p. 228.

[3] Bostick (1988) analyzes obstacles to land title reform in the U.S. Despite its comparatively inefficient system of property rights, the U.S. has vibrant markets for real estate. Individuals and business simply absorb the additional cost. Weak systems of governance for radio may be much more costly.

[4] Glaeser and Shliefer (2001b) argue that regulation replaced litigation as the principal mechanism of governance in the U.S. in the beginning of the twentieth century because regulators in the U.S. were less vulnerable to subversion by the rich and politically powerful. In the transition from central planning to a free economy in Poland and the Czech Republic, more aggressive administrative regulation of financial markets in Poland seems to have produced better results. See Glaeser, Johnson, and Shleifer (2001).

**A. Long-Run Decline in Administrative Enforcement**

For insights into radio regulation with a significant private radio industry and an independent regulatory authority, U.S. experience is particularly useful. The U.S. has had both of these institutional features since 1927. Improvements in radio technology and expansion of radio use are more clearly visible over a longer time horizon. U.S. experience provides insight into the evolution of regulatory activities over a time span that highlights important, ongoing changes.

This section will consider regulatory activities in four categories. The first category is administrative decisions, i.e. the regulator's decisions regarding radio use and the licensing of radio users. The second is administrative enforcement, i.e. the regulator's monitoring and enforcement of statutory radio law and administrative rules. The third is judicial process, meaning the development of public case law governing radio use and adjudication of disputes through the public judicial system. The fourth is private regulation, meaning private entities working out among themselves radio use norms, agreements, and operating capabilities. How does U.S. radio regulation in these categories compare between the early 1930s (when good data becomes available) and about the year 2002?

In the early 1930s, the U.S. radio regulator, the Federal Radio Commission (FRC), focused on administrative decisions. Those decisions predominately concerned AM radio broadcasting, although marine, amateur, and telegraphic uses of radio occasionally required administrative decisions. The largest functional units within the FRC were the Licensing Division (30 persons), the Engineering Division (26 persons), and the Legal Division (19 persons). As of September, 1931, the FRC employed 131 persons.[1]

Before 1933, the Radio Division of the Department of Commerce handled administrative enforcement of non-content-related radio regulations. The Radio Division divided the U.S. into nine districts, each with a district headquarters. An additional eleven branch offices were spread throughout the U.S. There was also a central frequency monitoring station in Grand Island, Nebraska. In July of 1932, the Radio Division employed about 187 persons.[2]

[1] Schmeckebier (1932) pp. 70-3. See also FRC (1931) p. 1. Perhaps three or four persons within the Administrative Section of the Legal Division worked on complaints associated with non-content related radio spectrum rules (unlicensed or unauthorized operation of radio stations). See FRC (1932) pp. 11-15.
[2] Radio Division (1932) pp. 6, 11. Employees in the DC office are estimated at 27 based on total salary for the DC office and the same average pay as for the field employees.

| Table 1<br>Licenses and Inspections, 1932 | | | |
|---|---|---|---|
| Radio Service | Licenses | Inspections (fiscal year) | % Licenses Inspected |
| Amateur | 30374 | 696 | 2% |
| Ship | 2160 | 1275 | 59% |
| Land (fixed) | 1165 | 1184 | 102% |
| Broadcasting (AM radio) | 625 | 1193 | 191% |
| Aircraft | 355 | 266 | 75% |
| Source: Radio Division (1932) p. 7. Government stations excluded. Land (fixed) includes geophysical, general and special experimental. | | | |

The Radio Division extensively monitored compliance with statutory and administrative rules. In 1932, the radio division carried out 11,125 inspections of radio equipment on ships departing subject to radio requirements. Those inspections covered 76% of ship departures subject to the inspection regulation.[1] The Radio Division carried out an additional 3,352 inspections of departing ships voluntarily equipped with radio.[2] These inspections reflect private demand that private institutions did not satisfy, at least at the price and with the expertise that the Radio Division offered. Table 1 shows inspection rates for other classes of radio stations. Officials attempted to inspect every broadcast station semi-annually. Other stations were inspected on an annual or non-calendar basis. In addition to inspecting stations, the Radio Division also monitored frequency use through over-the-air frequency measurements. Table 2 shows frequency monitoring statistics for 1932, along with investigations that the FRC pursued on its own initiative. Station inspections and frequency monitoring predominately focused on (AM radio) broadcasting. This was the most contentious area of radio regulation.

[1] Ibid, p. 1. Under U.S. radio legislation, steamships licensed to carry more than fifty persons and undertaking voyages on the oceans or the Great Lakes were required to have radio equipment meeting defined standards, as well two skilled operators. See Ship Act (1910). The inspection figures refer to fiscal year 1932.

[2] Radio Division (1932) p. 1.

| Table 2<br>Monitoring and Investigations, 1932 | | |
|---|---|---|
| Radio Division – Frequency Monitoring | Broadcast | Non-Broadcast |
| Licensed stations | 625 | 34054 |
| Frequency measurements | 66895 | 17738 |
| Stations found deviating | 491 | 2138 |
| % stations found deviating | 79% | 6% |
| FRC Investigations | Broadcast | Non-Broadcast |
| Operating rules (non-content) | 54 | 152 |
| Programming or messages | 156 | 24 |
| Financial issues | 32 | 6 |
| Total investigations | 242 | 182 |
| % investigations/licensed station | 38.7% | 0.5% |
| Source: Radio Division (1932) p.9; FRC (1932) pp. 12-3. | | |

Enforcement actions focused on education, cooperation, and compliance rather than on punishment and deterrence. Radio Division annual reports mention nothing about fines. License revocation was an extreme, rare penalty for not complying with license terms. Enforcement seems to have been largely informal and focused on technical advice to correct detected problems. To foster radio skills and knowledge, the Department of Commerce established a licensing scheme for radio operators under a 1910 radio law that required ocean steamers to have radio equipment "in charge of a person skilled in the use of such apparatus."[1] In 1932 the Radio Division examined and licensed 27,211 radio operators.[2] Unlawful radio use was associated with persons who did not know what they were doing.

From the early 1930s to the early 2000s, the balance among administrative decisions, administrative enforcement, and private regulation changed significantly. The FCC became the only government body regulating non-government radio use. According to a simple estimates based on publicly available data, the FCC in fiscal year 2001 employed 409 persons to support licensing and radio rights management activities related to domestic radio services[3] and about 168 persons to support administrative enforcement of non-content-related rules for domestic radio.[4] These figures indicate that, from the early

[1] Ship Act (1910).
[2] Radio Division (1932) p. 11.
[3] This figure is the sum of actual fiscal year 2001 full-time equivalents (FTEs) in licensing and spectrum management associated with mass media, wireless telecommunications, and engineering and technology activities. See FCC (2002), pp. 71, 75.
[4] The Enforcement Bureau's "Progress Report Year Two" indicates about 138 persons employed in enforcement field offices. See http://www.fcc.gov/eb/reports/progpres.pdf . Enforcement Bureau enforcement of radio service and spectrum use rules concerns the field offices and two of the four functions listed for one of the four divisions of the Enforcement Bureau. See the Enforcement Bureau's website, http://www.fcc.gov/eb/. FCC (2002) p. 73 lists 380 full-time FTEs under the Enforcement Bureau heading

1930s to the early 2000s, the number of regulatory staff employed in administrative enforcement of non-content-related domestic radio rules decreased slightly. The number working on administrative decisions regarding domestic radio approximately tripled.

The number of radio licenses has increased dramatically with improving radio technology and expanding radio use. The number of administratively recognized radio services has approximately tripled, from 30 in 1932 to 106 in the year 2002.[1] Table 3 shows the number of U.S. radio licenses in April, 2002. Rather than organizing the data according to the many administratively defined radio service categories, Table 3 summarizes license counts using the categories in Table 1, plus new categories for new land mobile, FM radio, and television services. Over the past seventy years the number of radio licenses increased about fifty-fold. In measuring the administrative licensing task, the number of licenses should be discounted by license duration, which increased approximately ten-fold from 1932 to 2002.[2] Given that the number of regulatory staff only tripled, productivity in making administrative decisions is about two-thirds greater in 2002 than in 1932.[3] This productivity improvement may reflect clarified and streamlined regulatory processes, better administrative tools, and increased staff effort.[4]

---

of the "Enforcement Activity." Based on the Enforcement Bureau's organization, FTEs associated with the Enforcement Bureau's domestic radio enforcement activities is calculated as (380-138)/8+138=168.

[1] Compare service categories listed in FRC (1932) p. 6, to the number of radio services listed at http://wireless.fcc.gov/uls/radioservices.html

[2] On license durations about 1932, see FRC (1931) pp. 2, 15, and FRC (1932) pp. 7, 10.

[3] The calculation is as follows (50x licenses)/ (10x license duration)/ (3x workers).

[4] Government workers often have a strong ethic of public service and attempt to do whatever is necessary to get the job done. Consider this description of the work of a Radio Inspector in 1924:

> *He comes to the office, not refreshed by a restful night's sleep, but dog-tired from a four or five hour vigil the night before.... Not once in a while but every night, does he do this; not occasionally does he receive an irritating communication... but he gets numbers of them daily. And you, in the comfort of your fireside, complain bitterly at a few annoying splashes of static or an occasional ship transmittal which interferes with your pleasure....And between [amateur radio operators and radio broadcasters], fired at from both sides with no support, stood the radio inspector, sleepless and irritated beyond description, but still struggling to bring peace into this big new family that had been suddenly placed under his wing.... The devotion to duty of the men in the service is remarkable.... The salary is insignificant. Much more has been tendered the inspectors by outside firms, but the majority prefer to stay and conquer your problems and to take such satisfaction as they may find in the fact that they are beyond a doubt doing more to give you better radio than any other individual or group in the art. Think of them as human, and think twice before you write a hastily worded and sarcastic letter.*

See Pyle (1924). While regulatory tasks have changed, many women and men at the FCC still show such dedication to its important mission.

**Table 3**
**Number and Term of License, 2002**

| License Category | Active licenses | Ave. License Term (years) |
|---|---|---|
| Amateur | 728,742 | 10 |
| Land (mobile) | 546,245 | 7 |
| Ship | 276,038 | 10 |
| Land (fixed) | 111,951 | 10 |
| Aircraft | 64,129 | 10 |
| Broadcast (FM radio) | 11,885 | 8 |
| Broadcast (television) | 9,084 | 8 |
| Broadcast (AM radio) | 4,727 | 8 |
| All licenses | 1,752,801 | 9 |

Sources and Notes: See Appendix A.

In addition to calling forth greater administrative productivity, radio regulation also responded to increased licensing demands by eliminating some licensing requirements. In 1938, the FCC allowed low-power radio devices to be operated without individual licenses.[1] Subsequent FCC decisions expanded the scope of this freedom.[2] Many devices that emit radio waves now operate under device class rules.[3] In 1983, the FCC eliminated individual licenses for the Radio Control Radio Service and the Citizens Band Radio Service. At that time, there were about six million licenses for these services.[4] In 1996, the FCC eliminated individual licenses for certain non-mandatory radios carried on ships and aircraft. There were then about 581,000 ship radio licenses and 131,000 aircraft radio licenses of these types.[5] Thus administrative decisions have limited the licensing task.

The regulator's role in inspecting for compliance with licensing and other radio regulations has decreased dramatically over time. In 1932, the Radio Division inspected 1275 ship radios, and in doing so inspected 59% of those licensed (see Table 1). In mid-1990s, the FCC inspected annually about 1600 ship radios, but there were about 581,000 licensed ship radios subject to inspection at that time.[6] FCC orders in the late 1990s privatized all ship radio inspections.[7] In the second half of 1998, the FCC conducted 68 random inspections of ship radios to monitor the private inspection process.[8] Inspections of ship radios and for all other types of radio services are now done mainly in reaction to complaints or particular concerns.

[1] FCC Docket No. 5335. For some background, see FCC (1989), para. 2.
[2] See, for example, FCC (1979).
[3] See 47 CFR, Part 15.
[4] FCC (1983) para. 4.
[5] FCC (1996b) para. 4.
[6] The FCC required that certain types of radios on certain types of ships be inspected at specific intervals. See FCC (1995c) para. 6.
[7] FCC (1996c) and FCC (1998).
[8] FCC (1998b).

Other aspects of administrative enforcement have also become less significant. Table 4 summarizes Notices of Violation that the FCC issued in the year 2001 regarding non-content-related rules for domestic radio use. Violations of antenna structure rules (primarily structure registration) and emergency address system rules accounted for 72% of Notices. Only 165 Notices (15%) concerned rules related to free-to-air radio signals. In 1932, the FCC did not have rules concerning antenna structure registration, which is important for aviation safety, and there were no rules concerning emergency address systems, which are important for civil defense. But recorded rule enforcement actions were more numerous in 1932 than in 2002. Recorded enforcement actions in 1932 addressed 2835 violations of non-content-related operating rules.[1] The decline in administrative enforcement since the early 1930s is particularly remarkable given the huge increases in radio service use and associated administrative rules since then.

**Table 4**
**Notices of Violation, 2001**

| Reason for Notice | Notices Issued | % of total |
|---|---|---|
| Antenna structure rules | 562 | 51% |
| Emergency address system rules | 229 | 21% |
| Free-to-air radio signal rules | 166 | 15% |
| Cable system standards | 95 | 9% |
| Protocols and procedures | 53 | 5% |
| Total Notices of Violation | 1105 | 100% |

Source and Notes: See Appendix A.

Enforcement often involves informal requests for action to comply with rules. An FCC report indicates that over 180 unlicensed ("pirate") broadcasters were shut down in the year 2000.[2] However, in that year the FCC field offices issued only 24 Notices of Violation for unlicensed or unauthorized stations violating the FCC's general licensing authority (47 U.S.C. §301).[3] Only 6 orders are given in a list of Enforcement Bureau-level actions against unauthorized broadcast stations.[4] For comparison, in fiscal year 1935, the FCC Field Section reported 441 complaints about unlicensed radio stations, of which 66 were broadcast stations. That year the FCC Hearings and Trial Section in Washington reported 10 persons indicted and convicted for violating radio law and 20 radio stations closed without prosecution but with promises from the operators not to engage in further unlicensed activity.[5] Because enforcement occurs in different ways, statistics on enforcement should be analyzed with care. But the big picture is clear:

[1] The cited figure is the sum of stations found deviating from assigned frequencies and investigations of operating rules, as given in Table 2 above.
[2] FCC Enforcement Bureau Progress Report, Year One (2000), available at http://www.fcc.gov/eb/reports/yearone.html
[3] See data at http://www.fcc.gov/eb/rfo/actions.html
[4] See listing at http://www.fcc.gov/eb/tpsd/ulo.html
[5] See FCC (1936) pp. 17, 64.

enforcement of radio regulation is now a much less significant regulatory activity than it was in the early 1930s.

One focus of current FCC enforcement effort is shutting down certain unlicensed radio broadcasters. These broadcasters, called "pirates," are typically unusually motivated individuals operating small, non-commercial stations far inland.[1] U.S. pirates, when fined several thousand dollars, typically claim that they are too poor to pay the fine. Stations called "pirates" in Europe have had much different characteristics. European pirate broadcasters have been large commercial operations broadcasting to Europe from ships or islands outside of European countries' territorial waters. Drawing entrepreneurial capital especially from the state of Texas, pirates off the coast of the UK ran operations in the mid-1960s that involved construction costs of about a half million dollars, reached an audience of about 2-10 million listeners per week, and generated profits that may have been as high as a quarter million dollars a month.[2] These pirate broadcasters raised difficult, contentious political and economic issues for European countries. With little financial or political support, "pirate" broadcasters in the U.S. have been much easier targets for regulatory action.

Monetary penalties imposed in radio regulation enforcement have been relatively insignificant. In the year 2001, the FCC imposed 19 forfeitures related to unlicensed or unauthorized radio use. The average size of these forfeitures was about $8,900, and their total amount was about $188,000. These forfeitures are small relative to the total amount and sizes of forfeitures that the FCC has imposed in other areas. FCC-ordered forfeiture for all causes in 2001 totaled $3.7 million, with the largest forfeitures being over a hundred thousand dollars for unauthorized switching of a customer's long-distance telephone provider ("slamming") and for failure to pay universal service levies.[3] Radio use has considerable value: about $25 billion was bid for auctioned radio use rights, and the total value of radio use is unquestionably much higher.[4] The penalties imposed for violations of radio regulations have been remarkably small relative to the value of radio use.

Some have questioned the value of administrative enforcement efforts relating to radio use. An FCC chairman in the mid-1990s recalled:

---

[1] Information on FCC proceedings against pirates is available at http://www.fcc.gov/eb/tpsd/ulo.html

[2] See Paulu (1967) pp. 21-5. Pirate broadcasters in Europe attracted advertising dollars from major corporations.

[3] Forfeiture data was compiled from the list of FCC Enforcement Bureau orders for 2001. See http://www.fcc.gov/eb/Orders/2001.html. A compilation of common carrier-rated enforcement action totals for 2000 indicates a figure of $29.3 million. See http://www.fcc.gov/eb/News_Releases/nrccea.html. This total includes a $6.1 million voluntary incentive payment from SBC to the U.S. Treasury for failure to meet performance goals established to gain FCC approval for SBC's merger with Ameritech. It also includes $16.6 million associated with consumer protection enforcement (e.g., slamming, misleading advertising, and unsolicited "junk" faxes) and $6.6 million associated with competition enforcement (e.g. local market opening requirements and regulations for local telephone companies). These figures include a variety of different types of values, such as voluntary incentive payments, payments associated with consent decrees, and figures associated with both notices of apparent liabilities and forfeiture orders

[4] For data on net bids in FCC auctions, see http://wireless.fcc.gov/auctions/summary.html. Note that some bidders have defaulted, and some of the net bid amounts are for re-auctioned licenses.

*I began the most aggressive effort to reduce the scope of the FCC's activities in agency history. My target was the FCC field offices. After study, I concluded that the principal and misbegotten purpose of the field offices was to do work for industries or other governmental agencies that should do the work on their own…. Blair and I proposed to close virtually all the field offices. The career bureaucrat who ran the group refused my direct order to recommend a major cutback. I issued the order to the next in command, who also refused. Finally, the third in command agreed to make the recommendation to the full Commission. Then the bureau chief relented. …We had won another battle.*[1]

As a result of this initiative, in 1995 the FCC closed its nine separate frequency monitoring stations dispersed across the US, as well as four additional frequency monitoring stations associated with field offices. Nine of the twenty-five field offices were reduced to two resident agents, and three of the six regional offices were closed. So little value was attached to radio enforcement skills that, for the first time in FCC history, involuntary separations were imposed on FCC staff. The regional and field office workforce was reduced by 120 persons. About 50 regional and field office workers lost their jobs through an associated, narrowly targeted, FCC reduction in force.[2]

While administrative enforcement of radio regulation has declined, the judiciary's role in regulating radio has remained negligible in areas other than broadcast programming.[3] U.S. courts have assumed that the FCC has exclusive jurisdiction over technical matters

---

[1] Hundt (2000) pp. 128-9.

[2] Statement of FCC Chairman Reed Hundt, Aug. 17, 1995, online at http://www.fcc.gov/Speeches/Hundt/spreh519.txt See also FCC (1995d). More recently the FCC has emphasized the importance of enforcement, but with emphasis on issues other than radio. See, for example, Kennard (2000). In 2001, the Commission requested from Congress increases in statutory caps on forfeitures (fines) that can be imposed on common carriers. See Powell (2001). In Sept. 2001, the FCC hired five new attorneys to support competition enforcement efforts. See http://www.fcc.gov/eb/News_Releases/newemp.html Radio enforcement has received some attention in the UK. A recent radio review notes:

> *…actual measurement information from the monitoring functions can be used to enhance the accuracy of the interference calculations discussed earlier. The RA [UK Radio Authority] has considerable experience in performing all these functions. A number of respondents to the review's consultation highlighted the growing importance of monitoring and enforcement and the need to maintain a credible and impartial service alongside moves to extend market mechanisms for spectrum management.*

See Cave (2002) p. 84. The UK Radio Authority currently employs 350 full-time equivalent staff in licensing and policy activities, and 130 in compliance. Ibid, p. 205. Note that these figures include staff involved in regulating programming and radio advertising, as well as in international issues related to radio spectrum.

[3] This is not just a peculiarity of the U.S. An Australian spectrum policy review recently noted:

> *A spectrum licensee may apply to the Federal Court for relief if a person causes interference by operating a radiocommunications device not in accordance with a license. The court has wide powers to address the interference, including issuing injunctions to stop the interference and granting damages. Although there have been situations where interference has occurred and parties have negotiated a solution, the ACA [Australian Communications Authority] is unaware of any civil actions to date [Feb., 2002 communication from ACA].*

See Productivity Commission (2002) p. 171. Since 1992 civil actions for spectrum interference have been available in New Zealand, but none have been undertaken. Ibid.

associated with radio use other than by the federal government.[1] Thus state courts have refused to consider common law tort claims in radio interference disputes arising from alleged lack of adherence to FCC rules.[2] Similarly, a federal court found that state courts cannot consider nuisance claims relating to radio interference created in violation of FCC standards.[3] Under §207 of the Communication Act, a person can claim damages against a common carrier either at the FCC or in a federal district court, but not both. Many important types of radio devices, such as microwave ovens, computers, certain types of private land mobile services, and radio broadcasting, are not classified as common carrier services.[4] A search in a major legal database turned up no case in which a U.S. court considered, on its merits, a claim against a radio user for violating FCC rules unrelated to the content of broadcast communications.[5] Courts have, however, judged radio frequency interference disputes in relation to private agreements.[6]

Private regulation of radio use is becoming more important. In 1971, the FCC established a private coordination rule in conjunction with technical regulations for common carrier point-to-point microwave radio services (CCPMRS).[7]
An insightful FCC study of this experience noted:

> *Perhaps the most interesting aspect of the CCPMRS technical regulations is what they do not contain. Except for the antenna pointing rule (and of course the general allocation constraint) there are no a priori restrictions on the selection, location or orientation of specific frequency assignments. There are no prior allotments of channels to markets, as in the broadcast services; no pre-channelization of the band, as in the private microwave and most other services; and no minimum mileage separations as in the private land mobile services.*

---

[1] The precedent for exclusive federal jurisdiction, though seemingly compelling, deserves further consideration. For more details, see Appendix B. The scope of the legal categories "radio" and "technical matters" are also unclear. See the introductory part of Section IV for a discussion of "radio." "Technical matters" might distinguish how communication takes place from what is communicated. But the how and what of communication are often deeply intertwined. This point is emphasized in folk maxims like "It's not what you say; it's how you say it," and "the medium is the message."

[2] Harbor Broadcasting, Inc. v. Boundary Waters Broadcasters, Inc., 636 N.W.2d 560 (2001); Monfort v. Larson, 257 A.D.2d 261, 693 N.Y.S.2d 286 (1999); Fetterman v. Green, 455 Pa.Super. 639, 689 A.2d 289 (1997).

[3] Broyde v. Gotham Tower, Inc., 13 F.3d 994 (1994). See also Blackburn v. Doubleday Broadcasting Company et. al., 353 N.W.2d 550 (1984) and Still v. Michaels, 166 Ariz. 403, 803 P.2d 124 (1991).

[4] Particularly with respect to radio services, whether a service is a "common carrier service" is often not obvious legally or economically. Legislation and rule-makings have addressed this issue. See FCC (1994), esp. para. 20-1.

[5] The direction of case law is to protect radio users from legal claims and obligations outside of FCC processes. Courts have pre-empted local zoning laws placing restrictions on radio frequency interference (In re Freeman, 975 F.Supp. 570 (1997) and pre-empted local regulations forbidding interference with public safety communications (Southwestern Bell Wireless Inc. v. Johnson County Board of County Commissioners, 199 F.3d 1185 (1999)).

[6] An eviction action for creating radio frequency nuisance under an occupancy agreement forbidding unreasonable nuisance was held to come under the subject-matter jurisdiction of a state circuit court. See Winfield Village Cooperative v. Ruiz, 181 Ill.App.3d 745, 537 N.E.2d 331, 130 Ill.Dec. 264 (1989). An eviction action served to a radio tower tenant for, among other causes, creating radio frequency interference for other tower tenants was similarly held to come under the subject-matter jurisdiction of a state court. See Western Cities Broadcasting, Inc. v. Schueller, 830 P.2d 1074 (1991).

[7] FCC (1971).

*Perhaps most notable of all is the absence of even a working definition of harmful interference. Individual licenses are allowed to set their own protection ratios.*[1]

Historically influential arguments have stressed that, without centralized administrative control of radio use, chaos, or at least failure to use radio effectively in the public interest, would result. But this didn't happen. In 1985, there were 15000 coordination actions for this radio service. Only 100 active protest cases were before the FCC.[2] The FCC study observed:

*According to information in the case folders and as confirmed by the Commission's licensing staff, most of these protests are eventually resolved informally through private negotiations between the parties and involve only minimal intervention by the Commission. The Commission gets involved only if one of the parties requests it, and then functions more as a mediator than a judge. However, since the Commission has the ultimate power to mandate a solution, a solution recommended by a member of the licensing staff no doubt carries considerable weight with licensees.*[3]

Thus administrative power to make and enforce rules, though still important as a clear point of reference, matters much less than private regulation in ordinary regulatory activities.

Private negotiations to address interference are now well institutionalized. Since 1983, FCC rules for private land mobile radio services have required applications for new frequency assignments, changes to existing facilities, or operation at temporary locations to include a recommendation from a private frequency coordinator. For frequencies below 470 MHz or above 512 MHz, a particular private frequency coordinator is specified for particular uses, while any frequency coordinator can be used for services provided on frequencies from 470 MHz to 512 MHz.[4] The FCC and three frequency coordinators have entered into Memoranda of Understanding regarding the frequency coordinator's actions "protecting public safety and other land mobile operations from harmful interference" and "promoting fair competition in the land mobile communications marketplace through the expedited enforcement of FCC rules and regulations."[5] In private land mobile radio services, private entities have thus taken on the tasks that once were central aspects of public regulation.

Private regulation also plays an important role for many other radio services. For example, an FCC rule for public domestic fixed multipoint microwave service (multipoint distribution service) states:

---

[1] Williams (1986) p. 16.
[2] Ibid, p. 31, fn. 19.
[3] Ibid, p. 31.
[4] See http://wireless.fcc.gov/plmrs/coord.html
[5] See http://www.fcc.gov/eb/interference/plmic.html. The quotes are from the second paragraph of the Memorandum of Understanding (MOA) with the Association of Public Safety Communications Officials. The MOA with the Industrial Telecommunications Association uses the same language as in the second quote.

> *Licensees, conditional licensees, and applicants are expected to cooperate fully in attempting to resolve problems of potential interference before bringing the matter to the attention of the Commission.*[1]

Many other FCC radio service rules also require private coordination and cooperative, good faith private negotiations to resolve technical problems.[2] Particular types of amateur radio services have developed well-organized, voluntary, private coordination mechanisms.[3] The FCC urges compliance with amateur frequency coordinator plans. When interference arises between a coordinated and non-coordinated amateur repeater or auxiliary station, FCC rules assign the non-coordinated station primary responsibility for resolving the interference.[4]

Private regulation tends to favor narrow service categories. The FCC currently recognizes 106 different radio services.[5] These radio service categories facilitate private regulation by grouping organizations using radio in similar ways. Organizations that operate in similar ways can more easily coordinate radio use among themselves. Moreover, linking particular radio uses to particular radio frequencies helps to control the possibility for competition in a particular radio use. Increasing industry hostility to "spectrum flexibility" seems to reflect at least in part an increased awareness of the role of service categories in supporting private regulation.[6]

Across seventy years, with an independent regulator and a large private industry, the balance of powers in U.S. radio regulation has changed significantly. The number of regulatory staff employed in making rules and licensing users has approximately tripled. The number of licenses processed per year has increased five-fold. In contrast, the number of regulatory staff employed in administrative enforcement of non-content-related radio regulations has decreased slightly. Other measures of such enforcement

---

[1] See 47 CFR § 21.902(a).

[2] See 47 CFR § 22.907(a), § 24.237, § 80.513, § 87.305, § 90.175, § 95.111, § 101.103.

[3] For example, the Utah VHF Society provides coordination for amateur VHF repeaters in Utah. See http://www.ussc.com/~uvhfs/frqcoord.html

[4] See 47 CFR § 97.201(c) and § 97.205(c). See also http://www.ussc.com/~uvhfs/fcc_coord.html

[5] See http://wireless.fcc.gov/uls/radioservices.html

[6] The trade press reported a participant in a wireless industry conference in the U.S. on April, 2002 as noting "Flexibility has gone from 'being a sexy buzzword to being almost a word of criticism during the conference.'" See Communications Daily, Apr. 21, 2002, "Spectrum Flexibility is Buzzword at CTIA Show, To Mixed Reviews." As a spectrum management specialist has pointed out regarding 3G services in the UK:

> *Owners of biased licenses [licenses favoring a particular service or technology] tend to lever an assumed right to limit competition off the relevant Government decision to bias the licenses, and the high prices paid for the UK 3G licenses are likely to have ramifications for policy development for 3G services in other bands well into the future.*

See Futurepace Solutions (2001) p. 13. In other words, 3G licenses are likely to want the regulator not to allow other companies to deliver 3G services using other "non 3G" radio rights. In contrast, the recent UK radio review strongly supported generic licenses for radio use, stating:

> *Existing licenses should be amended to remove restrictions which are not needed for reasons of international co-ordination or interference management, and new licenses should be issued with the minimum number of restrictions possible.*

See Cave (2002) Recommendation 7.2. The actual evolution of service categories in the UK remains to be seen.

have fallen absolutely, and, on a per license basis, dramatically. Judicial proceedings have never played an active role in non-content-related radio regulation. Private regulation, which tends to be neglected in policy analysis, has grown enormously.

## B. Limited Information Matters

*[Warning: This section contains material that some may find intimidating. Be not afraid! Nothing more is required than rudimentary knowledge of addition and subtraction, along with a willingness to read some algebraic notation. Even if you skip this section, you still know much about the world that is relevant to radio regulation.]*

Currently, in the U.S. and in other countries, the most important types of radio regulation are administrative decisions and private regulation, while the role of administrative enforcement and judicial processes are generally negligible. Simple ideological interpretations of institutional and functional forms of regulation would link administrative decisions with administrative enforcement, or private regulation with judicial proceedings. But those aren't the combinations observed. This section presents a simple model for understanding the advantages and disadvantages of administrative decisions and private regulation in a way that also explains how these types of regulation interact.[1]

Important aspects of administrative decisions are the limited information available to regulators and the bounded rationality of regulation.[2] In defining and issuing licenses, regulators make decisions about the initial allocation of radio rights. Radio rights have often been defined in detailed service rules. To give radio users more freedom, rights more recently have been defined by a small number of parameters, such as frequency assignments, limits on emission out of the assigned frequency, and limits on radiation outside of an assigned area. But in either case, regulators generally lack important information about costs and benefits of new radio uses in different circumstances. Moreover, practical limits on the feasible complexity of regulation mean that regulators cannot exhaustively tailor rights to circumstances.[3] Thus licensing rules, even service-specific ones, necessarily entail decisions at a high level of generality.

Private entities can establish radio regulation more closely related to local information and particular circumstances. Weather, terrain, the location of incumbents, and the demand for different services all significantly affect radio use. These and other relevant factors vary significantly by location. Entrepreneurship – perceiving a multitude of new possibilities and responding experimentally to particular circumstances – is more characteristic of small, competing private entities than of a single, public regulator. While private entities may confront limits on innovation that only public action can address, no social process of aggregating information can transfer what private entities learn through freedom, activity, and experience.[1]

[1] This model is not an actual description of what happens in any particular case; rather, it is a tool for seeking a true understanding of reality.
[2] The model does not assume, on the basis of reason or faith, the existence of a single, coherent, universal rationality. The meaning of rationality and the bounds on it are defined within the model. The value of these definitions should be evaluated in terms of the model's ability to provide practical help in the eternal quest for truth and better regulation.
[3] Regulation is written, public law. Voluminous, incomprehensible regulation destroys the meaning of written, public law.

| Table 5<br>Model of Limited Information | |
|---|---|
| $v_1>c_1$ | $v_2>c_2$ |
| $v_3<c_3$ | $v_4<c_4$ |

To clarify these points, consider an applicant for a license to provide a new radio service. Both the value to the applicant of providing that service *v*, and the cost to incumbent radio users of the new radio use *c*, vary across four sets of circumstances. Social welfare, i.e. the public interest, is defined as the difference between the value and cost of services actually provided. The applicant and the incumbents know the value and cost in all circumstances, but the regulator can distinguish only between some of the circumstances.[2] Based on the values, costs, and the regulator's knowledge of them, assume that the four circumstances can be categorized as in Table 5. The regulator can distinguish only between column 1 circumstances and column 2 circumstances. Within a column, the regulator cannot distinguish between the circumstances in different rows.

The regulator can choose not to license the applicant (licensing decision 0), to license the applicant for just the circumstances in column 1 (licensing decision 1), or to license the applicant for the circumstances in columns 1 and 2 (licensing decision 2). Define the regulator's rationality as making licensing decisions so as to maximize the aggregate difference between value and cost for its licensing choices (the aggregate difference will be called service value, *V*). Assume that the regulator enforces its licensing decision. Then service value with no license issued is $V_0=0$. Under licensing decisions 1 and 2, service values are, respectively

(1) $$V_1=(v_1-c_1)+(v_3-c_3)\,,\;\; V_2=\sum_{i=1}^{4}(v_i-c_i)$$

The regulator makes its licensing decision so as to achieve the maximum in $\{V_0, V_1, V_2\}$.

The regulator's rationality also includes an approach to justice. Assume that the regulator requires the applicant, if licensed, to compensate incumbents for the average cost imposed on them under the licensing scheme. Thus in licensing decisions 1 and 2, the compensation levels are, respectively

[1] This is a major theme of Friedrich Hayek's writings and the Austrian school of economics.

[2] The applicants and the incumbent cannot truthfully communicate all information to the regulator. Each party has different incentives in communication. Individual parties' incentives don't necessarily interact to produce the full truth through the working of an invisible mouth.

$$(2) \qquad t_1 = \frac{c_1 + c_2}{2}, \quad t_2 = \frac{\sum_{i=1}^{4} c_i}{4}$$

Compensation is paid, however, on a per circumstance basis, in only the circumstances in which the licensee chooses to operate. The compensation scheme, since it merely redistributes value, does not affect the regulator's optimal licensing choice.[1] In circumstances $i$, compensation may be more or less than $c_i$. Like the limited range of feasible licensing decisions, the limited range of compensation schemes reflects the regulator's lack of information and bounded rationality.

Consider the incentives of the applicant and the incumbents. The applicant, unlike the regulator, can distinguish circumstances across the rows of Table 5. Hence, given the freedom to use or not use its radio rights in particular circumstances, the value L of the license to the applicant under licensing decisions 1 and 2 are:

$$(3) \qquad L_1 = \max\{0, v_1 - t_1\} + \max\{0, v_3 - t_1), \; L_2 = \sum_{i-1}^{4} \max\{0, v_i - t_2\}$$

Since the licensee can choose to operate in all circumstances permitted under the license, $L_i \geq V_i$.[2]

The regulator's limited information and the licensee's information and freedom of action in particular circumstances can affect social welfare. A licensee might get the right to operate in a circumstance where doing so decreases social welfare, and the licensee might also have the incentive to operate there. That is the case, for example, when $V_1>0>V_2$, $c_3>c_1>0$, $c_3>v_3$, and $v_3-t_1>0$.[3] A licensee might get the right to operating in circumstances where doing so would increase social welfare, but the licensee might also have the incentive not to operate there. That is the case, for example, when $V_1>0>V_2$, $c_3>c_1>0$, $v_1>c_1$, and $v_1-t_1<0$.[4] In this situation an incumbent might also have an incentive to oppose a socially beneficial license ($v_3>c_3$ , $t_1<c_3$). In a different situation, despite the requirement to pay average compensation to incumbents, the applicant still might have an incentive to seek a license even when issuing a license is not socially beneficial.[5] These results illustrate a mismatch between private incentives and social welfare even with no externalities and an administrative attempt to provide full compensation.

---

[1] It does, however, affect the set of applicants that apply for licenses. See *infra*.
[2] If the licensee operates in each circumstance permitted under the license, it will earn $v_i$-$t_j$. Substituting with eqn. (2) gives eqn. (1). This result is merely an implication of average compensation across licensed circumstances.
[3] For a specific example, consider circumstance 3 when $v_1$=10, $c_1$=0, $v_3$=3, $c_3$=4.
[4] For a specific example, consider circumstance 1 when $v_1$=2, $c_1$=1, $v_3$=6, $c_3$=5. The incumbent would oppose the license because the applicant would operate only in circumstance 3 and pay only 3 as compensation.
[5] For a specific example, take $v_1$=3, $c_1$=2, $v_3$=10, $c_3$=12. If the applicant was granted a license, it would operate under circumstance 3, given the specified compensation scheme.

The rationality of the regulator's licensing decision is also bounded by the regulator not knowing what the applicant will do in particular circumstances after licensing. The regulator chooses the licensing rule to maximize service value within $\{V_0, V_1, V_2\}$. However, $V_1>V_2$ does not imply $L_1>L_2$. Thus the regulator's licensing choice might not maximize service value actually realized. Some regulators attempt to address this problem by imposing build-out requirements on licensees. Since $L_1\geq V_1$ and $L_2\geq V_2$, to the extent that such requirements matter, they generally reduce social welfare.[1]

Now suppose private negotiations over radio rights are costless. Then, without licensing, a company seeking to provide the new radio service would acquire rights from incumbent users in all circumstances where $v_i-c_i>0$. Rights would not be acquired if $v_i-c_i<0$. Private negotiations thus generate the maximum possible total social welfare. More generally, private negotiations are more attractive than licensing if the costs of private negotiation are sufficiently low.[2]

But the costs of private negotiations appear to be quite high in many circumstances. Most transfers of licenses occur without any change in the license.[3] Private negotiations for radio rights are almost never used as a means for enabling a new type of radio service.[4] Most private negotiations concern extending or modifying existing licensees' rights to do something similar to what has always been done with the license. Broadcasters' inefficient radio use shows the prohibitively high cost, thus far, of private negotiations to bring forward persuasive cases for changing broadcast service rules.[5]

---

[1] In some circumstances build-out requirements increase social welfare. Suppose $V_1>V_2>0$, $c_3>c_1>0$, $v_3=c_3$, and $t_1>v_1>c_1$. Then a licensee who could provide a socially valuable service would not do so unless a build-out requirement compelled the licensee to do so. But given this situation, the applicant would not have applied for the license in the first place. In a more complex model, other beneficial service opportunities might be sufficient to motivate the applicant to seek a license despite binding build-out requirements.

[2] Private negotiation over radio rights might be very costly if public radio regulation enforced restrictions against the desired radio use. Then private negotiations would have to encompass political action to change public radio regulation. But there are other reasons as well for high costs of private negotiations.

[3] In 2001, there were 3346 assignments of authorization and transfers of control among U.S. non-broadcast radio licensees included in ULS, compared to 916 requests for license amendments. The license amendments include changes in the license unrelated to operating conditions. Calculated from Wireless Bureau, FCC, ULS Data, *available at* http://wireless.fcc.gov/cgi-bin/wtb-datadump.pl (accessed April, 2002).

[4] An important instance of private negotiations changing radio use is the acquisition and deployment of radio dispatch licenses (SMR) to provide public mobile telephony. Thomas W. Hazlett, *The Wireless Craze, The Unlimited Bandwidth Myth, The Spectrum Auction Faux Pas, and the Punchline to Ronald Coase's 'Big Joke': An Essay on Airware Allocation Policy*, 14 Harv. J.L. & Tech. 426-7 (2001).

[5] On inefficient broadcast use of radio in the US, see Evan R. Kwerel & John R. Williams, FCC, *Changing Channels: Voluntary Reallocation of UHF Television Spectrum*, Office of Plans & Policy Working Paper No. 27 (1992); Thomas W. Hazlett, *The U.S. Digital TV Transition: Time to Toss the Negroponte Switch*, AEI-Brookings Joint Center for Regulatory Studies, Working Paper 01-15 (2001); Thomas W. Hazlett & Bruno E. Viani, *Legislators v. Regulators: The Case of Low Power FM Radio*, AEI-Brookings Joint Center for Regulatory Studies, Working Paper 02-1 (2002). FCC action to audit private land mobile radio use also suggests an awareness that private negotiations are not necessarily sufficient to produce efficient radio use. See Wireless Bureau, Private Land Mobile Radio Services, Construction and Operation Audit, *available at* http://wireless.fcc.gov/plmrs/audit.html (accessed Sept. 25, 2002).

The licensing process seems to play an important role in supporting private negotiation. Licensing brings interested parties together for extended discussions about interference issues. While these discussions tend to be contentious, they provide information and build relationships that facilitate private negotiations.[1] Licensing also facilitates private negotiations by narrowing the scope for negotiations. For example, the licensing process, both symbolically and through rules that are relatively easy to monitor and enforce, establishes the licensee's general right to use radio in a particular way. The details and boundaries of that use, articulated through the licensing process, then become the primary focus of negotiation. Moreover, technical coordination among licensees has a long tradition of cooperative, solution-oriented negotiations.[2] Thus post-licensing private negotiations take place in a low-transaction-cost context.

The model can help to analyze the interaction between licensing and private negotiation. Suppose the regulator granted a license for radio use under the circumstances in column 1 of Table 5. Suppose that radio regulation allows private negotiations, or that radio rights are not publicly enforced. Suppose that the new radio use in the top row of column 2 is socially beneficial ($v_2$-$c_2$>0), but new radio use in the bottom row of column 2 is not ($v_4$-$c_4$<0). Then after the licensing process, the new licensee and the incumbents are likely to be able to negotiate privately to allow the new licensee the additional use in the top row of column 2. The new licensee will recognize that it can not adequately compensate the incumbent for new use in the bottom row of column 2, so negotiation for such use will not be pursued.

Compare those results to the results of licensing the applicant for radio use in both column 1 and column 2 of Table 5. Most likely the magnitude of payments from the licensee to the incumbent would have been different. Moreover, because of the limitations of the compensation scheme that the regulator can impose, the new licensee might have been given an incentive to operate in the bottom row of column 2, or not operate in the top row. More extensive license rights for the new applicant only redistribute value and increase the possibilities for inefficient, i.e. not socially rational, radio use.

This model suggests how licensing and private negotiations might be closely related. Licensing is necessary to facilitate private negotiations. Private negotiations more effectively substitute for the substance of licensing. But licensing must retain enough substantial effect to attract interest in the licensing process. So, for example, allowing incumbents to have unlimited flexibility in radio use might not facilitate private negotiations over those additional rights. On the other hand, licensing broad "overlay" rights and administratively establishing rules for compensation for incumbents does not effectively distinguish between circumstances. It would provide bad incentives for radio use. One might hope to license overlay rights without any centralized, administrative scheme for compensation.[1] To the extent that could be done, the licensing process would do nothing to facilitate private negotiations, and it might create a more antagonistic environment. When regulation has been reduced to licensing and private negotiations, the best choice seems to be an awkward balance between the two.

---

[1] Persons value form and process in governance for reasons other than protection from arbitrary decisions. Lemley (2001) shows, with careful empirical analysis, that it is not socially cost-effective for the U.S. patent office to expend more resources to better evaluate patents. One interpretation of this finding is that there are significant motivations for filing patents other than exercising them, e.g. public affirmation of one's inventiveness. For an insightful, process-oriented perspective on tort law reform, see Wells (2001).

[2] An recent study of radiocommunications in Australia stated: "Anecdotal evidence suggests that the technical personnel of spectrum licensees often negotiate mutually acceptable resolutions to interference disputes." See Productivity Commission (2002) pp. 170-1. Licenses seem to provide license for technical personnel to negotiate disputes without the help of suits.

### C. Insecure Radio Rights?

Administrative decisions concern not just who gets radio rights in what circumstances, but also how those rights are defined. A recent UK review of market mechanisms in spectrum management noted:

> *It is possible to set boundary conditions, both geographic and adjacent band, for spectrum blocks. However, setting the boundary conditions at levels that prevent interference under all conditions is impractical (i.e. too constraining). Setting the conditions at levels higher than this leads to a potential for interference. Negotiation/co-ordination is therefore required in order to allow for satisfactory co-existence.*[2]

A spectrum management specialist explained the situation this way:

> *Importantly, the access conditions do not seek to fully manage interference, but draw 'a line in the sand' for the licensee to then manage the interference based on this known reference point.*[3]

Thus administrative decisions do not fix atomistic radio rights that are then bought and sold in private negotiations. Administrative decisions provide a reference point for the distribution of value in radio rights. Private negotiations reshape radio rights with respect to this distribution of value.

Because radio rights typically depend on complex administrative models and decisions, there is considerable scope to argue about the administrative reference point. Administrative decisions are not well-separated formally from private struggles over radio rights. Consider, for example, the FCC regulation that defines an important aspect of broadband PCS radio rights:

> *The predicted or measured median field strength at any location on the border of the PCS service area shall not exceed 47 dBuV/m unless the parties agree to a higher field strength.*[4]

---

[1] Kwerel and Williams (2001) presents an interesting plan that attempts to do that through two-sided auctions of radio rights that incumbents voluntarily offer for auction. Gerald Faulhaber, a professor at the Wharton Business School of Univ. of Pennsylvania and a former Chief Economist at the FCC, made a related presentation on spectrum management, property rights, and the commons to the FCC's Technological Advisory Council on June 12, 2002. That presentation drew on joint work with Prof. David Farber, a leading scholar in networking technology and a former Chief Technologist at the FCC.

[2] Burns et. al. (2001) p. 10.

[3] Futurepace (2001) p. 6.

[4] 47 CFR § 24.236). This is a typical specification for auctioned radio rights for geographic areas.

Predicting or measuring median field is not like checking the placement of a property marker. The Australian regulator forthrightly recognized the problem:

> *The [spectrum license] model also relies on imposing 'limits of power levels at the geographic and frequency boundaries', which poses a significant problem for the regulator! In the event of a complaint, the regulator has to establish* as a matter of fact *[emphasis in original] whether or not the power level has been breached. As our engineering and field technical officers were quick to point out, many phenomena in radiofrequency propagation lead to situations where power levels* cannot be measured accurately [emphasis in original]. *Indeed, there are situations where power levels measured only metres apart may be substantially different! The idea of absolute and measurable power levels at boundaries is unworkable.*[1]

Typical implementations have used engineering models that predict interference contours.[2] The practical regulatory significance of these models can best be understood by analogizing them to engineering cost models such as those used in universal service calculations and state rate proceedings.[3] The much higher level of cooperation in radio use, as compared to wireline network use, may help explain why models used in defining radio rights have been relatively uncontroversial thus far. With a more competitive environment and more intensive radio use, models and parameters used to define radio rights are likely to become the focus of regulatory battles in the same way that cost models are in regulators' rate proceedings.

The physical characteristics of radio signals make the problem more complex than cost analysis. Radio signal propagation is not only difficult to measure, but also difficult to model. Unexpected interference can occur even when users adhere to commonly accepted models and parameters for radio regulation. An Australian study noted:

> *Lawful interference can occur even where there has been full compliance with the spectrum management regime. Spectrum planning is not an exact science – a number of unpredictable factors, such as atmospheric conditions and malfunctioning devices, means that the actual propagation of a radiocommunications signal may not match that calculated by a spectrum planner. The Australian Communications Authority reported, for example, that one of its 'more unusual' investigations involved interference with a VHF and UHF emergency communications network in North Queensland from a HF broadcasting service situated in Victoria.*[4]

A UK study also grappled with the issue:

---

[1] Hayne (1997) p. 180.

[2] The extent of revelation and incorporation of company/device-specific information in these models is an important issue. See Cave (2002) para. 5.12 and 5.23, pp. 77, 80.

[3] The FCC HCPM/HAI Synthesis Cost Proxy Model is available at http://www.fcc.gov/wcb/tapd/hcpm/welcome.html FCC Section 271 proceedings document one particular use of the model. See, for example, the voluminous debate about cost models in the NY 271 proceedings. Filings can be obtained by entering "99-295" under "1. Proceeding" in the ECFS search, available at http://gullfoss2.fcc.gov/prod/ecfs/comsrch_v2.cgi About a year after Verizon's NY 271 application was approved, the New York State Public Service Commission in further proceedings reduced a variety of prices by about 30%. See http://www.dps.state.ny.us/fileroom/doc11086.pdf

[4] Productivity Commission (2002) p. 170, Box 8.3. Internal reference omitted.

> *One of the key questions that has to be resolved at the outset is whether the negotiation is based on calculation or measurement. While the closest to reality is measurement, it is not really practical to adopt this approach. If calculation is to be adopted then it will be essential that common data and common calculation tools are made available. There does however remain the question of what happens if an operator claims to be experiencing, and is able to measure, unacceptable interference even though the calculated interference levels indicate that the situation is satisfactory. It is considered that the likelihood of this is remote as experience tends to show that the engineering calculations used to assess interference are relatively conservative. However there would need to be a clear statement associated with the calculation tools that negotiated agreements reached on the basis of results obtained from the calculation tools do not completely guarantee interference free operation.*[1]

Not only is an objective measure of radio signal propagation elusive, but interference in an operational sense involves human perceptions, i.e. subjectivity. Radio use rights cannot simply be defined objectively.

A private legal system is unlikely to develop to provide independent support for radio use rights. Compare radio use to the wholesale cotton trade.[2] In the wholesale cotton trade, a private legal system has "…endured since the mid-1800's, surviving widespread social change, years of extreme price volatility, and substantial change in the background public legal regime."[3] The legal system governing cotton trade depends only on the most general aspects of public law. Radio use, in contrast, depends deeply on public administrative decisions. Moreover, compared to cotton traders, radio users are more diverse geographically, institutionally, and technically. While diversity makes gains from cooperation larger, it also makes cooperation harder to achieve. Private negotiations among radio users about radio rights are unlikely to separate themselves from public regulatory decisions and become institutionalized enough to form an effective private legal system.

On the other hand, private negotiations of radio rights do not provide a sound basis for pubic administrative enforcement. Some regulatory authorities do not require public recording of at least some types of private negotiations about radio rights.[4] Such an approach is conducive to cooperative, technical negotiations incorporating possibly sensitive company-specific information. In contrast, a recent radio review in the UK presented as a key element of radio regulation "a [public] database of deployed systems

---

[1] Cave (2002) p. 248-9.

[2] Bernstein (2001) provides a fascinating and thoroughly researched account of the private legal system governing wholesale cotton trade.

[3] Ibid. p. 1725.

[4] For example, a FCC service rule requiring broadband PCS licenses to protect incumbent fixed microwave licensees (47 CFR § 24.237(b)) states in part:

> *The results of the coordination process need to be reported to the Commission only if the parties fail to agree.*

There is no consideration of reporting requirements for other private negotiations permitted under 47 CFR § 24.236. See subsequent text.

and their technical parameters, and agreed changes to boundary conditions."[1] Under any disclosure requirements, private negotiations are likely to produce regulations more complex and more specific than administrative rules that would be applied to the same situations. Administrative enforcement is best suited to simple, clear, general rules. Making information available about the results of private negotiations will not produce regulation amenable to public administrative enforcement.

Private negotiations among licensees can also undermine regulatory expertise. The licensing process lowers private negotiation costs and thus creates a propitious environment for post-licensing private negotiations. But these private negotiations tend to convert substantive administrative decisions into mere indicators of the distribution of value between the new licensee and incumbents. Thus distributional considerations become a more central aspect of licensing. Distributional decisions relate to political priorities and strategic goals. Field-specific expertise tends to play a minor role in distributional decisions, while political choices, personal connections, lobbying, and happenstance are much more important. Regulatory expertise is wasted, and its credibility eroded, if it is used merely as a cover for distributional struggles.

Encouraging some interference disputes to emerge after licensing may help to create over time more secure radio rights by fostering a better separation and balance of powers in radio regulation. Interference disputes handled through an independent judiciary system would contribute to building a common law to help define and secure radio rights. Moreover, a judicial record of disputes would help regulators to identify clear, simple, and beneficial public regulations that would not be made subject to private negotiation. Such regulations would be a propitious object for administrative enforcement. Giving both an independent judiciary and administrative enforcement a greater role in radio regulation would help to foster more secure radio rights.

---

[1] Cave (2002) p. 83.

## III. Regulatory Geography

Radio regulation has often given little serious consideration to geography. A recent European Commission consultation put forward questions about harmonization, coordination, and the institutional framework for radio regulation. The consultation found:

> *It was generally considered that the distribution of radio spectrum to users should respond to local and national needs and would therefore best be carried out at national level. ....*
>
> ....
>
> *Most respondents tend to agree that international decision-making on the harmonisation of radio spectrum should be the rule so as to avoid practical problems at the level of radio spectrum management, but sufficient scope should remain for radio spectrum policy to meet national and local needs.*
>
> ....
>
> *The public consultation confirmed the objective to secure radio spectrum availability for pan-European radio systems, services, and equipment but this has to be balanced with national and local needs for radio spectrum.*[1]

This consultation confirmed the importance of balancing regulation across the international, pan-European, national, and local levels. The international level generally means regulation developed in the International Telecommunication Union (ITU). The pan-European level has meant regulation developed through the European Conference of Postal and Telecommunications Administrations (CEPT). The reference to "local" seems new, but as the last sentence quoted above indicates, it is not clear that local (sub-national) governance has been seriously considered. More significantly, despite dramatic developments in radio technology and services, the consultation shows remarkable conservatism. The consultation shows no significant advocacy for change in regulatory geography.

Changes in regulatory geography could be highly beneficial. A Green Paper in 1994 urged that licenses for advanced mobile services be awarded in the European Community "in a coordinated manner and/or at Community level."[2] The licenses, however, were awarded in an uncoordinated way at the national level.[3] Recent financial difficulties have lead to pressure for consolidation, as well as complaints about a particular license condition in Germany that makes mobile mergers costly.[4] Uniformity in regulation and ownership over a wide area can promote common user experiences and a greater scope of connectivity at lower cost. On the other hand, governance structures that promote wide-area uniformity can lead to less adaptation to local needs and less innovation in service offerings and business plans. These later factors seem increasingly important in light of pressing questions about mass-market demands for advanced mobile services.

---

[1] EC (1999) pp. 5, 12, 14.

[2] CEC (1994), Section VII, para. 7. Advanced mobile services are usually referred to as third-generation (3G) services, UMTS-2000 services or IMT-2000 services. The specific advanced mobile services that will be important to large numbers of persons are not yet known.

[3] For discussion and data on European advanced mobile services licensing, see Jehiel and Moldovanu (2001).

[4] If companies holding German spectrum licenses associated advanced mobile services merge, one such license must be returned to the German regulator.

In the 21'st century, short-range radio communications are likely to be very important. Home and neighborhood wireless networking has been attracting interest from a variety of community groups, small businesses, and large hardware and software providers.[1] Pico-radio networks, based on nodes no bigger than a shirt button and very short radio links, are an active area of research.[2] Many devices – refrigerators, toilets, cars, and perhaps children's clothes – may include radio communications to exchange relevant information with a ubiquitous network. The success of particular applications in particular places will depend strongly on local social, physical, economic, and demographic factors.

U.S. experience points to the potential value of local differences in radio regulation. Federal spectrum auctions in the U.S. allowed for simultaneous, coordinated bids for radio rights across the country. Table 6 shows the total paid for radio rights in auctions that offered radio licenses partitioning all of the U.S. into services areas roughly the size of states or smaller. Federal revenue acquired from those auctions seems large enough to attract attention. For example, in Illinois and Virginia federal auction revenue was about 5% of state general fund revenues for fiscal year 2000 and greater than all state revenues from corporate taxation.[3] However, in the US, as in Europe, raising revenue is widely considered to be an inappropriate primary objective for radio policy.

---

[1] Home and neighborhood wireless networking currently leans heavily on the 802.11b (Wi-Fi) protocol. This protocol has been built into Microsoft's Windows XP operating system. For a brief discussion of Wi-Fi networking, see http://news.com.com/2100-1033-273516.html For an example of a community network, see http://www.sflan.com

[2] On pico-radio, see Cameron (2002).

[3] For fiscal information on IL and VA, see http://www.ioc.state.il.us/Office/index.cfm and http://www.dpb.state.va.us/budget/faq/faq.htm

**Table 6**
**Federal Auction Revenue by U.S. State Geography**

| State/Territory | Total Rev. (US$, mil.) | Rev./Person (US$) | State/Territory | Total Rev. (US$, mil.) | Rev./Person (US$) |
|---|---|---|---|---|---|
| Alabama | $350 | $87 | Nebraska | $73 | $46 |
| Alaska | $18 | $33 | Nevada | $173 | $144 |
| Arizona | $492 | $134 | New Hampshire | $88 | $80 |
| Arkansas | $94 | $40 | New Jersey | $782 | $101 |
| California | $3,391 | $114 | New Mexico | $87 | $58 |
| Colorado | $343 | $104 | New York | $1,615 | $90 |
| Connecticut | $309 | $94 | North Carolina | $438 | $66 |
| Delaware | $54 | $82 | North Dakota | $22 | $35 |
| D.C. | $80 | $131 | Ohio | $768 | $71 |
| Florida | $1,678 | $113 | Oklahoma | $171 | $54 |
| Georgia | $931 | $144 | Oregon | $233 | $82 |
| Hawaii | $135 | $122 | Pennsylvania | $780 | $66 |
| Idaho | $65 | $65 | Puerto Rico | $356 | $101 |
| Illinois | $1,457 | $127 | Rhode Island | $82 | $81 |
| Indiana | $547 | $99 | South Carolina | $265 | $76 |
| Iowa | $147 | $53 | South Dakota | $31 | $45 |
| Kansas | $113 | $46 | Tennessee | $346 | $71 |
| Kentucky | $227 | $62 | Texas | $1,511 | $89 |
| Louisiana | $380 | $90 | U.S. Virgin Is. | $15 | $148 |
| Maine | $64 | $52 | Utah | $187 | $109 |
| Maryland | $547 | $114 | Vermont | $42 | $75 |
| Mass. | $537 | $89 | Virginia | $561 | $91 |
| Michigan | $554 | $60 | Washington | $577 | $119 |
| Minnesota | $275 | $63 | West Virginia | $76 | $42 |
| Mississippi | $224 | $87 | Wisconsin | $387 | $79 |
| Missouri | $406 | $79 | Wyoming | $30 | $67 |
| Montana | $28 | $35 | | | |

Source: See Appendix A. Total revenue above is $23.1 billion; average revenue per person, $91; median revenue per person, $81.

The more analytically important aspect of Table 6 is the range across states in revenue per person. These differences are not well explained by differences in the area of states. They relate to more complex aspects of economic and physical geography, such as income, industry types, land profiles, and weather. Using radio efficiently involves calculations of the opportunity cost of radio rights in trade-offs across various margins. Since the costs of radio rights vary by more than a factor of three across states, efficient radio regulation would involve significant regulatory variations across states.

Beneficial changes to regulatory geography depend on much broader considerations than just efficient radio use. The capabilities of real governing institutions of different geographic scope should be carefully considered. One should look for hidden capabilities that might be useful for promoting change and consider whether particular demands might usefully promote beneficial institutional growth.[1] Simple, clear, general rules can promote national uniformity and regulatory certainty. More complex, chaotic, and localized regulation can promote innovation, participation, and adaptability. Satisfaction with the status quo might not be sustainable, and less change now might mean more traumatic change latter. Moreover, as is regularly noted, communications plays an important part in shaping the democratic political process, culture, intellectual work, and personal relationships.[2] Given such scope for deliberation, any geographic scheme of governance might be successfully rationalized under particular conditions. This section offers, with faith, hope, charity, and some humor, not a particular program for reform but only some important history, facts, and insights for discussion.[3]

### A. Geography Truly Matters

Geographic categories are very important for understanding realities of human life. Weather and proximity to water have an astonishingly large effect on broad patterns of activity. At least 52% of world GNP in 1995 was produced in areas with humid/temperate climates located within 100 km of navigable sea. This area category includes only 8% of the world's inhabited land. GNP per capita in temperate areas within 100 km of navigable sea is more than six times greater than GNP per capita in tropical areas not within 100 km of navigable sea.[4] The significance of these geographic categories seems to relate to effects of climate on health and agricultural technology, and to the role of navigable water in facilitating intercourse of people, goods, and ideas.[5] Study of economic reality teaches that material circumstances depend strongly on geography.[6]

The importance of geography is also evident across 8,000 years of history in the area now called Japan. The most densely settled regions in the Jomon period (6,000 to 300 BCE) were also the most densely settled regions in 1998.[7] Consider as well the effects of U.S.

---

[1] Hirschman brought attention to these issues in writing about economic development. See, e.g. Hirschman (1958). These ideas are also important in considering regulatory reform in high-income countries.

[2] The EC public consultation on radio spectrum pointed to the objective of ensuring "the appropriate balancing of commercial and public interest in radio spectrum policy." CEC (1998) p. ii. The consultation found: "There appears to be tension between the objectives *to develop new services that meet consumer demand and to meet public policy objectives* [emphasis in original]." EC(1999) p. 13. It is unclear how increasing persons' capabilities to communicate freely, actively, and effectively relates to this balancing.

[3] An eminent U.S. constitutional lawyer has recently emphasized the virtue of humility. See Scalia (1997) pp. 72,137. I agree that humility is important, too. Moreover, as a low-ranking government bureaucrat, I am fortunate to have many opportunities to experience humiliation.

[4] Sachs (2001) p. 6-7. The climate categories are based on the Koeppen-Geiger classification system.

[5] Ibid., passim.

[6] This is also true within the U.S. Income per square kilometer in coastal counties of the U.S. (those with centers within 50 km of an ocean, the Great Lakes, or a commercially navigated river) is more than eight times that of "inland" counties. These coastal counties amount to 15% of U.S. land area but 60% of 1998 civilian income. Moreover, proximity to commercially navigated water has significant correlation with current productivity. See Rappaport and Sachs (2001), p. 1, and passim. Across the twentieth century, U.S. counties also display large and enduring differences in economic growth rates. See Rappaport (1999).

[7] Davis and Weinstein (2001) p. 14.

bombing of Japan in World War II.[1] The nuclear bomb that the U.S. dropped on Hiroshima, Japan's eighth largest city, killed 80,000 persons (20.8% of the city's population) and destroyed two-thirds of the built up area of the city. Kyoto, Japan's fifth largest city, was not bombed at all because of its cultural significance. The nuclear destruction of Hiroshima did not have an enduring effect on its attractiveness as a place to live. By 1975, the ratio of Hiroshima' population to Kyoto's was about the same as it was before the bombing.[2] The relative attractiveness of places to persons is a remarkably stable aspect of life.

Personal experiences of planes, trains, phones, the Internet, and other technologies, which didn't exist before, tend to foster a consensus that geography matters much less now than it did five years ago, or perhaps fifty years ago, or maybe two centuries ago. But reality is not always what most persons perceive it to be. Ultra-wideband communication between persons – meaning face-to-face or closer mutual physical presence in communication – is probably more important than any other means of communication even in materially wealthy groups of highly alienated and abstracted persons.[3] Wide-ranging studies of common property regimes show that local knowledge, local boundaries, and local collective action play an important role in good governance, as does respect for the rights of persons to increase their subjectivity by creating and sustaining local governing institutions.[4] Moreover, much current governing capability resides in institutions that have a strong geographic orientation.[5] Some might find fantasies about transcending body, place, or history inspiring or therapeutic.[6] But in policy deliberations and decisions, geographic considerations should play a major role.

### B. The US: More Centralized than Europe

Europe and the United States have significantly different geographic configurations of radio regulation. Early radio use was primarily for maritime communication.[7] Maritime issues have long concerned mutually recognized sovereign states. Such states worked out with each other early radio regulation.[1] Radio regulation in Europe developed through regional institutions using traditional inter-state mechanisms. In contrast, within the US, a single federal government acts as the sole regulatory authority. Radio regulation thus has been more centralized in the U.S. than in Europe.

---

[1] Ibid, Section III.
[2] This result holds true more generally for all the cities of Japan, many of which suffered tremendously from non-nuclear bombs. Incendiaries that U.S. bombers dropped on Tokyo on March 9, 1945 produced firestorms that killed more than 80,000 persons. Ibid. One might hope that changes in the way that persons communicate could contribute to a more peaceful world.
[3] See Galbi (2001b), Section III.
[4] See Ostrom (1990), Table 3.1 p. 90. Buck (2001) reviews Ostrom (1990) and related work and argues that a spectrum commons should replace spectrum auctions.
[5] National and sub-national governments and administrative districts (for water, electricity, etc) typically are strongly associated with particular territories. Moreover, such entities dominate many organizations, e.g. the UN and the ITU.
[6] On the theory and praxis of imagination, see e.g. Peale (1952), Unger (1987), and Resnick (2001). Cf. Silverstein (1974), p. 158, "The Little Blue Engine".
[7] The first widely recognized demonstration of radio communication was by Guglielmo Marconi in England in 1897. By 1903, Marconi, under the protection of patents he acquired, had installed 45 coastal station, major stations in England, Canada, and the US, 32 stations on board ships of the principal passenger lines, and stations on Italian, English, French, and American warships. Tomlinson (1938) pp. 12-3.

Different federal institutions in the U.S. have historically set out national regulatory schemes. To cope with interference among radio stations in the late 1920s, the Federal Radio Commission (FRC) established a comprehensive national plan that forced several hundred stations off the air.[2] Congress required in 1928, "as nearly as possible," an equal allocation of licenses among five geographic zones; among the states in a zone, licenses were to be allocated according to population.[3] The FCC, the successor to the FRC, established a comprehensive scheme for assigning television licenses to cities in 1952.[4] All radio use in the U.S. is subject to FCC licensing, and the FCC regulates radio under federal law.

In response to similar problems, European states created new regional institutions much less authoritative than the FRC and FCC. In 1925, European states formed the International Broadcasting Union (U.I.R) to help address interference problems. The U.I.R. and successor organizations have gathered facts, provided technical support, and helped to organize regional conferences and agreements. Negotiations have been continual on a wide range of issues. Agreements are specific, narrow, and need quasi-unanimity.[5] Interference problems are often not eliminated. In Europe in 1929, of 209 radio stations operating, 72 were not observing the agreed frequency plan.[6] In 1967, of 537 medium and long-wave radio stations in operation, 314 were not operating in accordance with the agreed frequency plan. Luxembourg has been a particularly egregious case. It has operated powerful radio and television stations transmitting programs in French, Dutch, German, and English to populations beyond its borders.[7] Europe has not been able to establish and enforce regional plans for radio use as successfully as the U.S. has.

In addition, a large category of radio uses in Europe fall outside the most developed framework for European regional radio regulation. Under international radio regulations, radio uses that are not "capable of causing harmful interference to the services rendered by the stations of another country" are not subject to international radio regulations.[1] The inter-state development of European radio regulation thus gives national regulators in Europe considerable regulatory freedom. Nonetheless, regulation of many local radio uses has been effectively coordinated across Europe.[2]

[1] The Berlin Conference of 1903, the first major radio regulation initiative, was an inter-state negotiation among Austria, France, Germany, Great Britain, Hungary, Italy, Russia, and Spain. It focused on ship-to-shore radio communication. For discussion of early radio regulation, see Codding (1952), Chapter II. At the Washington Conference of 1927, 64 countries, including 24 European countries and the United States, established expansive radio regulations and a table of frequency allocations.
[2] Rosen (1980), p. 128. The FRC plan was issued as General Order 40.
[3] This law, known as the Davis Amendment (45 Stat. 373), passed as part of the 1928 re-authorization law for the Federal Radio Commission.
[4] FCC (1952).
[5] Tomlinson (1938) pp. 179-212, Codding (1952) pp. 157-9, Paulu (1967) pp. 13-5.
[6] Tomlinson (1938) p. 181.
[7] Paulu (1967) p. 14-15, fn. p. 15.

Whether the more centralized U.S. radio regulation or the less centralized European approach would better serve the public interest should be considered. Certainly some aspects of European radio and television, such as the British Broadcasting Corporation, are highly respected worldwide. Europe also has been generally regarded as the leader in the development of mobile telephony. On the other hand, the U.S. has developed vibrant local, private radio and television broadcasters, a strong independent content-creation industry, and considerable technological dynamism in advanced wireless services. The issue is not whether the U.S. should be like Europe, or Europe like the US, or a third area like one or the other. The value of the contrast is to point to relevant facts and institutional possibilities that might stimulate more fruitful deliberation about regulatory geography.

### C. Barren Deliberation in the US

The U.S. provides a case study in deliberation that has failed to engender, with respect to regulatory geography, serious consideration of important facts, possibilities, and consequences. About 1927, when the Federal Radio Commission was created, U.S. legal scholarship dismissed all geographic boundaries in radio regulation (but one, usually) with appeals to obvious scientific and practical concerns:

> *In the present situation, unity of control is indispensible. Wave lengths must not conflict....National and uniform rules are necessary.*[3]
>
> *Radio communication cannot be confined by artificial state boundaries. It is essentially interstate in scope and character, broadcasting stations being so constructed that purely intrastate service is not only impracticable but all but impossible.*[4]
>
> *the practical advantages not to say necessity of some centralized control is apparent. ... The very nature of the scientific phenomenon made use of in radio communications demands centralized regulation as a condition of its advantageous exploitation.*[5]

[1] Radio Regs, ITU, 4.2 (2001). This provision dates back to the Washington Conference, 1927. *International Radiotelegraph Convention,* Washington, Gen. Regs., Art. 5 (1927). For discussion of its importance, see Codding and Rutkowski (1982) pp. 273-4.
[2] E.g. mobile telephony, and through CEPT, many other types of low-power, class-licensed devices.
[3] Lee (1925) p. 20.
[4] Zollman (1927) p. 121.
[5] "The Radio and Interstate Commerce" (1928) p. 921.

*That the federal government must control the broadcasting situation is generally admitted. The tremendous present importance and future possibilities of radio, the limitations upon the number of persons who may broadcast simultaneously without causing a chaos of interference, and the fact that radio waves are not confined within the bounds of a single state or nation, make obvious the necessity of unified federal control.*[1]

*If the air is to be used successfully by radio, it must be on the basis of a world utility, regulated by a world public service commission through agreement of the governments. ...it is hard to image a station that will not be strong enough to send a message over the boundary of a particular state.*[2]

This legal scholarship largely ignored amateur radio, it lacked insight into the subsequent trajectory of radio technology and radio uses (think, for example, of microwave ovens and garage door openers), and it failed to appreciate adequately then developing European examples of governance.[3] It foreclosed debate about regulatory geography with vague appeals to necessary implications of specialized, extra-legal knowledge.

Early U.S. radio law formally limited the scope of federal regulation. The first sentence of the Radio Act of 1912 specified:

*That a person, company, or corporation within the jurisdiction of the United States shall not use or operate any apparatus for radio communication as a means of commercial intercourse among the several States, or with foreign nations, or upon any vessel of the United States engaged in interstate or foreign commerce, or for the transmission of radiograms or signals the effect of which extends beyond the jurisdiction of the State or Territory in which the same are made, or where interference would be caused thereby with the receipt of messages or signals from beyond the jurisdiction of the said State or Territory, except under and in accordance with a license, revocable for cause, in that behalf granted by the Secretary of Commerce and Labor upon application therefore; but nothing in this Act shall be construed to apply to the transmission and exchange of radiograms or signals between points situated in the same State:* Provided, *That the effect thereof shall not extend beyond the jurisdiction of the said State or interfere with the reception of radiograms or signals from beyond said jurisdiction;*[4]

---

[1] "Federal Control" (1929) p. 245. The quoted sentences are the first two sentences of the article, which focused on determining which stations should be licensed.

[2] Chamberlain (1927) p. 343. Actually existing international radio regulation, on the other hand, recognized the possibility of radio use that does not cause harmful interference outside of a nation's boundaries. See above.

[3] In the U.S. on January 1, 1927, there were 14,768 amateur radio stations, 671 broadcast stations, and 583 other stations (transoceanic stations and domestic point-to-point stations). See Davis (1928) p. 3. Little centralized frequency planning and assignment was done for amateur stations or point-to-point stations.

[4] Radio Act (1912).

The contrast between this sentence, and an obvious, much simpler one, indicates at least a perceived need to describe limits on the scope of the law.[1] Even to a politically and rhetorically sophisticated person of that time, the natural sense of this sentence would have excluded weak radio emissions unrelated to commercial activity and not generally understood as communication, radiograms, or signals. Radio emissions that a home electrical generator might incidentally create are an example of such an exclusion. Many persons probably would have regarded the plain meaning of the sentence to imply additional exclusions as well.

The distinction between interstate radio communications and intrastate radio communications had little practical significance for early radio. Early radio uses – maritime communication, military communication, and "wireless telegraphy" – were closely associated with federal power. Private, non-commercial (amateur) radio users were interested in radio technology.[2] Since the ability to communicate over long distances was central to the perceived technological wonder of radio, only intrastate radio use was not an interesting possibility for amateurs. Moreover, influential figures in amateur radio strongly supported the Radio Act of 1912. The opportunity to get a federal license was hailed as a great victory for amateurs.[3] By the mid 1920s, most persons associated radio with AM radio broadcasting. Most AM radio broadcasts in the 1920s covered multi-state areas. Few persons in the 1920s cared about intrastate radio communications.

The Radio Act of 1912 was implemented in accordance with this predominate balance of interests. In an Annual Report submitted on November 13, 1912 to the Secretary of Commerce and Labor, the Commissioner of Navigation began discussion of the new radio act by explaining forthrightly what it meant:

> *In brief it [The Radio Act of 1912] prescribes that all apparatus and operators for radio communication within the jurisdiction of the United States (except*

---

[1] An example of a simpler sentence: "This Act prescribes that all apparatus and operators for radio communication within the jurisdiction of the United States (except Government stations and operators and those in the Philippines) shall be licensed by the Secretary of Commerce and Labor." See Annual Report (1912), under heading "Regulation of Radio Communication," quoted in text below.

[2] Private (amateur) radio grew rapidly on an unlicensed basis from 1900-1912. Around Boston alone 250 private stations were estimated to be in operation circa 1909. A significant number of those were asserted to be "equal or superior to those operated by the navy." A leading Navy operator was reported to have said that Navy radio stations were three years behind the leading radio technology and that this technological backwardness made Navy communications more susceptible to interference. The author of a long article on amateur radio in the magazine Electrician and Mechanic declared:

> *...legal action providing for the control by Government of wireless telegraph stations is at the present time immature and unnecessary. Wireless apparatus guaranteed to prevent interference has for some time been at the disposal of the Navy Department. Moreover, any attempt to eliminate amateur stations would simply ward off for a time a problem, the solution of which must finally be found by the scientist, not the lawyer.*

See Morton (1909). Army, Navy, and other government agencies presented private radio as a nuisance – mere entertainment for "amateurs" who interfered with government organs performing important national functions. In an article published nine months later, Morton was more deferential to the Navy and more accepting of licensing amateurs. See Morton (1910). Perhaps high Navy officials, after reading his earlier article, had dressed him down.

[3] See Section IV.A.1 of this paper.

*Government stations and operators and those in the Philippines) shall be licensed by the Secretary of Commerce and Labor.*[1]

The implementing regulations themselves were more legally fastidious. These regulations observed that the Act of 1912 limited federal authority to require licenses. The limit was recognized with a one-sentence regulatory provision:

> *The owner or operator of any apparatus who may be in doubt whether his apparatus, under [the first paragraph of the Radio Act of 1912], is exempt from license may write the facts to the Commissioner of Navigation, Department of Commerce and Labor, Washington, D. C., before applying for a license.*[2]

Less than a year later this issue of statutory construction and federal authority had devolved to a lower level of government:

> *The owner or operator of any apparatus who may be in doubt whether his apparatus, under this paragraph, is exempt from license may write the facts to the radio inspector for his district before applying for a license.*[3]

According to a scholarly article published in 1928, the Secretary of Commerce required all stations to be licensed.[4] The Radio Act of 1912 produced in implementation little deliberation about regulatory geography.

The Radio Act of 1927 did not significantly change the statutory description of regulatory geography. The geographic scope of regulation stated in the Radio Act of 1912 was more compactly stated in the introductory phrase of the Radio Act of 1927:

> *...this Act is intended to regulate all forms of interstate and foreign radio transmissions and communications within the United States, its Territories and possessions; to maintain the control of the United States over all the channels of interstate and foreign radio transmission;*

The Radio Act of 1927 included a qualified enumeration of powers with a qualified statement about preventing interference:

> *Sec. 4. Except as otherwise provided in this Act, the commission, from time to time, as public convenience, interest, or necessity requires, shall—*
> *(a)...*
> *(f) Make such regulations not inconsistent with law as it may deem necessary to prevent interference between stations and carry out the provisions of this Act...*
> ...
> *(k)...*

Most significantly, text in the first paragraph of the Radio Act of 1927 merely transformed the 1912 Act's stated limit on authority into an affirmative enumeration of authority:

> *...[a federal license is needed to] use or operate any apparatus for the transmission of energy or communications or signals by radio (a) from one place*

---

1 Annual Report (1912), under heading "Regulation of Radio Communication".

2 Dept. of Commerce and Labor (1912), Regulations, Part 1. A.

3 This transition was made in two steps. In regulations issued February 20, 1913 (Dept. of Commerce and Labor (1913a)), an owner or operator who might have doubts about whether he or she needed a license was instructed to contact a radio inspector or the Commissioner of Navigation. In regulations issued July 1, 1913, the quoted regulation was issued (Dept. of Commerce (1913b)).

4 Taugher (1928) p. 314.

*in any Territory or possession of the United States, or from the District of Columbia to another place in the same Territory, possession or District; or (b) from any State, Territory, or possession of the United States, or from the District of Columbia to any other State, Territory, or Possession of the United States; or (c) from any place in any State, Territory, or possession of the United States, or in the District of Columbia, to any place in any foreign country or to any vessel; or (d) within any State when the effects of such use extend beyond the borders of said State, or when interference is caused by such use or operation with the transmission of such energy, communications, or signals from within said State to any place beyond its borders, or from any place beyond its borders to any place within said State, or with the transmission or reception of such energy, communications, or signals from and/or to places beyond the borders of said State; (e) upon any vessel of the United States; or (f) upon any aircraft or other mobile stations within the United States,* [1]

To a sophisticated legal scholar, and probably also to some legislators, this statement implied in 1927 a broader scope for regulation than the same statement implied in 1912.[2] But to the typical U.S. voter in 1912 and in 1927, the plain meaning of the words most probably would be the same.

In Congressional testimony and deliberation preceding the Communications Act of 1934, state representatives showed little interest in radio services then extant. State regulators cared most about the kind of regulation that was most familiar. State commissions focused on rate cases. Most persons in the early 1930s understood radio to be freely available AM broadcasts. As the General Solicitor for the National Association of Railroad and Utilities Commissioners (NARUC) explained in 1934 to the House Committee on Interstate and Foreign Commerce:

*The particular interest of the State commissions is in the wire companies. Radio may become important to them from the point of view of regulation as the uses of radio increase. State representatives do not wish to surrender the future as to that industry, although the present prospect is that efficient Federal regulation will obviate occasion for State regulation, unless State regulation of intrastate rates shall some time become necessary. At present Federal regulation meets the need in the radio field.*[3]

Jurisdictional distinctions in radio regulation, distinctions with great significance for railroad and telephone regulation, thus attracted little deliberation.[4]

---

[1] Radio Act (1927).

[2] U.S. courts over that period were adopting a less restrictive interpretation of the Commerce Clause of the U.S. Constitution. See Cushman (2000). This change in judicial interpretation changes the significance of the sentence. For analysis of the Commerce Clause cases in relation to radio regulation, see Appendix B of this paper.

[3] Communications Bill, Hearings on H.R. 8301 Before House Comm. On Interstate and Foreign Commerce 135 (May 9, 1934) (statement of John E. Benton, General Solicitor for the National Association of Railroad and Utilities Commissioners), reprinted in Paglin (1989) p. 481.

[4] State commissioners expressed considerable concern that the Communications Act not "Shreveport" them out of regulating local telephone rates. See testimony of Kit F. Clardy, Chairman of the Legislative Committee of NARUC 70-74 (Apr. 11, 1934), Paglin (1989) pp. 416-20, esp. p. 419. Benton, the General Solicitor of NARUC, warned Congress against legislation encroaching on current state regulatory activities:

The statutory limits on federal radio regulation enacted in the Communications Act of 1934 duplicate with further emphasis those in the Radio Act of 1927. The first section of the Communications Act of 1934 established the FCC to regulate "interstate and foreign commerce in communication by wire and radio." Section 2(b) stated:

> *Subject to the provisions of section 301, nothing in this Act shall be construed to apply or to give the Commission jurisdiction with respect to (1) charges, classifications, practices, services, facilities, or regulations for or in connection with intrastate communications service of any carrier…*[1]

Section 301 contained the words of the enumeration of authority in the first paragraph of the Act of 1927, with only small changes concerning aircraft and mobile stations. Section 303 enumerated general powers, with the same qualifying heading as Sec. 4 of the Act of 1927, and §303(f) included the exact text of Sec. 4(f) of the Act of 1927. General powers enumerated in §303 included additional powers not included in Sec. 4 of the Act of 1927, but these additional §303 powers do not relate to the geographic scope of radio regulation. Thus the Communications Act of 1934 provided no more textual clarity about the scope of federal radio regulation than did the Radio Act of 1927, or the Radio Act of 1912.

U.S. courts did little to encourage deliberation about the geographic division of power in radio regulation. As Appendix B discusses, by 1928 several federal district court decisions stated that all radio communications are interstate. These decisions obliterated the distinction between interstate and intrastate radio. Moreover, these decisions ruled that federal regulation of all radio communications is constitutional under the Commerce Clause of the U.S. Constitution. Subsequent decisions of the DC Circuit and the U.S. Supreme Court followed the decisions of the district courts. Widely cited cases in these higher courts seemed to have relied essentially on *dicta* in earlier decisions, and the higher courts provided additional *dicta* on their own initiative. Qualifying language disappeared over time. By the end of World War II, courts seemed reluctant to examine carefully past precedent and the changing nature of radio communications. Moreover, developments in Commerce Clause law suggested that courts could not use that clause to provide judicial review of federal legislation concerning economic matters. Despite the great significance of communications to personal life, public life, and democratic deliberation, federal regulation of radio communication devices has been considered an "economic" or "technical" matter.[2]

---

> *[the paramount concern of State commissions is] that any legislation which may be enacted by Congress by so drawn that State regulation of intrastate communications shall not be broken down or hampered by the Federal law or by the operation of the Federal agency thereunder.*

Benton testimony, in Paglin (1989) p. 481. See also testimony of Andrew R. McDonald, First Vice President and Chairman of the Executive Committee of NARUC 131-4 (May 9, 1934), Paglin (1989) pp. 477-80.

[1] Communications Act of 1934, as enacted, Pub. L. No. 416, c. 652, 48 Stat. 1064 (June 19, 1934). Sec. 3 defined "interstate communication" or "interstate transmission" essentially using the terms and concepts in clauses (a), (b), and (c) of the first paragraph of the Radio Act of 1927.

[2] For analysis of the geographic scope of radio regulation permissible under the Commerce Clause, see Appendix B.

Further legislative activity after World War II seems to have been directed towards limiting the possibility of significant deliberation about the scope of federal radio regulation. In 1968, the Communications Act was amended to include a new §302. Section 302(a) stated:

> *The Commission may, consistent with the public interest, convenience, and necessity, make reasonable regulations governing the interference potential of devices which in their operation are capable of emitting radio frequency energy by radiation, conduction, or other means in sufficient degree to cause harmful interference to radio communications. Such regulations shall be applicable to the manufacture, import, sale, offer for sale, shipment, or use of such devices.*[1]

The text of §302(a) does not clearly indicate whether §302(a) is subject to limits stated elsewhere in the Communications Act. However, §2(b) of the Communications Act makes clear that §302(a) is subject to the provisions of §301 and should not be "construed to apply or to give the Commission jurisdiction with respect to" broad areas of intrastate radio communications.[2]

Legislative history has created some confusion about the statutory scope of radio regulation. With respect to the 1968 Communications Act amendment establishing a new §302, the Senate Report on the enacting bill stated:

> *The Federal Communications Commission presently has authority under section 301 of the Communications Act to prohibit the use of equipment or apparatus which causes interference to radio communications and, under section 303(f), to prescribe regulations to prevent interference between stations. Pursuant to this authority the Commission has established technical standards applicable to the use of various radiation devices. At the outset it should be emphasized, therefore, that this legislation is not primarily designed to empower the Commission to promulgate stricter technical standards with respect to radiation devices but to enable it to make these standards applicable to the manufacturers of such devices.*[3]

The legislative intent apparently was not to expand FCC concerns, but to give the FCC an additional regulatory tool for addressing concerns already within the scope of the Communications Act.

One might further consider, not the scope of the Communications Act, but what legislators thought was the scope of the Communications Act.[4] The 1968 Senate Report on §302 noted:

---

[1] Amendment to the Communications Act of 1934, Pub.L. 90- 379, 82 Stat. 290 (July 5, 1968).

[2] Recently, the U.S. Supreme Court addressed a somewhat related question about FCC jurisdiction. See AT&T Corporation v. Iowa Utilities Board, 119 S.Ct. 721 (1999). The issue here is simpler because the last sentence of §201(b) of the Communications Act does not apply to communication services not provided on a common carrier basis.

[3] 1968 U.S.C.C.A.N 2486, 2487. The Senate Report recommended passage, without amendment, of House bill H.R. 14910. The bill was thus passed and become Pub.L. 90- 379, 82 Stat. 290.

[4] As legislators are well aware, the public meaning of a law's text is not necessarily the same as its meaning to legislators (or, of course, its meaning in effect). Which meaning should be privileged depends on constitutional aspects of communications.

*An important example of interference to radio communications occurred in December 1965 at the time of the Gemini 7 space flight. The U.S. Government went into court and received a temporary restraining order against a manufacturing company in Corpus Christi, Tex., on the grounds that certain equipment at the plant, including the ignition system of a winch truck used for lifting steel, was interfering with the communications between a tracking station at Corpus Christi and the Gemini 7 spacecraft.*[1]

It seems implausible that legislators believed that the Communications Act gave the FCC authority to license the use and operation of winch trucks because they generate radio frequency energy.[2] On the other hand, legislators seem to have believed that the FCC had authority to regulate many types of communication, including local public safety communications as well as extra-earthly communications.[3]

Additional legislative history has been influential, but it does not seem to have much deliberative legitimacy. The Communications Amendment Act of 1982 was a collection of unrelated provisions, one of which amended §302(a) to include after the words "make reasonable regulations" an additional clause:

*(2) establishing minimum performance standards for home electronic equipment to reduce their susceptibility to interference from radio frequency energy.*[4]

This amendment gives the FCC authority over a certain class of equipment that might operate poorly due to reception of radio frequency energy. Of great significance has been the Conference Report's statement associated with that amendment:

*The conference substitute is further intended to clarify the reservation of exclusive jurisdiction to the Federal Communications Commission over matters involving RFI [radio frequency interference]. Such matters shall not be regulated by local or state law, nor shall radio transmitting apparatus be subject to local or state regulation as part of any effort to resolve an RFI complaint. The conferees believe that radio transmitter operators should not be subject to fines, forfeitures or other liability imposed by any local or state authority as a result of interference appearing in home electronic equipment or systems. Rather, the conferees intend that regulation of RFI phenomena shall be imposed only by the Commission.*[5]

This statement has been cited repeatedly, and rather loosely, in FCC orders, FCC letters, court cases, on an influential web site, and in the most comprehensive recent article on jurisdiction in radio regulation.[6] Note, however, that the statement goes far beyond clarifying the text of the statutory amendment associated with it. Moreover, the Conference met, wrote its report, and the report and the law were passed, all in one day.

---

[1] 1968 U.S.C.C.A.N. 2486, 2489.

[2] For an exploration of related issues, see Section IV

[3] Ibid. Cf. Sec. 1 of the Communications Act [47 U.S.C. 15] which present as a regulatory objective "a rapid, efficient, Nation-wide, and world-wide wire and radio communication service."

[4] Communications Amendments Act of 1982, PL 97-259, Sept. 13, 1982, 96 Stat 1087, Sec. 108.

[5] H.R. Conf. Rep. 97-765, 33, 1982 U.S.C.C.A.N. 2261, 2277.

[6] See, for example, FCC (1985) para. 5, FCC (1987) para. 8,, FCC (1990), FCC (1994b), Broyde v. Gotham Tower, 13 F.3d 994, 997 (1994), Southwestern Bell Wireless v. Johnson County Board, 199 F.3d 1185, 1191 (1999), and Freeman v. Burlington Broadcasters, 204 F.3d 311, 320 (2000). See also http://www.arrl.org/FandES/field/regulations/rfi-legal/#pl_97_259, and Brock (1999) p. 21.

The law passed on a voice vote that dispensed with reading the Conference Report.[1] While the Conference Report's statement is more closely related to §302(a)(1), it seems not to provide a correct description of the legislative intent in establishing that provision. In any case, surely legislative history from 1982 is weak evidence for legislative intent in 1968.

A few small edits buried in the middle of the Communications Amendment Act of 1982 made significant changes to statutory language concerning regulatory geography. First, the purpose of federal radio regulation in the introductory clause of §301 was expanded. The phrase "to maintain the control of the United States over *all the channels of interstate and foreign radio transmission*" became "to maintain the control of the United States over *all the channels of radio transmission*" [emphasis added].[2] Second, the clause describing jurisdictionally distinctive places, §301(a), was transformed into a clause describing every place. In particular, "from one place *in any Territory or possession* of the United States or in the District of Columbia to another place *in the same Territory, possession,* or District" became "from one place *in any State, Territory or possession* of the United States or in the District of Columbia to another place *in the same State, Territory, possession,* or District" [emphasis added].[3] These edits thus provided much stronger statutory support for federal regulation of intrastate radio communications.

Many members of Congress may have been unaware of the formal significance of these changes. The amendments were passed as a few lines of edits, not understandable on their own, placed in the middle of a law spanning thirteen pages and a wide variety of concerns.[4] Within the Communications Act as a whole, the edits make §301(d) redundant and heighten the contrast between the introductory clause of §301 and phrases in §1 and §2(a).[5] The Conference Report described these changes as helping to avoid

---

[1] See below.

[2] Change effected through PL 97-259, Sept. 13, 1982, 96 Stat 1087, Sec. 107 (1).

[3] Change effected through ibid., Sec. 107, (2),(3), and (4).

[4] The law is a collection of sections related only by their relevance to the Communications Act. The full text of the section entitled "Jurisdiction of the Commission," which comes in the middle of the law, is:

> *Sec. 107. Section 301 of the Communications Act of 1934 (47 U.S.C. 301) is amended –,*
> *(1) by striking out "interstate and foreign";*
> *(2) by inserting "State," after "any" the third place it appears therein;*
> *(3) by inserting a comma after "Territory" the first place it appears therein; and*
> *(4) by inserting "State," after "same".*

[5] Section 301(d) gives the FCC authority over radio:

> *within any State when the effects of such use extend beyond the borders of said State, or when interference is caused by such use or operation with the transmission of such energy, communications, or signals from within said State to any place beyond its borders, or from any place beyond its borders to any place within said State, or with the transmission or reception of such energy, communications, or signals from and/or to places beyond the borders of said State;*

As edited, §301(a) includes the situations described in §301(d).

Section 1 of the Communications Act describes the purpose of the Act. It begins: "For the purpose of regulating interstate and foreign commerce in communication by wire and radio…." Thus the purpose of the Act describes interstate radio.

Section 2 is titled "Application of Act". Section 2(a) begins:

> *The provisions of this Act shall apply to all interstate and foreign communication by wire or radio and interstate and foreign transmission of energy by radio, which originates and/or is received* wasteful proceedings when the FCC prosecutes Citizens Band radio operators transmitting in violation of FCC rules. The Conference Report also stated that the amendments make §301 "consistent with judicial decisions holding that all radio signals are interstate by their very nature."[1]

Congress itself had little time to ponder the significance of these amendments. The Communications Amendments Act of 1982 was introduced in the Senate on August 18, 1982 as a substitute for all but the title of a much different House bill. There was unanimous consent to dispense with reading the bill, and the Senate passed it straight away.[2] On August 19, the House requested a conference, the conference met, agreed to minor changes in the Senate bill, reported to both chambers, and both chambers agreed to the conference report. In short, legislation significantly changing the statutory basis for regulatory geography was introduced and passed in two days, with no deliberation in the legislature.[3]

Eventually federal legislation preempted even the radio issues that most interested state regulators. The federal statutory basis for authority over intrastate radio services was strengthened in 1982, as described above. In 1983, cellular telephony was offered to customers in Chicago. Over the next ten years, some states regulated some cellular phone rates. Then the Omnibus Budget Reconciliation Act of 1993 preempted, with little fanfare, state regulation of rates and entry for commercial mobile services.[4] Preempting state regulation of rates and entry for commercial mobile radio may have been a sound regulatory choice. The point is that it was relatively easy to do. State regulators in 1934 did not want to surrender their voice about governance of radio communications. But over time the natural functioning of the national political process seems to have foreclosed much needed deliberation about regulatory geography.

The tension between real needs and the deliberative status of federal radio regulation is evident in recent legislation. On November 22, 2000, a federal law authorized state and local governments to enact laws prohibiting violations of FCC rules governing interference from Citizens Band radio. The authorization was carefully limited to specific FCC regulations pertaining to Citizens Band radio. Services that the FCC licenses under §301 were explicitly privileged against sub-national regulation. The FCC was authorized to hear appeals of sub-national government's actions. In addition, the law declared:

---

> *within the United States, and to all persons engaged within the United States in such communication or such transmission of energy by radio, and to the licensing and regulating of all radio stations as hereinafter provided;*

Thus § 2(a) describes the application of the Act in terms of interstate radio and subsequently described rules for radio stations.

[1] H.R. Conf. Rep. No. 765, 97'th Cong., 2nd Sess., printed in Cong. Record, vol. 128, part 16 at 22132 (Aug. 19, 1982). The judicial citation is "See, e.g., Fisher's Blend Station Inc. v. Tax Commission of Washington State, 297 U.S. 650, 655 (1936)." For discussion of this case and relevant U.S. case history, see Appendix B.

[2] See Cong. Record, vol. 128, part 16 at 21826 (Aug. 18, 1982).

[3] The bill was presented to the President Ronald Reagan on September 2. He signed it on September 13, 1982. Reagan tends to be thought of as a president who sought to restrain the (federal) government.

[4] PL 103-66 (Aug. 10, 1993), 107 Stat 312.

*Nothing in this subsection shall be construed to diminish or otherwise affect the jurisdiction of the Commission under this section over devices capable of interfering with radio communication.*[1]

Overall, the legislation illustrates the practical importance of sub-national regulation. It also shows the national political concern that such regulation not have any legal significance for federal jurisdiction.

Concern over a possible reduction in federal jurisdiction largely shaped the legislative process. On Aug. 2, 1996, a bill was proposed in the Senate to give sub-national governments police powers to resolve interference relating to CB radio. The bill gave FCC concurrent jurisdiction over such issues and explicitly reserved the FCC's exclusive jurisdiction over radio interference falling outside the scope of the bill.[2] That bill was redrafted to retain and emphasize FCC authority over all radio interference. The "non diminish" clause quoted above was added. This new bill was proposed in the Senate on Apr. 17, 1997.[3] A bill introduced in the House on June 24, 1999, was similar to the Senate bill from 1997. The House bill included additional minor edits that emphasized FCC authority.[4] It also included a new sub-section requiring "probable cause" in state or local enforcement action against Citizens Band radio equipment aboard commercial motor vehicles.[5] The bill that finally passed the Senate (Oct. 31, 2000) and the House (Nov. 13, 2000) included further minor edits that again emphasized FCC authority in regulating radio interference.[6] Thus the national political process produced more than four years of deliberation about a possible, small reduction in federal jurisdiction over a particular, relatively unimportant radio use.

---

[1] State and Local Enforcement of Federal Communications Commission Regulations on Use of Citizens Band Radio Equipment, PL 106-521 (HR 2346) Nov. 22, 2000, 114 Stat 2438. The quotation is Sec. 1(6).

[2] The bill was S. 2025. See 142 Cong. Rec. S9555-02.

[3] The bill was S. 608. See 143 Cong. Rec. S3349-02.

[4] The bill was HR 2346; see 1999 Cong U.S. HR 2346, 106th Congress, 1st Sess. Sec. 302(f)(2) was extended to require that a state or local government statute or ordinance identify that radio stations licensed by the FCC "pursuant to section 301 ((47 USCA 301)) in any radio service for the operation at issue" are exempt from the state or local law. The FCC licenses CB radios on a class basis, rather than for particular operations. In addition, in Sec. 302(f)(5), the phrase "The enforcement of a regulation by a State or local government" was changed to "The enforcement of a statute or ordinance that prohibits a violation of a regulation by a State or local government". The latter awkward phrase does not seem intended to address state and local governments violating regulations. In light of the legislative history, the change is best interpreted to emphasize that the FCC writes the regulations, and state and local governments are only to pass laws to enforce those FCC regulations.

[5] See ibid, Sec. 302(f)(7).

[6] The bill became law as PL 106-521, 114 Stat 2438 (Nov. 22, 2000). Relative to HR 2346 as introduced in the House on June 24, 1999, PL 106-521 has a few minor changes. The heading "Section 1. Enforcement of Regulations Regarding Citizens Band Radio Equipment" was changed to "Section 1. State and Local Enforcement of Federal Communications Commission Regulations on Use of Citizens Band Radio Equipment." In reference to enforcement in Sec. 302(f)(4)(A) and (D), "State or local government" was changed to "State or local government agency". Other minor edits reduced FCC responsibilities to state and local governments. Reference to FCC technical guidance in defining "probable cause" was eliminated (Sec. 302(f)(7)). FCC responsibility to provide State and local governments with technical guidance concerning detecting and determining violations was qualified with the phrase "to the extent practicable" (Sec. 302(f)(3)).

U.S. experience highlights significant deliberative failure in the national political process. The weak policy, statutory, and constitutional basis for federal control over all radio use has not been considered in an open, substantive way. Regulatory geography for AM radio broadcasting in the late 1920s and early 1930s probably didn't matter much relative to the political, economic, and social problems of that time. But radio communications is much more important to life in the 21'st century. Getting better regulation requires seeking truth and sincerely evaluating current beliefs.[1] With respect to regulatory geography, U.S. experience thus far shows little evidence of these crucial aspects of policy deliberation.

**D. Not Whether But Where to Set Boundaries**

Private, area-based regulation of radio rights provides another perspective on possibilities for regulatory geography. Both Australia and the U.S. have auctioned rights to regulate privately radio use in areas defined by federal (public) regulation. Study of the boundaries defined for these auctions provides considerable insight into the actual considerations that have determined regulatory geography. The possibilities and benefits that private area-based regulations highlight are possibilities and benefits that should also be considered in determining the geography of public regulation.

To define area-based licenses, the Australian Communications Authority (ACA), a federal agency, delineated the Australian Spectrum Map Grid. This grid consists of 21,998 "squares" of the following sizes:[2]

> *...5 minutes of arc (approximately 9 kilometres) on the eastern seaboard and in Adelaide, Perth and Darwin, 1 degree of arc (approximately 100 kilometres) in regional Australia and 3 degrees of arc (approximately 400 kilometres in remote Australia.*[3]

The Australian Spectrum Map Grid is a geographic formalism that separates boundaries in radio regulation from all other geographic boundaries. Along with the specification of minimum bandwidth within a particular frequency range, this grid defines standard trading units (STUs). Trading in radio rights can occur only in whole STUs. Private regulation of radio rights within STUs is subject to publicly regulated STU boundary conditions where different right holders share a common STU boundary. These conditions include limits on out-of-area power emissions.[4] The parameters of these regulations are uniform throughout Australia for a given frequency band or service type.

---

[1] Seeking truth and sincerely evaluating current beliefs are not sufficient to produce good regulatory performance. Other activities. such as understanding different views, encouraging deliberation, and making persuasive arguments, also matter.

[2] The earth is not flat. It's not a sphere, either. It's a slightly lumpy, somewhat squashed spheroid. Grid cells are defined with reference to the Australian National Spheroid. See http://www.environment.sa.gov.au/mapland/sicom/sicom/tp_scs.html

[3]The ACA has used this grid for all frequency bands. See the Australian Communications Authority, "Introduction to Spectrum Licensing," online at http://www.aca.gov.au/licence/spectrum/index.htm

[4] See, for example, the core conditions defined in ACA (1996).

With some regard for geographic particularities, the ACA chooses collections of STUs as radio licenses to be auctioned. Consider the 500 MHz band, the band in which mobile voice services were first offered in Australia. To define new licenses in this band, the ACA considered "a population density model, the digital elevation model (RadDEM), existing radio sites and propagation models of typical transmitters operating from those sites."[1] The ACA defined licenses about the Australian capital (Canberra) and about the capitals of the seven other states and territories. The license covering Sidney (capital of New South Wales) also extended to the city of Wollongong, the license covering Hobart (capital of Tasmania) included the rest of Tasmania as well, and the license for Perth (capital of Western Australia) included all of a region in the south west of that state. Licenses were also defined about the city of Townsville, on the coast of the state of Queensland, and about the city of Newcastle, on the coast of New South Wales. Additional licenses were defined for the Northern Rivers region in New South Wales, the Central Western region of New South Wales, a coastal region of Victoria, and the Pilbara region of Western Australia.[2]

The geography for the 2 GHz band was somewhat different. This band is associated with advanced mobile services. A working paper considering licenses for this band focused on cities, their population, and the geography of city central business districts.[3] The licenses actually auctioned included licenses defined about Canberra and about the capitals of the seven other states and territories. The areas included in these licenses were generally smaller than the areas included in the corresponding licenses in the 500 MHz band. In addition, the ACA defined licenses about three cities in the state of Queensland, three cities in the state of New South Wales, and regions in the states of Victoria, Tasmania, South Australia, and Western Australia. The latter regions excluded the areas defined for the state capitals, did not include all remaining area in the corresponding state, and included some areas in other states.[4]

The different geographies of the 500 MHz and 2 GHz licenses show that some location specific factors have been considered in choosing license boundaries. The boundaries the ACA chose about Perth and about Sidney differ significantly between the 500 MHz band and the 2 GHz band. Other license areas about cities also have different shapes, as do the set of cities that have city-area licenses. The wide area licenses and the over-all area covered by licenses also differ.[5] However, the underlying Australian Spectrum Map Grid, which only crudely incorporates economic and population geography, is the same for all bands. It limits the scale at which location-specific factors can affect license geography.

---

[1] Hayne (1997) p. 189.

[2] For geographic descriptions of radio licenses in the 500 MHz band, see ACA (1996).

[3] RCC (1999), esp. Appendix 7.

[4] For geographic descriptions of radio licenses in the 2 GHz band, see ACA (2000).

[5] Fixed point-to-point radio communications fit poorly within a geographic framework for licensing. Incumbent services of this type in particular areas have limited geographic licensing. More generally, existing radio uses have made licensing slower, more complicated, and more incomplete than originally anticipated. See Productivity Commission (2002), Chapter 11.

To define sub-national licenses, the FCC uses geographies based on independent socio-economic analysis. Table 7 shows the geographies the FCC has used for area licenses. Metropolitan Statistical Areas (MSAs) are a well-established geography that the Office of Management and Budget, a separate federal executive agency, defines. MSAs, which cover the main urban areas of the US, are used in national censuses and government statistical studies.[1] RSAs were defined by the FCC to form with MSAs a complete geography for the U.S.[2] The Basic Trading Area (BTA) and Major Trading Area (MTA) geographies are from Rand McNally's *Commercial Atlas and Marketing Guide*.[3] They are based on flows of commerce. Economic Areas (EAs) are defined by the Bureau of Economic Affairs in the Department of Commerce. EA geography is based primarily on commuting patterns. EAs are defined to encompass, as much as possible, the workplace and residence of their populations.[4] As Table 7 shows, the FCC has defined additional aggregations of these geographies. The FCC has also added some elements to these geographies to cover distinctive areas.[5]

**Table 7**
**FCC License Geographies**

| Geography | Creator | Services/Auctions | Sub-units | Number of Areas |
|---|---|---|---|---|
| MSA/RSA | OMB/FCC | Cellular, paging, 2,12 | counties | 733 |
| BTA | Rand McNally | 5,6,10,11,17,22,23,35 | counties | 493 |
| EA | BEA | 16,18,21,24,30,34,36, 39,40,42,43 | counties | 175 |
| MTA | Rand McNally | 4,7,41 | BTA | 51 |
| MEA | FCC | 14,26,33,38,40 | EA | 51 |
| VPC | FCC | 20,39 | EA | 42 |
| REAG | FCC | 14 | MEA | 12 |
| EAG | FCC | 18 | EA | 6 |
| PCS Regions | FCC | 3 | MTA | 5 |

Notes: See Appendix A.

FCC decisions provide few facts and no compelling, consistent arguments for choices among MSAs/RSAs, BTAs, MTAs, BEAs, MEAs and other geographies. For example, the BTA geography was chosen for one range of PCS frequencies and the MTA geography for another range of PCS frequencies. In the four paragraphs explaining its decision, the FCC noted that MSAs/RSAs would fragment "natural markets," pointed to benefits of large service areas, and declared "…we believe that a combination of both

[1] For an interesting discussion of metropolitan area definitions and further references, see DOC (2001).
[2] See Cellular Market Areas at http://www.fcc.gov/oet/info/maps/areas/
[3] McNally (1992).
[4] See DOC (2001) and Johnson (1995).
[5] For details and citations, see http://www.fcc.gov/oet/info/maps/areas/

MTA and BTA service areas would maximize the benefits of having both large and small service areas." EAs and LATAs, other geographies offered by commenters, were not evaluated at all.[1] In 1995, the FCC changed its previous conclusion that MTAs were the best geography for 800 MHz SMR licenses. It chose instead EAs, arguing that these smaller areas would result in "a more diverse group of prospective bidders" and that EAs "reflect the actual coverage provided by 800 MHz SMR systems more accurately than MTAs…." BTAs were rejected because they might not "be sufficiently large to create a viable wide-area service."[2] Overall, the FCC has treated choices among geographies as a relatively unimportant licensing convention.

In the US, license boundaries generally have significance outside of radio regulation. All the geographies used in systematic licensing plans (Table 7) are partitions of counties. That means the boundaries of their elements are county boundaries. County boundaries within states are set under the authority of state legislatures. All state boundaries are also county boundaries. County boundaries thus include boundaries with deep legal, political, economic, and physical significance. Thus license boundaries in the US, at least initially, are not merely a convention of radio regulation.

The FCC slowly gave licensees the right to partition geographically their licenses and transfer parts of licenses to other parties. Allowing partitioning does not justify the FCC's meager consideration of geography. License partitioning has occurred infrequently. Moreover, the FCC's partitioning rules allow partitioning independent of county boundaries.[3] This means that the partitioning rules do not formally recognize the public significance of any (sub-national) boundaries. Thus the real significance of county boundaries in U.S. radio regulation does not reflect a well-considered regulatory choice.

Well-established political boundaries should be taken more seriously in radio regulation. Both Australia and the U.S. have a federal system of government with a written constitution that distributes power between the federal and state governments. The Australian constitution gives the Australian federal government enumerated powers that include the power to make laws with respect to "trade and commerce with other countries, and among the States," "postal, telegraphic, telephonic, and other like services," "astronomical and meteorological observations," and "foreign corporations, and trading or financial corporations formed within the limits of the Commonwealth."[4] The U.S. constitution gives the U.S. federal government enumerated powers that include the power "To regulate Commerce with foreign Nations, and among the several States, and with the Indian Tribes."[5] In Australia and the US, all radio uses within a state are not clearly subject to federal power under the constitution.[1] What is clear is that state boundaries are highly significant boundaries in these federations.

[1] FCC (1993) pp. 7729-34 (explanation in four paragraphs). For another example of mixing geographies, but with larger elements (MEAs and REAGs), see FCC (1997) pp. 10813-6.
[2] FCC (1995b) pp. 1483-4 (explanation in two paragraphs).
[3] FCC (1996), pp 21847-8.
[4] See Sec. 51(i.), (v.), (viii.), and (xx.) of the Australian Constitution, online at http://www.aph.gov.au/senate/general/constitution/
[5] See Article I, Sec. 8, Clause 3 of the U.S. Constitution, online at http://www.house.gov/Constitution/Constitution.html

Many radio license boundaries are wholly within important political boundaries. In Australia, the licenses about Perth and Sidney in both the 500 MHz band and the 2 GHz band are well within the state boundaries of Western Australia and New South Wales. More generally, the Australian Spectrum Map Grid defines 21,988 STUs.[2] Most of Australia is covered by six states and two territories. Thus most STUs must fall within a single Australian state or territory. For the U.S., Table 8 shows the extent to which different license areas cross state boundaries. Although area boundaries were defined with no respect for state boundaries, more than 75% of the areas in the MSA/RSA and BTA geographies do not cross state boundaries.

**Table 8**
**FCC License Geographies in Relation to State Geography**

| Areas with parts in | MSA/RSA | BTA | EA | MTA | MEA |
|---|---|---|---|---|---|
| 1 state | 697 | 377 | 78 | 9 | 9 |
| 2 states | 31 | 97 | 66 | 16 | 13 |
| 3 states | 5 | 16 | 23 | 9 | 7 |
| 4 or more states | 0 | 3 | 8 | 17 | 22 |
| total areas | 733 | 493 | 175 | 51 | 51 |
| % areas within 1 state | 95% | 76% | 45% | 18% | 18% |
| % areas within 3 or fewer states | 100% | 99% | 95% | 67% | 57% |

In Australia, the US, and elsewhere, significant governance capabilities exist outside of federal governments. State, regional, and city governments address broad areas of public welfare and safety, including land use, local public infrastructure, education, and crime.[3] Federal area-based radio licenses give private organizations responsibility to manage interference within their license areas and with respect to federally defined boundary conditions. Given the chance, state and local public governance could do likewise. Sub-national public governance institutions could help radio regulation respond to the particular conditions within sub-national political boundaries.[4]

---

[1] The High Court of Australia and the Supreme Court of the United States have power to decide whether particular acts of their respective federal legislatures are constitutional. For discussion of the federal-state balance in U.S. radio law, see Appendix B.

[2] Hayne (1997) p. 183.

[3] In Australia and the US, state governments hold residual powers under the federal constitution and are subject to enumerated limits on their power under state constitutions.

[4] This point relates to both administrative and judicial decisions. Legal cases concerning ownership, nuisance, and interference are typically heard in sub-national court systems. This is the type of law relevant to adjudicating radio rights.

The geographic scope of corporations using radio is not a good policy reason for ignoring sub-national political boundaries in radio regulation. Radio technology has an increasingly wide range of applications. National firms are no more likely to dominate radio use than to dominate commercial real estate. Moreover, promoting the development of small and medium-sized enterprises is widely seen as a key to job creation, innovation, and social and economic dynamism. More divided political power in radio regulation might lessen the influence of large corporations and create more opportunities for small and medium-sized enterprises with strong local ties and local knowledge.

Auction results in the U.S. show that, in appropriate circumstances, a large number of companies are interested in radio licenses within only a few states. Table 9 shows the geographic scope of winning bidders' licenses in the four auctions based on BTA geography. About two-thirds or more of winning bidders acquired licenses in three or fewer states. A highly competitive, global industry produces a wide range of radio devices. Companies do not need to manufacture their own radio devices in order to provide customized radio services. Radio services are services. Service companies do not have to provide services nationally in order to have a viable business.[1]

**Table 9**
**State Geography of License Buyers**
(from auctions using BTA geography)

| Buyers with licenses in: | Auction 5 | Auction 6 | Auction 11 | Auction 17 |
|---|---|---|---|---|
| 1 state | 36 | 27 | 40 | 46 |
| 2 states | 15 | 10 | 25 | 12 |
| 3 states | 16 | 7 | 12 | 19 |
| 4 states | 4 | 4 | 15 | 6 |
| 5 states | 5 | 3 | 5 | 4 |
| >5 states | 12 | 15 | 25 | 15 |
| Total buyers | 88 | 66 | 122 | 102 |
| % in 5 or fewer states | 86% | 77% | 80% | 85% |
| % in 3 or fewer states | 76% | 67% | 63% | 75% |
| % in only 1 state | 41% | 41% | 33% | 45% |

Source: Calculated from FCC ULS data.

Localism in radio regulation could foster the development of radio services better adapted to local conditions and needs. In considering television license allocations in 1952, the FCC stated:

[1] Internet service providers in the U.S. are a good example of small but vibrant communications service companies.

*In the Commission's view as many communities as possible should have the opportunity of enjoying the advantages that derive from having local outlets that will be responsive to local needs.*[1]

Localism then was understood as standard radio technology delivering localized programming. Centrally planned radio regulation was widely considered to be the best, or only, way to organize the necessary infrastructure for localized programming. The situation today is rather different. Both radio technology and radio regulation now offer much greater possibilities and much greater uncertainty. Consider, for example, that the U.S. Congress recently passed a law authorizing twelve specific low-power television stations across the U.S. to provide digital data services as a pilot project to explore providing Internet access to unserved areas.[2] Such experimentation might be more widespread if it could be enabled at a local level. Area-based radio licenses demonstrate that boundaries in radio regulation can be drawn locally. To encourage the development of new radio technology and to promote beneficial competition among jurisdictions, radio regulation should recognize political geography.

[1] FCC (1952) p. 624.

[2] See Consolidated Appropriations–FY 2001, PL 106-554, Dec. 21, 2000, 114 Stat 2763. AccelerNet, a Houston-based broadband provider founded in 1995, owns or has an interest in ten of the stations benefiting from PL 106-554. See http://www.accelernet.net/corp/PressRelease012901.html

## IV. Personal Freedom and Licensing

Whether persons need to be licensed to use radio has often been considered with misplaced emphasis. Consider this recommendation from the European Radiocommunications Committee (ERC) on exemption from individual licensing:

> *Licensing is an appropriate tool for Administrations to regulate the use of radio equipment and the efficient use of the frequency spectrum. However, the technical characteristics of radio equipment require less intervention from the Administrations as far as the installation and use of equipment is concerned. Administrations and especially users, retailers and manufacturers will benefit from a more deregulated system of authorising the use of radio equipment.*
>
> *There is general agreement that when the efficient use of the frequency spectrum is not at risk and as long as harmful interference is unlikely, the installation and use of radio equipment can be exempt from a licence. ...*
>
> *When radio equipment is subject to an exemption from individual licensing, anyone can buy, install, possess and use the radio equipment without prior permission from the Administration. Furthermore, the Administration will not register the individual equipment. The use of the equipment can be subject to general provisions.*[1]

The recommendation first describes licensing as a tool for regulators (Administrations), one that might be used less while still meeting regulatory objectives. Further on, the recommendation describes consequences for personal freedom in a paragraph whose main subject is radio equipment. The recommendation, read narrowly, points in a sensible policy direction. But this recommendation, and radio regulation more generally, needs to better appreciate physical truth and human freedom.

Electromagnetic radiation is a fundamental aspect of the physical world and human freedom. All bodies with temperature above absolute zero radiate some energy across the whole electromagnetic spectrum.[2] When human beings use fire to keep warm, to dispel darkness, or to communicate, they use electromagnetic spectrum. Putting on a wool sweater or working across a thick rug can generate electromagnetic waves, as can a variety of other human activities. The freedom to be warm, to shine light, or to create

---

[1] ERC (1995) p. 1.

[2] Thermal radiation is proportional to temperature in Kelvin and the bandwidth measured. At typical room temperatures, human beings are net electromagnetic radiators. A rough estimate of human power is straight-forward. Approximate a human being as a sphere of water with 1 meter diameter. Since water is close to a black body, the human body is essentially a black body. With these approximations, the Stefan-Boltzmann law then implies that a non-feverish human being sitting in a typical office is equivalent to a (point) radiator with net output about 260 Watts. Most of this power is radiated in a way invisible to supervisory personnel, i.e. the power is radiated in the infrared part of the electromagnetic spectrum. My skills as an electrical engineer have faded badly; you should verify this calculation yourself. All necessary information is readily available. See, for example, http://www.eng.auburn.edu/~wfgale/usda_course/section1_4_page2.htm

electric sparks in the world is not contingent on the efficient use of electromagnetic spectrum.

The definition of radio communications subject to regulation must be understood as subordinate to the understanding of human freedom. While the first two international radiotelegraph conventions defined radio communications implicitly as wireless telegraphy, the U.S. proposal to the International Radiotelegraph Conference in Washington in 1927 took a different approach. The proposal defined "radio communications":

> *The term "radio communication" as used in this convention means the transmission of intelligence, photographs, reproductions, or any other subject matter, without connecting wires, by radiated electromagnetic energy.*[1]

Light is a form of radiated electromagnetic energy. Thus the plain language of the above definition covers all forms of light-mediated communication – reading texts; observed motion, form, and gesture; and explicit signaling with light, such as hanging lanterns in a church to convey signals of public importance. If this proposal had not been limited *de facto* by an understanding of freedom, it would have represented a more dramatic denial of human rights than the most oppressive governments in the world have ever proposed.

Definitions of "radio communications" adopted in international radio regulations made the controlling position of human freedom less obvious but no less important. The Washington Conference of 1927 adopted this definition of "radio communications":

> *The term "radio communication," applies to the transmission by radio of writing, signs, signals, pictures, and sounds of all kinds by means of Hertzian waves.*[2]

The convention did not define "radio" or "Hertzian waves."[3] In the Madrid Conference of 1932, the relevant definitions in international radio regulations became:

> *Radio communication: Any telecommunication by means of Hertzian waves.*
>
> *Telecommunication: Any telegraph or telephone communication of signs, signals, writings, images, and sounds of any nature, by wire, radio, or other systems or processes of electric or visual (semaphore) signaling.*[4]

Again there was no definition of "radio" or "Hertzian waves," but some types of light-mediated communication clearly fell within the definition of telecommunication. The Atlantic City Conference of 1947 added a definition of Hertzian waves:

> *Hertzian Waves: Electromagnetic waves of frequencies between 10 kc/s and 3 000 000 Mc/s.*[5]

---

[1] U.S. (1927) Article 2, Section 1. The lowest frequency category in the proposed frequency allocation was "10 to 75" kilocycles (KHz); the highest "18,100 and above" KHz. See Table 1, p. 589.

[2] RR (1927), Article Zero. Article 5 defined frequency categories ranging from "10 - 100" KHz to "Above 60,000" KHz.

[3] These formal weaknesses are also strengths. The adopted definition of "radio communication" provided more interpretative space in the legal field for incorporating common understandings of human freedom into the legal definition of radio communication.

[4] RR (1932) Annex to Convention (p. 85). .

[5] RR (1947) Chapter 1, article 1, section I, paragraphs 2,4, and 5. The lowest frequency category in the Table of Frequency Allocations was "10-14" KHz, and the highest "Above 10 500" MHz. See Chapter III,

This definition of Hertzian waves excludes the electromagnetic waves that the human eye normally processes (electromagnetic waves with frequencies between 810 terahertz and 1620 terahertz, i.e. light). By 1959 international radio regulations had equated radio waves and Hertzian waves, and defined both together with reference to a specific medium of propagation. The definition also included additional frequencies:

> *Radio Waves (or Hertzian Waves): Electromagnetic waves of frequencies lower than 3 000 Gc/s [3 000 000 Mc/s], propagated in space without artificial guide.*[1]

The Geneva Conference of 1979 modified the definition to recognize its arbitrariness:

> *Radio Waves or Hertzian Waves. Electromagnetic waves of frequencies* ***arbitrarily*** *[emphasis added] lower than 3 000 GHz, propagated in space without artificial guide.*[2]

This definition is currently embedded in the national law of many countries, including U.S. administrative law.[3] The incoherence, instability, and arbitrariness of the definition of radio communications have attracted little attention. That indicates confidence that radio regulation will defer to established understandings of human freedom.

Such confidence has some factual support. Consider some aspects of U.S. experience. While it is common knowledge that high-power electric lines can cause interference to radio and television signals, the FCC only regulates limited aspects of electric utilities.[4] A single automobile with its engine running creates at 10 meters' distance a field strength greater than the limit that defines auctioned spectrum boundary rights in the frequency range 700 and 800 MHz.[5] Nonetheless, the FCC does not regulate automobiles. A U.S. government study showed that in worst cases windmills produce objectionable distortions

---

art. 5. The nomenclature of frequencies went from "VLF (Very Low Frequency) : Below 30 kc/s" to "EHF (Extremely High Frequency) : 30 000 to 300 000 Mc/s". See Chapter 2, art. 2.

[1] RR (1959), Art. 1, Sect. 1, para. 7. The nomenclature of frequencies was revised, and the new system included a new category "300 to 3 000 Gc/s (GHz) or 3 Tc/s (THz) : Decimillimetric waves". See Art. 2, Sect. III (112).

[2] Final Acts of the World Administrative Radio Conference, Geneva, 1979 at 32 (1980).

[3] For the US, see 47 CFR § 2.1(c). For industrial, scientific, and medical equipment, 47 CFR §18.107(a) defines:

> *Radio frequency (RF) energy. Electromagnetic energy at any frequency in the radio spectrum from 9 kHz to 3 THz (3,000 GHz).*

The FCC's constituting statute defines "radio communications" in a way that does not explicate the meaning of "radio". Specifically, 47 USC 153(33) defines "radio communication" as "transmission by radio of writing, signs, signals, pictures, and sounds of all kinds," as well as associated services and facilities.

[4] On common knowledge of power lines causing radio and television interference, see Southern Indiana Gas and Electric Company v. Gerhardt, 241 Ind. 389, 172 N.E.2d 204 (1961). State courts have exercised jurisdiction over disputes concerning electromagnetic radiation resulting from power lines. See Central Kansas Telephone Co. v. Kansas Corporation Commission, 113 P.2d 159 (1941); Floyd v. Ocmulgee Electric Membership Corp., 16 S.E.2d 208 (1941); and Hale v. Farmers Electric Membership Corp., 99 P.2d 454 (1940). On FCC regulation of electromagnetic waves over power lines, see FCC (1999a). The FCC regulates some aspects of some types of lights. See FCC (1999b). New Zealand's Radio Spectrum Management Group recently found that an electric power crisis that led to serious power outages had a dramatically beneficial effect in reducing noise in frequency bands used for wireless services. The study did not consider the value of activities that create electromagnetic waves. See RSMG (1998).

[5] Compare Skomal (1978) Fig. 2.3, p. 27 to 47 CFR § 27.55(b).

of TV reception within a few miles' distance.[1] Yet the FCC never tilted its regulatory power toward windmills. Free space optical communications systems are now being implemented for terrestrial services and for communications among satellites.[2] The FCC has chosen not to regulate these systems, and generally does not regulate light or the generation of light.[3] The FCC has provided a comprehensive regulatory scheme for radio use. But in important, practical ways the meaning of those words has been subordinate to ideas of human freedom, understood with respect to true descriptions of persons and the world.[4]

Distracted by the many pressing tasks of the day, regulators and other human beings can fail to act in accordance with what is most important. Promoting every person's freedom to be who she or he truly is, in the world as it really exists, surely is one of the most important objectives of communications policy. What freedom means should not be taken for granted.[5] The freedom that comes from being alone, lost in a cave, facing certain death in a few days, differs from the freedom that comes from an ever-loving wife who has a secure, well-paying job.[6] The physical freedom of a drunk differs from that of a well-trained actor, and the intellectual freedom of the learned is not the same as that of the innocent. One must confront real, historical experience of freedom and reflect on what one finds, on what one knows to be true, and on how to address any contradiction between the two – you, now, here, in this field.

---

[1] Senior, Sengupta, and Ferris (1977) p. 11. Recently, The FCC's Technical Advisory Council (TAC) was asked to assess the current state of knowledge on electromagnetic noise levels. See http://www.fcc.gov/oet/tac/requests.pdf, p. 2. The TAC issued its final report on this issue on June 12, 2002. See http://www.fcc.gov/oet/tac/.

[2] See, for example, the offerings of Terabeam (http://www.terabeam.com). Optical systems are being implemented for inter-satellite links. See FCC (2001b).

[3] In considering optical inter-satellite links, FCC (2001b) para. 16, states:

> *Optical beam communications are not considered a type of radio communication since they operate at frequencies above 300 GHz, and they are not within the jurisdiction of the Communications Act* [footnote to 47 USC §§ 152, 153(33)].

The relationship between this finding and the definition of radio waves is unclear. See 47 CFR § 2.1(c). For an FCC concern with lights, see FCC (1999b). Light interference (light pollution) is a major public concern. See, for example, material available from the International Dark-Sky Association (http://www.darksky.org) and the New England Light Pollution Advisory Group (http://cfa-www.harvard.edu/cfa/ps/nelpag.html). In 1999, Texas Governor George W. Bush signed a law to reduce light pollution. See discussion and links at http://www.limber.org/ida/lp.html.

[4] An FCC staff member with over twenty years of service fondly remembers being shown, as part of his new employee orientation, the thriller, "Patrolling the Either" (MGM, 1944) The heroes in this movie are agents in the FCC's Radio Intelligence Division, which sought to uncover domestic spy transmitters in World War II. This movie was the first made-for-television drama and was shown simultaneously on three television stations in April 10, 1944. The FCC has long recognized the importance of defending freedom. The challenge is to do that most effectively for everyone in current circumstances.

[5] Medieval Jewish and Christian thinkers struggled extensively with the issue of freedom, while, for complicated reasons, most Islamic thinkers did not. See Fradkin (2002). Modern thinkers, both within Islam and outside of it, should not presume that similar thinking about freedom is not relevant to their lives and not worthy of their attention.

[6] The latter insight seems to be largely lacking in academic literature on political philosophy and critical theory. A jazz musician shared it with me in a bar in Boston.

## A. Different Kinds of Freedom: Hams, Hackers, and Yackers

Amateur radio, the Internet, and commercial wireless services are fields dominated by radically different understandings of personal freedom. Hams, hackers, and yackers, the most distinctive characters in those respective fields, share common biological, social, and cultural features. They are members of the same species, they may be neighbors, or even the same person engaged in different activities in different times and places. In each of these fields, person-to-person communication for non-commercial purposes comprises an important part of activity in the field. Each of these fields involves similar, general-purpose information and communications technologies. This technology is amenable to personalization, diversity, and decentralized evolution fueled by user creativity. It is also amenable to abusive use, interference, and breakdowns in operating standards and cooperation. This section will show that, despite their similar characteristics and possibilities, the fields of amateur radio, the Internet, and commercial wireless services have incarnated much different ideas of freedom.

The facts about these fields should serve as a policy warning. The range of real possibilities in regulation is enormous. Yet freedom that has merely a conventional meaning is not a worthy guide to policy. Deliberation about freedom needs to evolve as a working consensus in tension with a search for truth about what persons do and the way the world is. This section provides material for such deliberation.

### 1. Amateur Radio

Amateur radio is a field that grew with the development of radio technology early in the twentieth century. Amateur radio users are called amateurs or "hams."[1] *Now You're Talking*, an amateur radio association's guidebook for aspiring hams, describes amateur radio as follows:

> *Ham radio offers so much variety, it would be hard to describe all its activities in a book twice this size! Most of all, ham radio gives you a chance to meet other people who like to communicate. That's the one thing all hams have in common. You can communicate with other hams on a simple hand-held radio that fits in your pocket.*[2]

Public service is also an important part of hams' self-understanding. Hams have long provided emergency communications for local community events and in response to natural disasters. Hams have also contributed significantly to advancing radio technology, such as pioneering early high-frequency communications and popularizing packet radio. Some hams currently communicate via Morse Code and single-sideband voice communications, methods of communications that have been in use for over half a century. Other hams explore digital signal processing, software-defined radios, moon-bounce communications, and communications using specially designed earth-orbiting satellites. The three-million hams worldwide form a community with a strong sense of tradition and identity. Amateur radio has created for hams opportunities for social interaction, for serving the public, and for exercising engineering creativity. Hams' appreciation for the freedom they have found in amateur radio, and their dedication to preserving it, cannot be doubted.

---

[1] The name "amateur" radio for independent, non-commercial radio users probably evolved from an implicit comparison between these users and army and navy radio operators. Independent, non-commercial radio users also tended to be called "boys." Anecdotal evidence suggests that most "amateurs" were young and male, but the average age or sex composition of early amateurs has not been well-established. For a wonderful, personal account of a "girl amateur," see Radio Amateur News (1920).

[2] ARRL (2000) p. 1.

To a remarkable degree, hams understand their activities to be dependent on government license. This view goes all the way back to the beginning of radio. The development of radio implicitly raised the question of whether persons have some natural rights to communicate by radio. Most governments, and amateurs, seemed to assume that persons do not. Nonetheless, absent effective means of suppression, amateur radio developed naturally, through personal curiosity and creativity. About 1912, there were roughly 8000 amateur radio users and 230 amateur radio clubs in the U.S.[1] The U.S. Radio Act of 1912 was hailed as a great victory for amateurs. An amateur activist/magazine publisher declared, "The amateur had at last come into his own…. Uncle Sam has set his seal of approval upon the amateur's wireless...", and of course, "the entire credit for obtaining the amateur's rights belongs to [the author]…"[2] Here is what the Act said regarding amateurs' rights:

> *No private or commercial station not engaged in the transaction of bona fide commercial business by radio communication or in experimentation in connection with the development and manufacture of radio apparatus for commercial purposes shall use a transmitting wave length exceeding two hundred meters, or transformer input exceeding one kilowatt, except by special authority of the Secretary of Commerce and Labor…*[3]

Ten years later a historian asked, without even a whiff of self-consciousness, "What has the amateur done in the past ten years, to justify the privileges granted him by his government?"[4]

Over time amateurs' rights have been further elaborated through national and international regulations. In these regulations, amateurs' rights to communicate internationally depend not just on the amateur's government but on mutual international agreement. International radio regulations require governments not to permit amateurs to communicate with amateurs in a country whose government objects to such communication. Government suppression of amateur communications in one country thus creates a reciprocal obligation for like suppression by other countries.[5]

Amateur privileges do not include privacy. Communications between two hams are potentially accessible to anyone with appropriate operating equipment and within the geographic range of the signal. International amateur regulations enforce this lack of privacy by declaring that amateur stations should transmit their call signs (station identification) "at short intervals" and by stating that international amateur communications should be "in plain language."[1] In the US, FCC regulations require that call signs be transmitted at the end of each communication and at least every ten minutes during communications.[2] The FCC provides on the Internet a searchable licensing database that allows a call sign to be linked to the licensee's name and address.[3] FCC regulations prohibit international *or* domestic amateur communications "in codes or ciphers intended to obscure the meaning thereof."[4] Controversy over encryption technology, police access to personal communications, and unauthorized collection of personally identifying information has been bluntly pre-empted in amateur radio regulation.

---

[1] Gernsback (1913).
[2] Ibid.
[3] Radio Act (1912) Sec. 4, fifteenth article. Thus the rights of amateurs were recognized implicitly by absence of complete restriction.
[4] Taussig (1922) p. 191.
[5] RR 25 I §1.

International treaties and national regulations also govern the content and purpose of amateur communications. Under international radio regulations, communications between amateurs in different countries must be:

> *...limited to messages of a technical nature relating to tests and to remarks of a personal character for which, by reason of their unimportance, recourse to the public telecommunications service is not justified. It is absolutely forbidden for amateur stations to be used for transmitting international communications on behalf of third parties.*[5]

FCC regulations prohibit amateur communications containing "obscene or indecent words or language," "false or deceptive message," or music.[6] The amateur radio service cannot be used for "communications for hire or for material compensation," "any form of broadcasting," "one-way communications" except of limited types, "activity related to program production or news gathering for broadcasting purposes," retransmissions "from any type of radio station other than an amateur station," or "[c]ommunications, on a regular basis, which could reasonably be furnished alternatively through other radio services."[7] Although regulation specifically forbids amateurs from engaging in "any form of broadcasting," amateur radio is considered similar to broadcasting for content regulatory purposes.[8] The regulations on the content and purpose of amateur radio underscore the absence of any presumption of free communications in amateur radio.

Under international and national regulations, persons must demonstrate technical qualifications to be licensed to engage in amateur communications. An ITU recommendation passed in August, 2001 indicates:

> *...at minimum, any person seeking an amateur license should demonstrate theoretical knowledge of specific topics in the areas of radio regulations, methods*

---

[1] RR 25 I §2 (1) and 25 I §5(2) .
[2] 47 CFR §97.119.
[3] See http://wireless2.fcc.gov/UlsApp/UlsSearch/searchLicense.jsp
[4] 47 CFR §97.113(a)(4).
[5] RR 25 I §2 (1).
[6] 47 CFR §97.113(a)(4).
[7] 47 CFR §97.113(a)(2), 97.113(a)(5), 97.113(b)(a) .
[8] See FCC (1987). The number of amateurs and the allocation of frequencies among amateurs are not regulated. Many of the arguments used to justify broadcast content regulation do not apply to amateur radio.

*of radiocommunications, radio system theory, radio emission safety, electromagnetic compatibility, and avoidance and resolution of radio frequency interference.*[1]

Amateurs who want to make use of amateur frequencies below 30 MHz are required under international regulations to be able to transmit and receive in Morse Code.[2] In the US, FCC regulations link a ladder of three licensing tests to the extent of amateur frequency privileges.[3] The connection between passing a more difficult licensing test and getting more extensive frequency privileges seems to relate not to necessary technical knowledge, but to creating objective support for group boundaries and privilege hierarchies.

Well-established amateur radio associations strongly support amateur radio freedom defined by international and national authority. The International Amateur Radio Union (IARU), created in 1925, actively participates in ITU conferences that define international radio regulations through international treaty.[4] The American Radio Relay League (ARRL), founded in 1914 for U.S. amateurs, serves as the International Secretariat for the IARU.[5] For the aspiring ham, the ARRL sells for $19 a three-hundred page guidebook, *Now You're Talking*, subtitled, *All You Need For Your First Amateur Radio License*. Chapter 1 (30 pages) is "Federal Communication [sic] Commission's Rules." The underlying theme of this chapter is that hams have been given operational

[1] IARU (2001) describing ITU-R M.1544.

[2] RR 32 I §3(1).

[3] The lowest class of license requires passing a 35-question written examination, the middle class adds a five words per minute Morse Code test, and the highest class license requires passing a 50-question test and a five words per minute Morse Code test. The FCC recently restructured the set of amateur licenses and reduced the speed requirement for the amateur Morse Code tests. Morse Code proficiency at the level of 16 code groups per minute and 20 words per minute was retained for commercial radio operators. See FCC (2001c). The Morse Code test has generated considerable world-wide opposition among amateurs, and it is likely to be eliminated within a few years in most countries other than the U.S. For information, see the articles on the No-Code International website, http://www.nocode.org/articles.html. On the ARRL's position, see http://www.nocode.org/articles/NCI_ltr_to%20ARRL_re_Morse_res.html

[4] Membership of the IARU consists of member-societies, with only one member-society per country or separate territory, each having a single vote. The rights of a member-society may be suspended if it "has failed to fulfill its duties" under the IARU Constitution, "has acted contrary to the interests of Amateur Radio or the IARU," or "no longer adequately represents the interests of radio amateurs throughout its country and/or separate territory." A two-thirds majority is required to terminate a member-societies membership in the IARU. See IARU Constitution, online at http://www.iaru.org/iarucnst.html.

[5] The organization serving as International Secretariat has a powerful position in the IARU. Under the IARU Constitution, a member society serving as International Secretariat (in this case, the ARRL) continues to serve until a successor is elected (the time period or procedure by which such an election might be held is not specified). The policy and management of the IARU is carried out by the Administrative Council, which consists of a President, Vice President, Secretary and six additional members, two from each of three regions of the world. Single candidates for President and Vice President are nominated by the International Secretariat, subject to agreement from the (existing) Administrative Council that the proposed candidates are "suitably qualified." Candidates for these offices are ratified by a simple majority vote of Member-Societies. The President and Vice President serve terms of five years and continue in office until the nomination of a successor has been ratified. The International Secretariat designates the Secretary of the Administrative Council, who serves at the will of the International Secretariat. See IARU Constitution, ibid.

freedoms by government and that these freedoms are secured by hams submitting to government authority. These excerpts indicate the orientation:

> *You should also post your original license, or a photocopy of it, in your station after it arrives in the mail. You will be proud of earning the license, so display it in your station. A copy of the license on the wall also makes your station look more "official."*
>
> *Amateur licenses are printed on a laser printer, and issued in two parts (Figure 1-2). One part is small enough to carry with you; the other can be framed and displayed in your shack. This means you can carry your license* and *display it! Laser ink can smear, so it's a good idea to have your wallet copy laminated. If the small part is too large for your wallet, you can make a reduced-size copy on a photocopier. If you carry a copy, you can leave your original safely at home. Although you can legally carry a copy, your original license must be available for inspection by any U.S. government official or FCC representative. Don't lose the original license!*
>
> …
>
> *Keep in mind that the FCC has the authority to modify the terms of your amateur license any time they determine that such a modification will promote the public interest, convenience, and necessity.*
>
> …
>
> *Consider laminating your original license, as the ink sometimes lifts off the paper, even if you place the license behind glass in a frame.*
>
> …
>
> *Suppose you receive an official notice from the FCC informing you that you have violated a regulation. Now what should you do? Simple: Whatever the notice tells you to do.*[1]

Thus the authority of national regulations, re-enforced by international treaties, defines what amateurs understand as their freedom.

Interference among hams is not a significant public policy issue. Amateur radio has been assigned rights to use only limited radio frequencies. In addition, one ham's use of a particular frequency at a particular place and time could potentially interfere with another ham's use of that frequency. The number of hams is not limited, nor are particular frequencies assigned to particular hams. Interference among hams undoubtedly occurs. Nonetheless, formal, authoritative regulation of interference among hams is essentially non-existent. Hams seem to cope with interference through personal courtesy, operational adaptability, and consensus-based coordination mechanisms.

### 2. The Internet

The Internet consists of interconnected but somewhat autonomous electronic information and communications systems. The Internet began as a system for sharing computational resources among geographically dispersed academic and government researchers. By the late 1980s, e-mail had inadvertently become an important use. In the mid-1990s, a simple interface for sharing text and images (the Web) spurred dramatic Internet growth. Hackers, the community of experts in computer programming and electronic networking, are acutely conscious of their particular freedoms. Hackers have played a major role in making the Internet what it is.[1] Other Internet users more passively make choices among the capabilities that the Internet offers. Habitually exercised and thus naturalized, such choices also inculcate the Internet field's understanding of freedom.

---

[1] ARRL (2000) pp. 1-6, 1-16, 1-17, 1-25.

Compared to freedom in amateur radio, freedom in the Internet field takes on a radically different meaning. *A Declaration of the Independence of Cyberspace*, a document issued on the Internet in 1996 and subsequently widely discussed, opened with these words:

> *Governments of the Industrial World, you weary giants of flesh and steel, I come from Cyberspace, the new home of the Mind. On behalf of the future, I ask you of the past to leave us alone. You are not welcome among us. You have no sovereignty where we gather.*[2]

An informally recognized hacker-leader performed another dramatic challenge to authority:

> *In 1977, while attending a science-fiction convention, [Richard M. Stallman] came across a woman selling custom-made buttons. Excited, Stallman ordered a button with the words "Impeach God" emblazoned on it.*
>
> *…Stallman wore the button proudly. People curious enough to ask him about it received the same well-prepared spiel. "My name is Jehovah," Stallman would say. "I have a special plan to save the universe, but because of heavenly security reasons I can't tell you what the plan is. You're just going to have to put your faith in me, because I see the picture and you don't. You know I'm good because I told you so. If you don't believe me, I'll throw you on my enemies list and throw you in a pit were Infernal Revenue Service will audit your taxes for eternity."*
>
> *Those who interpreted the spiel as a word-for-word parody of the Watergate hearings only got half the message. For Stallman, the other half of the message was something only his fellow hackers seemed to be hearing. One hundred years after Lord Acton warned about absolute power corrupting absolutely, Americans seemed to have forgotten the first part of Acton's truism: power, itself, corrupts. Rather than point out the numerous examples of petty corruption, Stallman felt content voicing his outrage toward an entire system that trusted power in the first place.*[3]

---

[1] A leading scholar has stated:

> *Hackers build the Internet. Hackers made the Unix operating system what it is today. Hackers run Usenet. Hackers make the World Wide Web work.*

Raymond (2001), from section "What is a Hacker?"

[2] Barlow (1996).

[3] Williams (2002) Chapter 4, beginning sixth paragraph from the end (pp. 55, 56). Stallman's spiel seems more like a loose and obtuse adaptation of Job 38-41. The rest of the story may differ superficially from Job 42, but there are also similarities. See Williams (2002) Chapter 14. Note as well that the paragraph following the last one cited quotes Stallman as saying, "The way I see it, any being that has power and abuses it deserves to have that power taken away." This appears to be a different perspective on power than the one in the immediately preceding paragraph. Stallman's home page, www.stallman.org, presents

Elsewhere, this hacker-leader presents himself as a general, writing:

> *...some of my cities have fallen. Then I found another threatened city, and got ready for another battle. Over time, I've learned to look for threats and put myself between them and my city, calling on other hackers to come and join me.*
>
> *...We can't take the future of freedom for granted. Don't take it for granted! If you want to keep your freedom, you must be prepared to defend it.*[1]

Another hacker-leader wrote and made freely available on the Internet a fourteen-page article, "How to Become a Hacker." The substance of the article begins this way:

> *Hackers solve problems and build things, and they believe in freedom and voluntary mutual help. To be accepted as a hacker, you have to behave as though you have this kind of attitude yourself. And to behave as though you have the attitude, you have to really believe the attitude.*[2]

The article then offers a modern Zen poem for inspiration, and advises: "So, if you want to be a hacker, repeat the following things until you believe them…." Although exactly what should be repeated is not entirely clear, this appears to be the litany:

> *The world is full of fascinating problems waiting to be solved.*
> *No problem should ever have to be solved twice.*
> *Boredom and drudgery are evil.*
> *Freedom is good.*
> *Attitude is no substitute for competence.*

The article then goes on to advise on how to acquire hacking skills and on socially valued uses of these skills.

Hackers and others actively engaged with the Internet would consider government licensing of Internet users to be an outrage. This is so even though hackers seem to be historically linked to amateur radio users, and a leading hacker is also a leader in amateur radio.[3] This is so even though the challenge of initiating new users and the problems of abusive use, interference, and breakdowns in operating standards and cooperation are similar in amateur radio and on the Internet. Nonetheless, government licensing of

---

many different concerns about justice and mercy. Some of these concerns are of dubious moral and intellectual quality. Stallman's significance as an ethical thinker seems to come from his awesome programming achievements and his clear operating principles as a programmer.

[1] Stallman (1998), from the last two paragraphs. Stallman presents himself as a big man. U.S. government officials before WWI tended to referred to individual, non-commercial radio users as "amateurs" or "boys." The tone of friendly condescension is particularly apparent in a Congressional Report from April, 1910:

> *The committee means fair play for industrious, inventive American boys, and, without consulting with him, ventures the statement that there is no other man in the United States in more cordial accord with their ambitions than the President of the United States…*

House Report (1910) p. 7.

[2] Raymond (2001), from section title "The Hacker Attitude."

[3] See Raymond (2000), section 2. The entry for "hacker" in Jargon File, version 4.3.1, 29 Jun 2001, states, "We have a report that it [the term "hacker"] was used in a sense close to this entry's by teenage radio hams and electronics tinkerers in the mid-1950s." See Raymond (2001b). Bruce Perens, one of the founders of the Open Source movement and an important hacker, is also an active ham. He founded No-Code International and also is a member of ARRL. Perens has stated publicly that he contributed $1000 to ARRL's frequency defense fund in the year 2001. Perens website is http://perens.com. For some of his views on amateur radio, see http://perens.com/Articles/NCI-reform.txt

Internet users would be abhorred as a violation of God-given inalienable rights. Or abhorred as a violation of natural human rights. Or abhorred as a violation of what most persons, deliberating under appropriately specified circumstances, would come to agree to regard as rights persons should have irrespective of decisions made by duly constituted governing authorities. Freedom to use the Internet is not understood as a privilege granted by national governments and international treaties. In the Internet field, freedom means capabilities that persons should *personally* recognize, cultivate, and defend.

Internet users have recognized, cultivated and defended capabilities that some governments would prefer to suppress. Supporters of democracy in China and adherents of the Falun Gong movement have vigorously sought to communicate with persons in China. When the founder of China's first human-rights website was arrested, supporters copied his website to a server in the U.S. and the contents of the website remained accessible to persons in China.[1] In contrast to amateur radio communications, international law does not require the U.S. or any other country to shut down Internet communications that the Chinese government does not want to occur.

Hackers and others actively engaged in shaping the Internet have sought to promote capabilities for privacy. In 1993, a programmer in Finland implemented in his spare time an anonymous re-mailer (a means for e-mail anonymity) that by 1996 had more than half a million users worldwide.[2] By early 1999 there were about 40 anonymous re-mailers accessible on the Internet.[3] Services have been developed to allow users to browse the Web anonymously. Encryption technologies are readily available on the Internet. Internet activist have put strong pressure on the U.S. government to relax its restrictions on the export of advanced encryption software, and they have strongly resisted law enforcement initiatives to increase capabilities to monitor Internet communications. In contrast to amateur radio users, Internet users probably would not warmly embrace government regulations requiring them to identify all their communications with a national identification number linked to their names and addresses in a national, publicly available database.[4]

The Internet recognizes relatively few restrictions on purpose of use or content. The Internet was designed as a general-purpose medium. It has been used for person-to-person mail, instructional services, transaction systems, monitoring and tracking services, telephony, text, audio and video broadcasting, and a variety of other purposes. Content on the Internet spans the full range of human expression. Poetry, personal diaries, independent newsgathering and reporting, political commentary from widely different perspectives, false and deceptive information, and racist, xenophobic, misogynistic and misandristic tracts all can be found readily on the Internet. Photographs and videos of naked human sexual acts, which many would regard as obscene or, alternatively, prurient, are also widely available. When the U.S. government in 1996 passed a law regulating indecent content on the Internet, many high-traffic websites temporarily suspended normal communication as an act of protest. A broad coalition of Internet and civil liberties groups challenged the law in court, and the U.S. Supreme Court ruled that such regulation is unconstitutional.[1] Other governments and judicial systems have different standards for regulating the content of communications. Yet for most users around the world, the Internet offers access to a wider range of information and communications capabilities than is available through other media.

[1] See Neumann (2001).
[2] The re-mailer, at http://www.penet.fi, was closed in 1996. See http://www.penet.fi/pres-english.html.
[3] See Lohr (1999).
[4] Internet domain names (top level "web addresses") are associated with a registrant and physical address through the WHOIS protocol. For an example of a WHOIS service, see http://www.geektools.com/cgi-bin/proxy.cgi . The number of Internet users who have their own domain name is a small subset of all Internet users. Moreover, much Internet use of Internet users who have a domain name takes place without reference to the user's domain name. The policy implications of the WHOIS database are insightfully analyzed in Mueller (2002), Section 11.2.2.

Interference on the Internet is a major public policy issue. The large volume of unsolicited email ("spam") distributed on the Internet essentially creates noise in Internet users' mailboxes and causes inefficient use of personal attention, a scarce resource.[2] Completely eliminating such noise is widely recognized to be not only infeasible but also undesirable.[3] Nonetheless, Internet users have looked to government for help, and governments around the world are actively seeking to regulate e-mail interference in ways that produce net benefits.[4] Interference in domain name addresses has also emerged as a major public policy issue. As is the case with interference in U.S. patent applications, special institutions and regulations have been created to govern domain name interference disputes. Domain name interference regulation has emphasized "private sector leadership," a slogan popular in the U.S. government in the late 1990s. However, national governments and international treaty organizations, acting in ways that obscure political accountability, have strongly influenced domain name regulation. Careful analysis also indicates that the scope and the significance of interference problems have been exaggerated.[5]

### 3. Commercial Wireless Services

Commercial wireless services are provided in ways commonly appreciated for many goods. That means a person, in the role of customer or consumer, buys the goods that she wants. Freedom in this context is about having money. Freedom also depends on the scope of opportunities to buy and on the prices at which goods are available. Competition among profit-seeking firms within a well-functioning capital market is widely considered to promote consumer freedom.

---

[1] Reno v. American Civil Liberties Union, 521 U.S. 844, 117 S.Ct. 2329.

[2] Daum Communications, Korea's largest e-mail service provider, recently implemented a "pay-for-service" system for the transmission of more than 1000 e-mails. Daum has also sued three companies in Seoul District Court for sending more than 6 million messages a month to Daum subscribers. See http://www.koreaherald.co.kr/SITE/data/html_dir/2002/05/23/200205230020.asp On the economics of personal attention, see Galbi (2001a).

[3] Most persons would like to receive an unsolicited e-mail from a long lost friend.

[4] For example, European Parliament (2002).

[5] See, generally, Mueller (2002). He notes:

> *Much attention has been devoted to the threat of cybersquatting. Less attention has been paid to the danger that measures to control it are expanding property rights to names at the expense of free expression, privacy, and competition.*

Consider, for example, the multi-national corporation Orange. In 1994, Orange, established as the fourth mobile service provider in the UK, began selling. By early 1996, Orange served about a half-million customers and was on the FTSE-100 list of the UK's largest companies. Among its many offerings, Orange pioneered pre-pay service, per second billing, and caller ID as a standard feature. By late 1999, Orange, then serving about 3.5 million customers, was bought by Mannesmann, a Germany corporation. Early in 2000, Vodaphone, a British company, bought Mannesman, and the European Commission required the divestiture of Orange. While prices of mobile calls continued to drop, Orange developed additional services, including conference calling, vocal e-mail delivery, and voice recognition technology. In mid-2000, France Telecom bought Orange from Vodaphone. The new company, named Orange SA and headquartered in Paris, served in 2000 about eighteen million French customers, eight million British customers, and nine million customers in wholly owned subsidiaries in seven other countries. Now included in the CAC40, a list of France's forty largest corporations, Orange aims to provide service in 50 countries by 2005.[1]

Orange presents its commercial wireless services as promoting freedom. Orange calls its services wirefree™ communications, rather than wireless communications, and thus emphasizes freedom from being "encumbered by wires." Considerable thought went into choosing the name Orange:

> *The team brainstormed names and refined the core brand proposition from four options to a composite of three ideas (my world, manager, my friend). The composite idea was "It's my life." ....*
>
> ....
>
> *"Orange" was the word that best represented their ideas, with its connotations of hope, fun, and freedom.*[2]

In 1999, Orange launched a £12 million UK advertising campaign emphasizing, "Orange enables you to communicate wherever, whenever and however you want."[3] This freedom is important to users:

> *Text messaging may be the ultimate street language for today's teenagers, but many are using this new way of communicating as a way to combat nerves when embarking on new relationships according to a recent survey conducted by Orange. Over 30% of 16-18yr olds are sending an average of over 20 messages a week and many respondents are relying on texting to avoid those tricky face to face conversations.*
>
> ...
>
> *Denise Lewis, [Orange] Group Director of Corporate Affairs, stated: "Text messaging provides yet another way to share our thoughts with others. Where shyness used to prevent some from communicating their feelings, text messaging*

---

[1] See Orange.com, sections "investor relations>about orange>history" and "corporate affairs>about orange>orange firsts". Oftel figures show UK mobile prices to have fallen about 60% from 1996/97 to 2000/01. Tracking price in highly competitive service markets is difficult, and thus such figures should be analyzed carefully. See Oftel (2001) Fig 2.4. and Appendix A.

[2] See http://www.orange.com/English/corporate/Orangestory.asp?bhcp=1

[3] "Wirefree Communication, Naturally with Orange," Orange News, Marketing, 1 Sept. 1999; online at http://www.orange.co.uk/cgi-bin/news/search.pl.

*has fully opened the gates; the buzz of receiving text messages goes on the anytime, anywhere scenario. Orange text messaging is the ideal way to keep in touch!"*

*Text messaging is now more popular than ever. A total of 373m text messages were sent over the Orange network (UK and France) in January this year, an 86% increase in the last six months.*

*To cater for the increasing text messaging phenomenon, Orange has launched Orange Out Here, a new mobile phone package that offers five free text messages a day. The package also includes up to 2 reserve calls which can be used when call credit has expired, as well as an additional £5 free airtime on top of the £5 already available on Just Talk.*[1]

Orange offers a variety of other products to increase users' freedom of action and expression. Users can buy a mobile phone featuring always-on connectivity to the Internet, wireless local connectivity to laptop computers and personal digital assistants, and tri-band operation allowing roaming across five continents (£179.99, monthly payment available). Orange has worked with a British bank to allow customers to do banking through their mobile phones. Orange implemented a service that allows users to bet on horse-races through their mobile phones. Users have a wide range of choices of phone styles and accessories. For example, users can buy a Mexican influenced cactus phone cover, an oriental Dragon design, or pop art female and male faces (£19.99 each).[2] Users satisfied with a standard black phone are also free to choose that.

While commercial wireless services are associated with a widely practiced freedom, this type of freedom has significant limitations. Orange offers on a profit-seeking basis the above choices to customers in the UK. Orange seeks to provide similar choices to customers in France, where its headquarters and top management reside, and in the many other countries. But every country, like every person, is special and unique. National laws, typically considered to be related to national security or cultural integrity, limit the operation of companies like Orange.[3] Laws and government practices that limit private, profit-seeking corporations' willingness to invest capital also limit the scope of commercial wireless services.[4] The extent of consumer spending power is an additional constraint on commercial wireless services.

Freedom associated with commercial wireless services does not encompass important aspects of persons and societies. Policy issues such as international justice, privacy, and use rights are treated as just another product attribute that an individual chooses, like animal testing for perfume or the organically grown status of vegetables. But in most societies, politics and policy decisions are about more than just marketing products. The composite idea of Orange, “It’s my life,” presents an anonymous speaker articulating “it,” “my,” and “life.” One feels impelled to ask: “Who are you?”; “Are you alienated from your life?”; “Is that it?” An understanding of freedom merely concerned with choices for having or owning misses important aspects of human being.

---

[1] "THE SURVEY SAID: TEXT MESSAGING THE ULTIMATE FLIRTING TOOL!" Orange News, Marketing, 26 March 2001; online at http://www.orange.co.uk/cgi-bin/news/search.pl. To make the news release easier to read, I have added four commas, changed "have" to "has" in the first sentence of the last quoted paragraph, and capitalized the initial letters in the words "out here" in that sentence.

[2] The above products are on offer at http://shop.orange.co.uk/NASApp/esales/esales.

[3] With respect to restrictions in the US, see, for example 47 U.S.C §310 and FCC (2001f) (FCC's approval of radio license transfers in conjunction with Deutsche Telekom's acquisition of VoiceStream).

[4] Poor regulation and weak protection for investments are widely cited as problems inhibiting the development of commercial services.

Concern about interference in commercial wireless services focuses on the interaction of different service providers’ radio signals. Under current radio regulations, commercial wireless service users seldom experience such interference. Service qualities that users experience typically depend much more significantly on network build-out and network technology. However, before constructing a network, commercial wireless service providers typically seek a license to use specific frequencies, and the license is often associated with particular network technologies. For example, Orange in the year 2000 bought for £4.1 billion a license to provide UMTS service in the UK.[1] Acquiring a license greatly reduces the extent to which Orange has to manage radio signal interference as a business risk and an operating concern. On the other hand, the value of Orange’s UK UMTS license is likely to depend strongly on whether other companies can acquire licenses to provide services similar to UMTS in the UK using alternative radio rights. Orange’s primary business focus is not on governing radio frequency use but on providing wireless services to customers. Under current radio regulations, radio signal interference in commercial wireless services is only a minor business concern outside of the licensing process. Under different radio regulations, such as an unlicensed use regime, radio signal interference might significantly affect service qualities that customers perceive. In such a situation, dealing with radio signal interference would be another aspect of serving customers.

Interference among commercial wireless service providers’ radio signals is a rather different issue in low-income countries than in high-income countries. Low-income countries typically have little military use of radio; underdeveloped, public radio and television broadcasting; and much greater demand for low-quality radio services at correspondingly low prices. International radio regulations and the global radio equipment industry may not produce regulations and equipment appropriately adapted to the relatively low opportunity cost of radio use in low-income countries. Although low-income countries face different circumstances, their commercial wireless services may be effectively constrained by interference concerns in high-income countries.

### 4. Freedom around the World

To better understand freedom, consider what communications capabilities persons actually exercise. Table 10 summarizes amateur radio, Internet, and mobile telephony users per thousand persons in 164 countries, grouped by World Bank income categories. Looking across all countries, the median figures show about three times as many mobile users as Internet users, and about a thousand times as many mobile telephony users as amateur radio users. These data indicate that the commercial wireless services field has been relatively successful in bringing its type of freedom to persons around the world. These facts do not necessarily imply that commercial wireless services are much more important to personal welfare and the common good than are the Internet and amateur radio. But they do describe important aspects of persons' communications activities in relation to much different fields of freedom.

[1] Orange also owns UMTS licenses in the France, the Netherlands, Belgium, Denmark, and Switzerland. See Orange (2002) p. 119.

**Table 10**
**World Communications Capabilities By Income**
(median for countries in income class)

| | | Users Per Thousand Persons | | |
|---|---|---|---|---|
| Income Class | Number of Countries | Amateur Radio | Internet | Mobile Telephony |
| 1. High | 39 | 0.7660 | 310 | 711 |
| 2. Upper Middle | 32 | 0.3320 | 82 | 212 |
| 3. Lower Middle | 47 | 0.0801 | 26 | 67 |
| 4. Low | 46 | 0.0022 | 3 | 10 |
| All Classes | 164 | 0.0951 | 35 | 105 |

Comparing Income Classes

| | | Ratio of Users Per Thousand | | |
|---|---|---|---|---|
| Ratio of Income Classes | Number Ratio | Amateur Radio | Internet | Mobile Telephony |
| 1/2 | 39/32 | 2.3 | 3.8 | 3.3 |
| 2/3 | 32/47 | 4.1 | 3.2 | 3.2 |
| 3/4 | 47/46 | 36.9 | 9.8 | 7.0 |

Comparing Fields

| | | Ratio of Users Per Thousand | | |
|---|---|---|---|---|
| Income Class | Number of Countries | Amateur/ Internet | Amateur/ Mobile | Internet/ Mobile |
| 1. High | 39 | 0.25% | 0.11% | 43.54% |
| 2. Upper Middle | 32 | 0.41% | 0.16% | 38.42% |
| 3. Lower Middle | 47 | 0.31% | 0.12% | 38.42% |
| 4. Low | 46 | 0.08% | 0.02% | 27.38% |
| All Classes | 164 | 0.27% | 0.09% | 33.07% |

Sources and Details: See Appendices A and C.

The extent of communications capabilities has major significant for emergencies and disasters. One approach to emergency response and disaster communications is to have a small number of communicators who have special disaster response skills. The admirable services of amateur radio users are well-recognized in this regard. But having more dispersed communications capabilities can also play a critical role in emergency response. The presence of mobile phones among the passengers on a hijacked U.S. airline, along with the courageous and decisive action of a few strong men, probably prevented the U.S. Capitol from being destroyed and many additional persons killed.[1] Increasing persons' capabilities also increases the possibilities for evil acts. Commercial wireless services, the Internet, and even amateur radio can be used for good or for evil. When confronted with those two general possibilities, liberal democracies usually place their faith in the good and promote freedom, while making prudent preparations to confront evil acts that might occur.

Difference in income levels of countries are strongly associated with differences in realized communications capabilities. The number of users per thousand persons drops sharply with country income class for amateur radio, the Internet, and mobile telephony. The median share of amateur radio users and Internet users is much less than 1% of the population in low-income countries. The share of mobile telephony users is only about 1%. Given that about 40% of people live in countries in the low-income category and about another 35% live in countries in the lower-middle-income category, many persons around the world have not realized important communications capabilities.

Some types of freedom are more easily realized than others in low-income countries. Commercial wireless services have done relatively well in low-income countries. On the other hand, amateur radio fares particularly badly in low-income countries and in Africa, Asia, and the Middle East (see Table 11).[2] Amateur radio understands freedom as closely tied to government authority. Good government is not easy to establish and maintain. It is particularly lacking in low-income countries and in many countries in Africa, Asia, and the Middle East. Amateur radio needs to better adapt to conditions in low-income countries. More generally, communications capabilities in low-income countries develop more quickly in fields less closely tied to government authority.

[1] For eulogies for these men and biographical details, see http://www.letsrollheroes.com/memorialsites.shtml
[2] For a complete list of countries, along with category indicators and users per thousand, see Appendix C.

**Table 11**
**World Communications Capabilities By Region**
(medians for countries in region)

| Region | Number of Countries | Users Per Thousand Persons: Amateur Radio | Internet | Mobile Telephony |
|---|---|---|---|---|
| Western Europe | 22 | 0.8400 | 300 | 748 |
| Africa | 45 | 0.0024 | 3 | 15 |
| Middle East | 10 | 0.0376 | 69 | 209 |
| Caribbean | 14 | 0.3181 | 51 | 117 |
| North and Central America | 10 | 0.1602 | 34 | 122 |
| South America | 12 | 0.2057 | 50 | 162 |
| Asia | 19 | 0.0164 | 26 | 81 |
| Oceania | 11 | 0.1923 | 22 | 18 |
| Central and Eastern Europe | 21 | 0.2602 | 57 | 188 |
| World | 164 | 0.0951 | 35 | 105 |

Comparing Fields

| Region | Number of Countries | Ratio of Users Per Thousand: Amateur/ Internet | Amateur/ Mobile | Internet/ Mobile |
|---|---|---|---|---|
| Western Europe | 22 | 0.28% | 0.11% | 40.2% |
| Africa | 45 | 0.08% | 0.02% | 19.1% |
| Middle East | 10 | 0.05% | 0.02% | 32.9% |
| Caribbean | 14 | 0.62% | 0.27% | 43.4% |
| North and Central America | 10 | 0.48% | 0.13% | 27.4% |
| South America | 12 | 0.41% | 0.13% | 31.0% |
| Asia | 19 | 0.06% | 0.02% | 32.5% |
| Oceania | 11 | 0.87% | 1.09% | 125.0% |
| Central and Eastern Europe | 21 | 0.46% | 0.14% | 30.0% |
| World | 164 | 0.27% | 0.09% | 33.1% |

Sources and Details: See Appendices A and C.

**B. Natural Freedoms and Equal Rights**

Around the world many persons now firmly believe in ideas of natural freedom and equal rights. The exact meaning of these beliefs and the extent to which they are fundamentally justified animates much scholarly discussion. The current state of scholarly literature suggests that the epistemic status of such beliefs is essential mysterious. Nonetheless, most persons do not ground their beliefs in detailed study of political and critical philosophy.[1] Disciplined by broad discussion and open-minded observation and analysis, mere intuitions of truth about natural freedom and equal rights can make major contributions to practical public decisions. Try it and see whether this is good for radio regulation.

Natural freedom seems to mean that persons should not be made to do something, or prevented from doing something, without true reasons. Consider the FCC's decision in 1996 to eliminate individual licensing for certain non-mandatory radios on ships and aircraft. The FCC explained that licenses are not needed:

> *Individual licensing also is unnecessary for any of our regulatory purposes. We perform our regulatory responsibilities for the Maritime and Aviation Services primarily through the rulemaking process to allocate spectrum, to implement requirements for license eligibility, and to define types of communication that may be transmitted. In addition, all channels are shared by all licensees so spectrum management occurs through channel sharing, in real time, or through control exercised generally by the FAA or Coast Guard.*[2]

The "overwhelming support" that the "vast majority" of commenters provided made eliminating individual licensing an easy decision for the FCC.[3]

Intuitions about natural freedom can provide motivation for seeking truth and acting on it, even despite a contrary consensus and satisfaction with the status quo. The FCC's proposal in 1982 to eliminate individual licenses for Citizens Band (CB) radio and radio control devices encountered "almost unanimous opposition" from commenters. The FCC noted:

> *Many of the commenters argue that if our position that we do not accomplish spectrum management through the licensing function in these services is true, then it is equally as true for other services, and not a valid reason to eliminate individual licenses for R/C [radio control radio] and CB [radio].*[4]

This argument warned the FCC that challenging the consensus in this area could have consequences elsewhere. The FCC was not afraid:

> *This argument is not valid. A license in the R/C or CB services permits operation on every channel in each service. A license in the GMRS (General Mobile Radio Service) or in the private land mobile radio services authorizes operation only upon the frequency (ies) or frequency pair assigned....*[5]

Commenters argued that licensing was what made the FCC matter:

> *Many of the comments stated that licensing in these services serves the function of informing users that there is an FCC and that there are rules the user must obey.*

[1] And even great learning does not prevent gross misjudgments. See Berkowitz (2000).
[2] FCC (1996b) para 8.
[3] See ibid., para. 5.
[4] FCC (1983) para 17.
[5] Ibid.

*Commenters stated that licensing reminds an operator that transmitting is a regulated privilege.*[1]

The FCC had enough institutional self-confidence to dismiss this argument:

*There are few CB'ers who haven't heard of "Uncle Charlie." The existence of a license does not play an essential role in this process…*[2]

Commenters argued:

*… since the majority of users apply for licenses and comply with our rules, licensing must work.*[3]

In response to this superficial appeal to the comfort of the status quo, the FCC provided facts and analysis:

*This merely begs the question, and does not consider whether other Commission rules or policies besides the licensing function are responsible for rule compliance. [presentation of data showing increasing number of CB interference complaints despite licensing]…*[4]

That almost all parties active in this proceeding opposed an extension of their freedom is worth contemplating. That the FCC nonetheless gave them that additional freedom is worth savoring.

One can also find instances of radio regulations that are shockingly contrary to intuitions about natural freedoms. An FCC decision in August, 2000, reconsidered and confirmed in August, 2001, provides a good case study.[5] In November, 1999, the FCC received a petition to eliminate its rules forbidding CB radio users from communicating with other CB radio users located more than 250 kilometers away or in countries other than Canada. CB radio signals can propagate unpredictably and capriciously under effects of sun spots and other natural phenomena. The question was whether CB users should be allowed to communicate – intersubjectively interact – using unusual, unintentional, objective interactions of their signals. Intuitively, one might ask, "Why should human beings not be allowed to acknowledge each other's presence?"

The FCC refused to allow this intuitive natural freedom. The FCC declared "inconsistent with the purpose of the CB Radio Service" communication using CB radio signals that happen to travel more than 250 kilometers. Such freedom would "fundamentally change the nature of [CB Radio Service]," which the FCC defined as a "short-distance voice communications service."[6] Thus restrictions on human freedom (intersubjective interactions) were preserved to preserve FCC service definitions against the occasional, objective, natural occurrence of long-distance CB signal propagation. That seems intuitively wrong. Similarly, the FCC did not give CB radio users the freedom to communicate with CB radio users in other countries. CB radio users were allowed only to "exchange messages with [foreign] government stations relating to civil defense and [to] other stations in the Citizens Radio Service."[1] Many persons around the world would intuitively believe that this CB radio "wall" surrounding the U.S. should be torn down.

---

[1] Ibid, para. 20.
[2] Ibid.
[3] Ibid para. 23.
[4] Ibid.
[5] FCC (2000), FCC (2001d).
[6] FCC (2000), from para. 6 and 7.

The scope of natural freedoms is generally thought to expand in extraordinary circumstances. Asked in August, 2001, "whether the Commission ever intended to actually place a limit on distance of communications in emergency situations," the FCC reiterated:

> *...we believe that the Commission intended the CB Radio Service to be used for the express purpose for which it was authorized. In this regard, we note that there is nothing in the rules that indicates the Commission intended to permit any communication between CB Radio Service stations over 150 miles apart....*
>
> *With regard to the Petition's request that we amend the rules to specifically permit emergency communications in excess of 155.3 miles, we do not believe this amendment is necessary. As an initial matter, we note that individuals who find themselves in emergency situations are likely to have stations in other radio services, such as amateur, marine, land mobile stations, or cellular or other wireless telephones, available either to them or to another individual close to the emergency location.*[2]

The FCC then went on to note comparative advantages of other radio services in emergency situations, including the benefits of Enhanced 911 systems in mobile telephony. With its reason firmly fixed on its regulations, the FCC ended the key paragraph in its ruling by noting:

> *...there is nothing in the CB Radio Service rules that prevents an individual who receives a message that contains a request for emergency assistance, regardless of how far away the transmitting station is, from using other communications services to inform public safety providers of the need for assistance.*[3]

A natural, first reaction to a person's cry for help is to communicate to her, directly, through the same means she communicated to you, recognition of her situation and her need for help. Only policy makers with no respect for intuitions about freedom would pretend that such a freedom does not exist.

Intuitions about natural freedoms and equal rights point to many additional, critical insights into radio regulation. Natural freedoms and equal rights are generally understood with respect to human beings, male and female. While human beings are animals in a simple-minded biological sense, even highly disabled humans are intuitively considered to be part of this highly privileged animal class: human beings understood as equal persons.[4] Most persons intuitively sense that human intentionality gives an action a greater claim to freedom and to equality in rights. Most persons also intuitively understand that sexual difference does not imply fundamental inequality but points to the creative potential of communication. Most persons intuitively sense that human expressiveness, even within the historically shallow conventions of programming languages, has a natural claim to freedom. These intuitions can help to improve radio regulation.

---

[1] Ibid, para. 8.
[2] FCC (2001d) from para. 8 and 9.
[3] Ibid, para. 9.
[4] While humans are intuitively privileged, animals might be considered to have rights, too. Animal rights are not considered in this paper, although they are related to practical ethics for scholarly advancement in high-income countries today.

### 1. Implications of Intentionality

Radio regulation often includes class licensing for categories of radiators. The rules in the US, known as Part 15, are well-articulated and hence provide rich material for analysis.[1] Part 15 categorizes radiators as incidental, unintentional, or intentional:

> *Incidental radiator. A device that generates radio frequency energy during the course of its operation although the device is not intentionally designed to generate or emit radio frequency energy. Examples of incidental radiators are dc motors, mechanical light switches, etc.*
>
> *Unintentional radiator. A device that intentionally generates radio frequency energy for use within the device, or that sends radio frequency signals by conduction to associated equipment via connecting wires, but which is not intended to emit RF energy by radiation or induction.*
>
> *Intentional radiator. A device that intentionally generates and emits radio frequency energy by radiation or induction.*[2]

At least formally, this categorization is based on intentionality, a central aspect of human consciousness.

Contrary to intuitions about natural freedom, human intentionality gives a radiator less claim to freedom in radio regulation. Incidental radiators are least intentionally associated with radiation. Regulations for these devices consist of just one sentence:

> *Manufactures of these devices shall employ good engineering practices to minimize the risk of harmful interference.*[3]

Unintentional radiators use radio frequency energy more intentionally than incidental radiators. Regulations for unintentional radiators consist of about forty-five pages of detailed technical regulations. Intentional radiators intentionally project radio waves out into the world beyond themselves. Doing this makes a device subject to even more extensive and restrictive regulation.[4]

Why would energy associated with human intentionality be more subject to regulatory suppression than similar energy that is meaningless? The conspiracy theory – radio regulation intentionally suppresses human freedom – seems implausible. To seek further understanding, consider this scenario. Suppose there were public concern about noise.

---

[1] 47 CFR 15. These rules can be accessed online from http://www.fcc.gov/oet/info/rules/

[2] 47 CFR 15.3(n),(z),and (o).

[3] 47 CFR 15.13.

[4] See 47 CFR 15.5(d). Additional comparisons between regulations for intentional radiators and regulations for unintentional radiators are difficult because the regulations for both categories are complex and differentiated.

Suppose local boys like to play outdoor jam sessions on improvised instruments. Suppose that overly sensitive car alarms unintentionally make as much noise as the boys. Shouldn't regulation attempt to restrain car alarms before it attempts to restrain boys? Perhaps it comes down to cost-benefit analysis: intentionality is inversely correlated with marginal suppression costs. It just turns out that it is cheaper, at the margin, to suppress boys rather than car alarms. But it is not surprising that authoritative regulation can more cheaply suppress intentionally created energy. Intentionality implies the possibility of change in response to authoritative communication. Just tell the boys to stop. This might be the general logic by which radio regulation suppresses classes of devices more closely associated with human freedom.

Contrary to intuitions about equal rights for human persons, increased human intentionality implies in radio regulation more unequal rights among similar radiators. All incidental radiators are treated equally. They can be made and used in any way consistent with general norms not defined by government. In this sense, public regulation respects the natural freedoms and equal rights of incidental radiators.

Unintentional radiators, more associated with human intentionality than incidental radiators, have unequal rights under radio regulation. Among unintentional radiators is a sub-category called "exempt devices." Exempt devices include "a digital device used exclusively…by a public utility," "a digital device utilized exclusively in an appliance, e.g. microwave oven, dishwasher, clothes dryer, air conditioner (central or window)," and other specifically enumerated devices.[1] Another distinction is between "Class A digital device" and "Class B digital device":

> *Class A digital device. A digital device that is marketed for use in a commercial, industrial or business environment, exclusive of a device which is marketed for use by the general public or is intended to be used in the home.*
>
> *Class B digital device. A digital device that is marketed for use in a residential environment notwithstanding use in commercial, business and industrial environments. Examples of such devices include, but are not limited to, personal computers, calculators, and similar electronic devices that are marketed for use by the general public. ...*[2]

Class A digital devices are subject to less restrictive regulations than Class B digital devices. These regulations show that class licensing does not necessarily imply equal rights under radio law.

Intentional radiators also seem to have unequal rights. Regulations for intentional radiators, grouped by that term's definition, include three subparts, "Subpart C-Intentional Radiators," "Subpart D – Unlicensed Personal Communications Services," and "Subpart E – Unlicensed National Information Infrastructure Devices." The latter two subparts represent more recent regulations. The presence of subparts D and E, which naturally fall under subpart C, suggest that service names, rather than intrinsic characteristics, matter in regulation. In a 1989 order reforming Part 15, the FCC noted:

[1] 47 CFR 15.103.
[2] 47 CFR 15.3(h) and (i).

*The provisions for new devices generally were adopted in response to petitions for rule making that requested authorization only for the specific device in question. This incremental method of adopting device-specific regulations resulted in rules that are lengthy and difficult for the public to understand. It has resulted in the adoption of standards that are overly complex and, in some cases, unnecessarily restrictive. There are also a number of apparent inconsistencies in the technical standards between Part 15 devices that have similar interference potentials. Early standards adopted to control interference are frequently significantly different from what is needed at the present time due to improvements in equipment, such as receiver sensitivity, the increased proliferation of both licensed and unlicensed operations, and changes to the frequency allocations of authorized radio services.*[1]

The FCC addressed these problems in 1989, but the regulatory dynamic that created them remains the same. That dynamic continually recreates unequal rights.

To better support natural freedoms and equal rights, radio regulation should limit radiator classes to a small number and should recognize a greater claim to freedom for radiation more closely associated with human intentionality. Clear principles can help in resisting practical pressures to define continually new radiator classes. Worth exploring is a scheme with just two classes: non-purposeful radiators and purposeful radiators. Non-purposeful radiators could be analogized to pollution sources. Insights and experience in environmental regulation could be applied to regulating these radiators. Purposeful radiators could be analogized to persons. Insights and experience in regulating personal activities such as walking and talking could be applied to regulating purposeful radiators. Such an approach would make radio regulation more consistent with intuitions about the implications of human intentionality.

### 2. Sexual Awareness in Radio Regulation

Aspects of transmitters dominate radio regulation. Radio regulation, both internationally and nationally, divides radio communication into transmission and reception. Transmitters are the primary focus of regulatory concern, while receivers are treated as a mere adjunct to transmission.[2] The subjectivity of receivers, important in practice, is generally ignored in discussion of radio regulation.[3] Communication tends to be viewed as a mechanical transfer of substance from transmitter to receiver. Thus radio regulation presents a fundamental inequality between transmitters and receivers, one that obscures important, practical aspects of communication.[4]

---

[1] FCC (1989) para. 4. In 1979, the FCC revised Part 15 for similar reasons. See FCC (1979).

[2] As Caves (2002) p. 77 notes:

> *In recent years, the importance of receive performance, however, has been sometimes underestimated, particularly in some European regulations [footnote to RTTE Directive (1999/5/EC)] which have tended to relegate receiver aspects to a secondary and non-essential status.*

[3] A much deeper understanding of reception has had great historical significance outside of radio regulation. See Luke 1:26-38.

[4] This inequality contributed to a recent dispute between a microwave licensee and a PCS licensee. The incumbent microwave user was required to be protected from interference. The microwave service rules

Intuitions about sex help to expose these weaknesses in radio regulation. Males and females, although they function differently in sex, are fundamentally equal as human persons. They should have equal roles in government. Moreover, sex is not well understood merely in terms of a division between transmission and reception. Sex is a creative, unitive act. Sex is about communication. Radio regulation should be about communication, too.

Communication and the subjectivity of receivers are central to one of the most successful areas of radio regulation. Under FCC rules, all receivers associated with a licensed radio service must be labeled with the following words (the first label):

> *This device complies with Part 15 of the FCC Rules. Operation is subject to the condition that this device does not cause harmful interference.*[1]

All other devices subject to FCC certification or verification, with one small exception, must be labeled with these words (the second label):

> *This device complies with Part 15 of the FCC Rules. Operation is subject to the following two conditions: (1) this device may not cause harmful interference, and (2) this device must accept any interference received, including interference that may cause undesired operation.*[2]

In a world in which all radio devices had an FCC label and acted in accordance with what the label communicates, the first label would suffice. There would be no need for the second label and its additional clause. Yet freedom and evil are part of the world and must be recognized. Clause (2) of the second label shows that radio regulation understands these facts and seeks to communicate them, too. The second label counsels the user that she must just accept certain events.

Communication should not just describe the world, but also help to change it. Under FCC rules, computers and peripherals marketed for use in a residential environment are required to have the above second label affixed to them. In addition, users of these devices are required, under FCC radio regulations, to be given instructions similar to the following:

> *If this equipment does cause harmful interference to radio or television reception, which can be determined by turning the equipment off and on, the user is encouraged to try to correct the interference by one or more of the following measures:*
> *– Reorient or relocate the receiving antenna.*
> *– Increase the separation between the equipment and receiver.*
> *– Connect the equipment into an outlet on a circuit different from that to which the receiver is connected.*
> *– Consult the dealer or an experience radio/TV technician for help.*[3]

---

specified a 5 MHz transmission channel, but did not specify the bandwidth of reception. The microwave receivers, equipment placed in service in 1980, accepted signals across 12 MHz, which led to interference being received from a new PCS licensee. The meaning and implications of this effect was disputed. See FCC (1998).

[1] 47 CFR 15.19(a)(1).

[2] 47 CFR 15.19(a)(3). The exception concerns cable input selector switches. See 47 CFR 15.19(a)(2).

[3] 47 CFR 15.105(b).

These instructions suggest that the equipment manufacturer does not have any responsibility for interference. In 1989, the FCC specifically declined to require that instructions include contact information for a manufacturer's representative who would address interference problems.[1] Yet such a requirement does not seem burdensome compared to the requirements imposed on transmitters in radio regulation.

Over the past two decades, these little noticed aspects of communication in radio regulation have significantly reduced interference concerns. Regulators, equipment manufacturers, and other organized groups have developed voluntary standards addressing interference perceived in reception.[2] In response to improved designs for receivers, the number of interference complaints to the FCC has dropped significantly since 1982.[3] Available public information indicates that FCC does not now consider radio frequency interference, as real persons actually experience it, to be a major public problem. On the FCC website in the late 1990s, descriptive material concerning consumer issues stated:

> *The FCC does not routinely investigate complaints of interference to telephones and home electronic or entertainment equipment. The FCC will only investigate where there is convincing evidence that results from a violation of rules and, then, only on a law priority basis.*[4]

---

[1] FCC (1989b) para. 123. From 1983 to 1985, the FCC required companies marketing RF lighting devices to provide customers with the name and address of a company official to contact about interference problems. This requirement was subsequently eliminated. See FCC (1988) para. 18.

[2] For information on the FCC's relationship with these voluntary efforts, see FCC (1996d) and FCC (1986).

[3] FCC (1991), ARRL Regulatory Information, on the web at http://www.arrl.org/FandES/field/regulations/rfi-legal/#pl_97_259. The Senate Report on the Communications Act Amendment of 1968 stated that the FCC "…received some 38,000 interference complaints during fiscal 1964." See 1968 U.S.C.C.A.N. 2486, 2488. A Senate Report in 1981 stated, "The Committee notes that several million interference complaints are received by the FCC each year." S. Rep. 97-191,8; 1982 U.S.C.C.A.N. 2237 2244. FCC (1996e) stated, "Radio frequency interference to telephone conversations result in an estimated 500,000 complaints annually to telephone companies and broadcast stations and 25,000 annual complaints to the FCC." This may be a high estimate if it actually just relates to telephones. Introducing a bill concerning CB radio interference (S. 608), the sponsor stated, "In all, FCC receives more than 30,000 radio frequency interference complaints annually – most of which are caused by CB radios." See 143 Cong. Rec S3349-02, Proceedings and Debates of the 105'th Congress, First Session, April 17, 1997. Other evidence indicates that, about a year earlier, the number of complaints received was "nearly 45,000 [interference] complaints annually." See 142 Cong. Rec. S9555-02, statement on S. 2025, Proceedings and Debates of the 104th Congress, Second Session, Aug. 2, 1996. FCC (1996f) stated, "Each year the FCC receives thousands of complaints of interference to televisions, radios, audio systems, telephones, and other home electronics equipment." A poster in an FCC lobby during Public Service Awareness Week, May 6-10, 2002, stated, "Last year the FCC resolved 2100 spectrum interference problems for the public." While the above evidence is not entirely consistent, radio frequency interference complaints to the FCC probably dropped from millions in the early 1980s to thousands in the year 2001. Given the growth in population and much more dramatic growth in numbers and use of radio frequency devices and home electronics equipment, the trend in absolute numbers reflects an even greater change in the significance of interference to persons or devices. FCC tracking of informal consumer complaints and inquiries currently does not include radio frequency interference complaints. This suggests that such complaints are now relatively unimportant. See http://www.fcc.gov/Bureaus/CGB/News_Releases/2002/nrcg0203.html

[4] This statement was made in the course of discussing "CB Violations" in a list of top consumer issues available from the FCC website inclusive of the period June. 17, 1997 to May 1, 1999. To see this, search for http://www.fcc.gov/cib/ncc/top50.htm in the Internet Archive Wayback Machine, available online at

The consumer section of the FCC website currently offers this advice to persons experiencing interference in radio or television reception:

> *...the source of the problem could be your home electronics equipment. It may not be adequately designed with circuitry or filtering to reject the unwanted signals of nearby transmitters. The FCC recommends that you contract the manufacturer and/or the store where the equipment was purchased to resolve the problem.*[1]

Intuitions about equality among communicating parties suggests that these relatively successful approaches to receiver regulation could be extended more equally to transmitters. Regulators could merely express an expectation that transmitters, like receivers, would function harmoniously and satisfactorily for most persons in the existing radio frequency environment. Regulation might just urge persons to contact service providers about any interference problems that might arise.

More extensive regulation of performance responsibility might be warranted. Suppose that radio regulation assigned manufacturers liability for secular declines in the performance of radio communications devices to the extent that radio frequency interference causes the decline in performance. Moreover, suppose that radio regulation required that a statement of this liability be included with the instructions for every radio communications device. The statement would include identification and convenient means to contact a manufacture's representative designated to address interference liability concerns. This statement would give customers some additional rights that, all else equal, might be reflected in higher prices to customers. Given rapid turnover in many radio communications devices, the prevalence of guarantees and warranties observed without specific requirements, and companies often expressed desires to establish relationships with customers, such a regulatory requirement does not seem unnatural or unduly burdensome.

Such a regulatory requirement could be highly beneficial to all parties. Use of old radio communications equipment often interferes with possibilities for new uses and types of radio communications. Much better than protecting users' radio rights is to make those rights, as embedded in the specific equipment used, vanish over time. Under a good regulatory regime, the price/performance ratio for radio equipment could easily shrink by half every two years. In such an environment, the cost of paying interference liability for each radio device after three years of use would add only 12.5% to the cost of the devices. Radio regulation that treated transmission and reception more equally and that attached more importance to communication could produce enormous benefits for persons, services providers, and equipment manufacturers.

---

http://www.archive.org. Introducing a bill concerning CB radio interference (S. 608), Mr. Feingold stated, "...the FCC indicated that due to a lack of resources, the Commission no longer investigates radio frequency interference complaints. Instead of investigation and enforcement, the FCC is able to provide only self-help information which the consumer may use to limit the interference on their own." See 143 Cong. Rec S3349-02, Proceedings and Debates of the 105'th Congress, First Session, April 17, 1997.

[1] See the answers to the question "What can I do about interference...?" at http://www.fcc.gov/cgb/broadcast.htm and http://www.fcc.gov/cgb/radio.htm

### 3. Software Rights for Adults

Human freedom increasingly relates to software capabilities. Most persons do not want software that arbitrarily rules over them. Free software/Open Source licenses give users, and software developers, the freedom to study, modify, and share software.[1] The growing significance of GNU/Linux points to the practical importance of these freedoms. Microsoft, in contrast, has been very successful in providing a Windows platform, which is based on proprietary code. The Windows platform gives software developers the capability to serve easily a large number of users. Building upon the Windows platform, independent software developers have created for users many software choices and capabilities.[2] These choices and capabilities are also an important aspect of freedom.

With hardware, certain kinds of freedom are difficult to suppress. FCC rules state:

> *The users manual or instructional manual for an intentional or unintentional radiator shall caution the user that changes or modifications not expressly approved by the party responsible for compliance could void the user's authority to operate the equipment.*[3]

A computer is categorized as an unintentional radiator. It seems implausible that computer manufacturers expressly approve many of the sorts of hardware changes (upgrades) persons make to their computers. Moreover, changes are likely to be detected only if they interfere with some other device. In addition, the FCC is likely to respond to such interference only if it relates to radio communications for critical public functions. Administrative action is a weak instrument for voiding users' natural freedoms with hardware.

Software defined radios (SDRs) are now attracting considerable interest. An SDR converts a wide bandwidth of radio frequency energy to a digital data stream.[4] Based on an agreed communications protocol expressed in software, digital signal processing equipment in the SDR extracts the intentionally radiated signal from the data stream. This architecture facilitates low-cost communication using many different communications protocols. As a leader in the SDR field explains:

> *Some believe that future radio services will per force provide seamless access across cordless telephone, wireless local loop, PCS, mobile cellular and satellite mobile modes of communication, including integrated data and paging. Anyone who needs to access even half that many radio modes at once clearly will have to move to software radio based infrastructure.*[5]

---

[1] See Stallman (1998) and Raymond (1997) for important presentations of free software/open source.

[2] A platform provider less capable of credibly providing independent software developers with significant technological and business freedom will, all else equal, attract fewer such software developers. An important public policy issue is the extent to which Microsoft has unfairly used the Windows platform to limit the freedom of independent software developers.

[3] 47 CFR 15.21.

[4] Mitola (1996) provides a review of SDRs by a leading authority in the field.

[5] Ibid, "Who Needs Software Radios?"

Moreover, since more effective communications protocols can be implemented simply by upgrading SDR software, SDRs encourage timely deployment of advanced communications capabilities.[1]

SDRs have different implications for freedom than do traditional radio hardware. In a recent order, the FCC stated:

> *…a means will be necessary to avoid unauthorized modifications to software that could affect the compliance of a radio. Because groups such as the SDR Forum and ETSI are sill in the process of developing standards for encryption and digital signatures that could be used in software defined radios, we decline to propose specific requirements for authentication. Instead, we propose a more general requirement that manufacturers take steps to ensure that only software that is part of a hardware/software combination approved by the Commission or a TCB [Telecommunication Certification Body] can be loaded into a radio.*[2]

The FCC established rules by which manufacturers get FCC authorization for the software used in their radios. The FCC established rules for authorizing changes in software, and the FCC also established rules relating to third-party development of software.[3] The FCC clarified that, in exercising its regulatory authority, it will distinguish between software applications and software that controls radio frequency operating parameters.[4] Compared to its control over unauthorized hardware changes, the FCC's control over freedom with radio software may be much greater.[5]

Having a single federal agency with broad discretion and powerful means to control an important field of software may not be sound policy. Innovation in software, like innovation in ideas and fashions, potentially can occur and disseminate quickly at low cost. Detailed administrative control stifles innovation. The expressive aspect of software, a language of control, also makes establishing stable, constraining categories of regulatory control more difficult than for other less expressive areas, such as supplying standard electric currents. Moreover, software users much more naturally create social and technical networks than hardware users.[6] Some programmers unquestionably will figure out how to get around FCC control of radio software. Creating a large space for extra-legal software could undermine more generally the rule of law in radio regulation.

Giving adults freedom to program SDRs is a more important public interest than avoiding some risk of radio signal interference. SDRs make speaking in electromagnetic spectrum equivalent to speaking a programming language. This capacity for human expressiveness has an intuitive claim to freedom analogous to free speech.[1] Freely programmable SDRs could more easily adapt to dynamic patterns of radio use in a given place. This makes individual freedom in radio use less socially threatening. Mature and civilized radio users could more easily and intentionally accommodate each others' behavior. Radio regulation might thus recognize human intentionality to be not a threat but a good. In addition, SDRs allow large-scale, low-cost production of radio equipment that could be modified locally to adapt to needs in low-income countries with challenging climactic and geographic circumstances. Promoting a propitious environment for SDRs' development would foster expansion of communications capabilities to persons in disadvantageous circumstances.

---

[1] This advantage has statutory significance in the U.S. See 47 U.S.C. §157, and the Telecommunications Act of 1996, PL 104-104, February 8, 1996, 110 Stat 56, §706. The GNU Radio Project shows how SDRs can contribute to technological progress. See http://www.gnu.org/software/gnuradio/gnuradio.html

[2] FCC (2001e) para. 30.

[3] Ibid. para. 12-15, 18-20.

[4] Ibid., footnote 42. The order did not indicate how the FCC would categorize or divide the operating system software.

[5] See Lessig (1999) for an insightful exploration of software's regulatory significance.

[6] See, for example, the GNU Radio page at http://www.gnu.org/software/gnuradio/gnuradio.html

### C. Freedom in the House

Some kinds of personal freedom are more conventional than others. Freedom within the home from certain types of searches and seizures has deep historical and sentimental roots:

> *The special, intimate function served by the home was first legally recognized in antiquity. Greek and Roman law accorded the home special protection as the living area of the family, and as a place of worship and refuge. Common law expressions such as "My house is my castle" reflect the enduring prominence of the home in the development of Western law. The right to privacy in the home was first recognized as a fundamental right in the Virginia Bill of Rights from 1770, and it is embodied in the Fourth Amendment of the United States Constitution. In Germany, the inviolability of the home was first recognized as a fundamental right in the Paulskirchen Constitution of 1849. The Weimar Constitution also guaranteed the privacy of the home, but, as explained above, it offered no means of enforcing that right.*[2]

A U.S. judge explained:

> *A sane, decent, civilized society must provide some such oasis, some shelter from public scrutiny, some insulated enclosure, some enclave, some inviolate place-which is a man's castle.*[3]

What does this have to do with radio regulation?

In the mid-eighteen century, the invention of the Leyden jar spurred explorations and demonstrations of the wonders of electricity. Leyden jars can discharge powerful electrical currents and sparks. They provided a tool for exercising reason to better understand nature, and hence attracted considerable attention:

> *The Leyden jar revolutionized the study of electrostatics. Soon "electricians" were earning their living all over Europe demonstrating electricity with Leyden jars. Typically, they killed birds and animals with electric shock or sent charges*

[1] The issue here is the extent to which the government should control programming SDRs. Whether it is better for SDR software to be proprietary software or free software/open source is another question.

[2] Killian (2000) p. 192, ft. 129. The quotation above omits internal citations to Herdegen, Matthias, *Artikel 13*, in *Kommentar zum Bonner Grundgesetz* 8 (1993).

[3] United States v. On Lee, 193 F.2d 306, 315-6 (1951) (dissent).

> *through wires over rivers and lakes. In 1746 the abbé Jean-Anoine Nollet, a physicist who popularized science in France, discharged a Leyden jar in front of King Louis XV by sending current through a chain of 180 Royal Guards. In another demonstration, Nollet used wire made from iron to connect a row of Carthusian monks more than a kilometer long; when a Leyden jar was discharged, the white-robed monks reportedly leapt simultaneously into the air.*[1]

Undoubtedly, scientists, electricians and other active persons of the time regularly created large sparks and surging currents using Leyden jars and other equipment set up within their homes. About 150 years later, Henrich Hertz recognized that these activities broadcast radio frequency energy.

Taking this long view points to an important question about freedom. Whether persons exploring electricity in the mid-eighteenth century created radio frequency energy *intentionally* is a question best left to some academic philosophers. Perhaps one might question whether, in the mid-eighteen century, experimenting with electricity should have required government license. While some government officials might answer that question affirmatively, such a view seems extreme and inconsistent with the sort of freedoms that naturally exist.[2] But suppose that in some common law country in the mid-eighteenth century persons were required use Leyden jars only in accordance with relevant government regulations. Persons who violated these regulations could be subject to criminal proceedings. Would a law enforcement official be required to have more specific authorization than a general warrant to inspect a house for a Leyden jar being used in violation of the law?[3]

Questions of this sort deserve some thought. In the US, the FCC seems to assert that its agents may, without a specific warrant, inspect a house for any equipment that radiates radio frequency energy. Inspections must be allowed "without unnecessary delay"; the FCC suggests that "[i]mmediate on-the-spot inspections are generally necessary."[4] Unauthorized radio station operation in the U.S. has on occasion been grounds for criminal charges.[5] FCC field agents use equipment not in general public use to monitor, among other things, radio frequency energy emanating from houses. To persons with little legal learning, such searches of the home seem to violate the Fourth Amendment of the U.S. Constitution.[1]

---

[1] Katz (n.d.).

[2] The governing principle of a totalitarian society might be summarized as, "Anything not explicitly permitted is forbidden." Fortunately, that principle is rather difficult to put into practice.

[3] Davies (1999) provides numerous references to evidence that might be relevant to considering this question. But it is apparent from page one that the author's rhetorically transparent "authentic history" contains gross, tendentious misinterpretations. I am not a lawyer. I enjoy neither reading nor writing such articles. While I am willing to make sacrifices to serve the public, I decided that there is no compelling reason for me to read that 203 page paper. Readers with a different sense of public needs and intellectual enjoyment might reach a different decision.

[4] See FCC EB Inspection Fact Sheet (July 2000), online at http://www.fcc.gov/eb/otherinfo/inspect.html

[5] "FM Station Operator Convicted on Criminal Charges for Operating Without FCC Authorization," FCC News Release, Sept. 6, 2001; online at http://www.fcc.gov/eb/News_Releases/kubweza1.html; "FCC Investigation Leads to Arrest of Unlicensed FM Radio Operator," FCC News Release, May 25, 2001, online at http://www.fcc.gov/eb/News_Releases/nrusany0105.html. While willfully and knowingly violating the U.S. Communications Act is a criminal offense (see 47 U.S.C. 501), the FCC itself does not have authority to conduct criminal prosecutions.

The relationship between radio regulation and freedom in the home may not have mattered for practical reasons in the past, but this relationship is likely to be stressed in the future. Few persons, from their homes, without prior regulatory authorization, create of their own initiative the magnitude of radio frequency energy that Benjamin Franklin did in Philadelphia in the mid-eighteenth century.[2] Most persons simply use devices that radiate radio frequency energy according to hardware that others design. The number and diversity of radio devices used within the home are likely to grow rapidly in the future. Moreover, radio devices, such as SDRs, will give at least some users effectively greater freedom to control devices' radio emissions. Courts will likely confront difficult questions concerning how law enforcement officials search for unlawful radio devices and enter houses to inspect radio equipment that legally relevant evidence indicates are present.

Rather than await long and contentious court battles to establish the parameters of freedom within the home from searches for radio devices, policy makers should on their own initiative change radio regulations to recognize more formally causes for inspections for radio devices. Causes might be explicitly related to evidence that legally operating parties are experiencing harmful interference, and that this harm is probably related to illegal radio operation by another party. No documented harm, no cause for law enforcement search of a home for radio devices. Causes might also need to be identified, at least transitionally, with respect to clearly enumerated public safety issues on particular frequencies.[1] Recognizing "fixture bands," radio frequencies with use rights geographically circumcised by real estate boundaries, represents a narrower perspective on freedom applied to a more encompassing category of space.[2] But freedom within the home with respect to radio means more than just use rights.

---

[1] The tone of the questions in the "frequently asked questions" section of the FCC EB Inspection Fact Sheet, *available at* http://www.fcc.gov/eb/otherinfo/inspect.html, reflect this perception. A recent U.S. Supreme Court decision defining search in relation to electro-magnetic radiation seems relevant here. Kyllo v. United States, 121 S.Ct. 2038 (2001).

[2] Franklin personally explored the nature and uses of electricity. His work attracted world-wide attention. In a letter, available online at http://www.dctech.com/physics/humor/ben_franklin.php, he described a tremendous electrical shock he received in the course of one of his demonstrations. Franklin, often regarded as the quintessential American, concluded his account by passing on an anecdote about the stupidity of two Irishmen.

[1] Any vulnerability to interference of critical public safety functions, such as air traffic control, should be treated as a problem in itself. Public safety functions should be capable of dealing with deliberate, hostile interference. Thus they should in general be more robust to interference than other applications.

[2] Chartier (2001) puts forward an argument for "fixture bands."

## V. What To Do?

Regulations long established should not be changed for light and transient causes. For many years, a large area of radio regulation has developed based on a consensus that it predominately concerns technical aspects of radio signal interference. Few persons are able to contribute to deliberations thus organized. Yet centuries of conversation and experience have explored, in ways deeply relevant to everyone, the meaning of interference and freedom. Now interactive, broadband and ubiquitous communications, which undoubtedly will depend heavily on radio, are expected to reshape personal activities and relationships. Radio regulation should no longer be a field ruled by a reason inaccessible to most persons. Now is the time for radio regulation to recognize, as most persons do, revolutionary ideas about government, persons, and freedom.

What is one to be doing? Discuss the separation and balance of powers in radio regulation. Discuss the geography of governance in radio regulation. Discuss what freedom and equality mean for radio regulation. This paper provides much relevant, field-specific information and many references. More important is to call forth more vividly true lessons from the collective memory of humanity. Persons and groups always face the risk of forgetting, or remembering but separating thought from life. Everyone around the world can help show each other what there is to see.

Persons outside Western Europe and North America offer much hope. Just as a person might come to take for granted how beautiful her or his most dear one is, those closest to the revolutionary ideas discussed here are most likely to forget them. Especially if it seems easier to do so. One can imagine, decades from now, missionaries from Africa and Asia bringing back to life throughout Western Europe an offer of living Christianity. Neither Western Europe nor the U.S. should have to wait that long for the ideas needed to reform radio regulation.

Western Europe, North America, and other high income regions could benefit from looking outward. The Islamic world's emerging struggle with the meaning of freedom and authority is directly relevant to discussions that need to take place about radio regulation. Persons in Central and Eastern Europe have broad experience of the type of regulatory transition that most countries now face in the field of radio regulation. Revolutionary changes in radio regulation in Guatemala already offer insights that might be hugely beneficial.

Most importantly, Western Europe, North American, and other high-income regions should look outward to recognize the good of others. Even without any new developments in radio, the material conditions of life in high-income countries are, for the most part, sweet and easy. That is not the case for about three-quarters of the world's population. Persons in low and middle income countries desperately need better communications capabilities. As the President of Uganda has forcefully declared, Africa can work itself out of poverty under the right conditions.[1] Shrewd leaders in Africa and in low-income countries elsewhere might choose radically more liberal policies for radio regulation than those currently in place in high-income countries.[2] High-income countries foregoing such policies should at least ensure that they do not prevent international capital, the international radio equipment industry, and international radio regulations from responding to support enormously important choices that low-income countries might make.[3]

[1] See Yoweri Museveni, "How America Can Help Africa," Wall Street Journal, Commentary, May 24, 2002.

[2] Ibargen (2001) points out (p. 19): "Many wealthy countries in the West may waste many [radio resources] with no apparent impact on their economies." In other words, the West may be able to afford not to reform. The situation is different in poor countries.

[3] Malaria, tuberculosis, and HIV/AIDS are inflicting tremendous suffering in Africa. Culturally appropriate, personal communications, which depend on widespread dispersal of two-way communications capabilities, are critical for incarnating across the population knowledge about the causes and means of prevention of these diseases. Providing money and skills, in a mode of service rather than domination, would also help. See Jeffrey Sachs, "Bononomics Rocks," Wall Street Journal, June 6, 2002 and William Easterly, "Tired Old Mantras at Monterrey," Wall Street Journal, March 18, 2002.

## References

Amateur Radio Relay League [ARRL} (2000), *Now You're Talking! All You Need to Get Your First Ham Radio License*, 4'th ed., Larry D. Wolfgang and Joel P. Keinman, eds. (Newington, CT: ARRL).

*Annual Report of the Commissioner of Navigation to the Secretary of Commerce and Labor for the Fiscal Year Ended June 30, 1912* (Submitted November 13, 1912 from Commissioner Eugene Tyler Chamberlain to Secretary Charles Nagel), pp. 34-43; online at http://www.ipass.net/~whitetho/1912arcn.htm

Archer, John (2000), "Sex differences in aggression between heterosexual partners: A meta-analytic review," Psychological Bulletin, vol. 126, no. 5 (Sept.) pp. 651-680.

Australian Communications Authority [ACA] (1996), *Radiocommunications Spectrum Marketing Plan (500 MHz)*; online at http://auction.aca.gov.au/auction_results/500mhz_results_page/index.htm

ACA (2000), *Radiocommunications Spectrum Marketing Plan (2 GHz Band)*; online at http://www.aca.gov.au/spectrum/auctions/pdf%20files/marketing_plan.pdf

Barlow, John Perry (1996), "A Declaration of Independence of Cyberspace"; online at http://www.eff.org/~barlow/Declaration-Final.html

Barnett, Randy E. (2001), "The Original Meaning of the Commerce Clause," *University of Chicago Law Review* 68 (Winter) pp. 101-47.

Berkowitz, Peter (2000), "And Lofty Flows the Don," *The New Republic* (Nov. 13); online at http://www.thenewrepublic.com/111300/berkowitz111300_print.html

Bernstein, Lisa (2001), "Private Commercial Law in the Cotton Industry: Creating Cooperation Through Rules, Norms, and Institutions," *Michigan Law Review* 99 (June) pp. 1724-88; draft online at http://papers.ssrn.com/sol3/papers.cfm?abstract_id=281437

Beyer, Dave, Mark D. Vestrich, and JJ Garcia-Luna-Aceves (1996), "The Rooftop Community Network: Free, High-speed network access for communities," paper presented at the Harvard/DOE conference, The First 100 Feet: Options For Internet and Broadband Access, Arlington, VA, Oct 29-30; proceedings published in by Deborah Hurley and James H. Keller (MIT Press, 1999); online at http://www.ksg.harvard.edu/iip/doeconf/first.html

Blondheim, Menahem (1994), *News over the wires : the telegraph and the flow of public information in America, 1844-1897* (Cambridge, Mass.: Harvard University Press).

Bonner, J. Rebekka S. (2002), "Reconceptualizing VAWA's 'Animus' For Rape In States' Emerging Post-VAWA Civil Rights Legislation," Yale Law Journal, 111 (April) pp. 1417-56.

Bostick, C. Dent (1988), "Land Title Registration: An English Solution to an American Problem," *Indiana Law Journal*, vol. 63 (Winter) pp. 55-111.

Brock, Ralph H. (1999), "The Ghost in the Computer: Radio Frequency Interference and the Doctrine of Federal Preemption," *Computer Law Review and Technology Journal*, Vol. 1999, No. 1 (Fall 1998-Spring 1999); online at http://www.sbot.org/docs/RFI.pdf

Brown, George D. (2001), "Constitutionalizing the Federal Criminal Law Debate: Morrison, Jones, and the ABA," 2001 *Univ. of Illinois Law Review* (Nov.) pp. 983-1024; draft online at http://papers.ssrn.com/sol3/papers.cfm?abstract_id=293202

Buck, Stuart (2001), "Replacing Spectrum Auctions with a Spectrum Commons," forthcoming in *Stanford Technology Law Review*; online at http://papers.ssrn.com/sol3/papers.cfm?abstract_id=268744

Burns, John, Paul Hansell, Helena Leeson, and Phillipa Marks (2001), "Implications of international regulation and technical considerations on market mechanisms in spectrum management," Report to the Independent Spectrum Review, 1320/AE/SMM/R/3 (6 Nov.); online at http://www.spectrumreview.radio.gov.uk/newsite/report.htm

Cahoon, Colin (1990), "Low Altitude Airspace: A Property Rights No-Man's Land," *Journal of Air Law and Commerce* 56 (Fall) pp. 157-198.

Cameron, David (2002), "Power Play," *Technology Review*, March 27, 2002; online at http://www.techreview.com/articles/print_version/wo_cameron032702.asp

Carter, Meg (1998), *Independent radio: the first 25 years* (London: RSA on behalf of The Radio Authority) online at http://www.radioauthority.org.uk/publications-archive/adobe-pdf/25yrshis.pdf

Cave, Martin (2002), *Review of Radio Spectrum Management: An independent review for Department of Trade and Industry and HM Treasury* (March); online at http://www.spectrumreview.radio.gov.uk

Chamberlain, Joseph P. (1927) "Department of Current Legislation: The Radio Act of 1927," *American Bar Association Journal*, vol. 13, p. 343-6.

Chartier, Mike (2001), "Enclosing the Commons: A Real Estate Approach to Spectrum Rights," paper presented at the AEI-Brookings Joint Center Confererence "Practical Steps to Spectrum Markets," Nov. 9.

Codding, George A. Jr. (1952/1972), *The International Telecommunication Union* (New York: Arno Press).

Codding, George A., Jr. and Anthony M. Rutkowski (1982), *The International Telecommunication Union in a Changing World* (Dedham, MA: Artech House Inc.).

Commission of the European Communities [CEC] (1994), *Towards the personal communications environment: Green Paper on a common approach in the field of mobile and personal communications in the European Union*, COM(94) 145.

CEC (1998), *Green Paper on Radio Spectrum Policy*, Brussels 09.12.1999, COM(1998)596; online at http://europa.eu.int/information_society/topics/telecoms/radiospec/radio/index_en.htm

Communications Act (1934) (as enacted), "An Act to provide for the regulation of interstate and foreign communication by wire or radio, and for other purposes," Public Law No. 416, June 19, 1934, 73d Congress; online at http://showcase.netins.net/web/akline/1934act.htm; also in Paglin (1989).

Constant, Benjamin (1819), "The Liberty of Ancients Compared with that of Moderns," online at http://www.uark.edu/depts/comminfo/cambridge/ancients.html.

Cushman, Barry (2000), "Formalism and Realism in Commerce Clause Jurisprudence," *University of Chicago Law Review*, 67 (Fall) pp. 1089-1150.

Davies, Thomas Y. (1999), "Recovering the Original Fourth Amendment," *Michigan Law Review*, Vol. 98 (Dec.) pp. 547-749.

Davis, Donald R. and David E. Weinstein, "Bones, Bombs and Break Points: The Geography of Economic Activity," NBER Working Paper 8517 (Oct.), online at http://www.nber.org/papers/w8517

Davis, Stephen (1927), *The Law of Radio Communication* (New York: McGraw-Hill Book Co.)

Department of Commerce, U.S. [DOC] (2001), *U.S. Statistical-Purpose Areas*, Conference of European Statisticians, Tallinn, Estonia, 25-28 Sept., Working Paper Nol 31, online at http://www.unece.org/stats/documents/2001/09/gis/31.e.pdf

Department of Commerce, Bureau of Navigation (1913b), *Regulations Governing Radio Communication*, edition of July 1,1913 (Washington: GPO); online at http://www.ipass.net/~whitetho/1913breg.htm

Department of Commerce and Labor, Bureau of Navigation (1912), *Regulations Governing Radio Communication*, edition of Sept. 28, 1912 (Washington: GPO); online at http://www.ipass.net/~whitetho/1912reg.htm

Department of Commerce and Labor, Bureau of Navigation (1913a), *Regulations Governing Radio Communication*, edition of February 20, 1913 (Washington: GPO); online at http://www.ipass.net/~whitetho/1913areg.htm

Derrida, Jacques (1988), *Limited Inc* (Evanston, IL: Northwestern University Press).

*Directive 2002/21/EC of the European Parliament and of the Council, on a common regulatory framework for electronic communications networks and services (Framework Directive)*, 7 March 20002, published in the Official Journal of the European Communities, 24.4.2002, L 108/33, online at http://europa.eu.int/information_society/topics/telecoms/regulatory/new_rf/index_en.htm

Downs, Charles F., II, "Calvin Coolidge, Dwight Morrow, and the Air Commerce Act of 1926," essay online at http://www.calvin-coolidge.org/pages/history/research/ccmf/downs.htm

Ellickson, Robert C. and Charles Dia. Thorland, "Ancient Land Law: Mesopotamia, Egypt, Israel," 71 Chi.-Kent L. Rev. 321 (1995).

European Commission [EC] (1999), *Next Steps in Radio Spectrum Policy: Results of the Public Consultation on the Green Paper*, Brussels, 10 Nov. 1999 COM(1999)538, online at http://europa.eu.int/information_society/topics/telecoms/radiospec/radio/index_en.htm

European Parliament (2002), *European Parliament and Council directive concerning the processing of personal data and the protection of privacy in the electronic communications sector* (A5-0130/2002); see http://www.eurunion.org/news/press/2002/2002034.htm

European Parliament (2002b), *Directive 2002/21/EC of the European Parliament and of the Council, on a common regulatory framework for electronic communications networks and services (Framework Directive)*, 7 March 20002, published in the Official Journal of the European Communities, 24.4.2002, L 108/33, online at http://europa.eu.int/information_society/topics/telecoms/regulatory/new_rf/index_en.htm

European Radiocommunications Committee [ERC] (1995), *Harmonised Regime for Exemption from Individual Licensing of Radio Equipment*, ERC/Recommendation 01-07 E (Bonn 1995, Helsinki 2000); online at http://www.ero.dk/doc98/official/pdf/REC0107E.PDF

Federal Communications Commission [FCC] (1936), *First Annual Report of the Federal Communications Commission to the Congress of the United States for the Fiscal Year 1935* (Jan. 6) (Washington: GPO); online at http://www.fcc.gov/mb/audio/decdoc/engrser.html#EARLY

FCC (1952), *Sixth Report on Television Allocations* (released April 14, 1952), Radio Regulations 91:601 (rel. Apr. 14).

FCC (1971), Establishment of Policies and Procedures for Consideration of Application to Provide Specialized Common Carrier Services in the Domestic Public Point-To-Point Microwave Radio Service and Proposed Amendments to Parts 21, 43, and 61 of the Commission's Rules, First Report and Order, Docket No. 18920, 22 Rad. Reg. 2d (P & F) 1501 (rel. June 3).

FCC (1979), Amendment of Part 15 to redefine and clarify the rules governing restricted radiation devices and low power communication devices, First Report and Order, Docket No. 20780, 79 FCC 2d 28 (rel. Oct. 11).

FCC (1983), Elimination of Individual Licenses in Radio Control Radio Service and Citizens Band Radio Service, Report and Order, PR Docket No. 82-799, 48 FR 24884 (rel. May 10, 1983).

FCC (1985), In the Matter of 960 Radio, Memorandum Opinion and Declaratory Ruling, FCC 85-578 (rel. Nov. 4).

FCC (1986), Amendment of Parts 15, 68, and 76 of the Commission's Rules to Require Labelling of Home Electronic Equipment and Systems Susceptible to Interference from Radio Frequency Energy, Memorandum Opinion and Order, 1 FCC Rcd. 289 (rel. Oct. 29).

FCC (1987), In the Matter of David Hildebrand, Memorandum Opinion and Order, PR Docket No. 81-302, 2 FCC Rcd. 2708 (rel. Apr. 29).

FCC (1987b), In the Matter of Mobilecom of New York, Memorandum Opinion and Declaratory Ruling, DA 87-1237, 2 FCC Rcd 5519 (rel. Aug. 31).

FCC (1988), FCC Regulations Concerning RF Lighting Devices, Memorandum Opinion and Order, Gen. Docket No. 83-806, 3 F.C.C.R. 6097 (rel. Oct. 20).

FCC (1989), Revision of Part 15 of the Rules regarding the operation of radio frequency devices with an individual license, First Report and Order, Gen. Doc. No. 87-389, 4 FCC Rcd 3493 (rel. Apr. 18, 1989).

FCC (1990), Letter from Robert L. Pettit, General Counsel, FCC, to Christopher D. Imlay, ARRL (Feb. 14); online at http://www.arrl.org/FandES/field/regulations/rfi-legal/pettit.html

FCC (1991), Letter from Robert H. McNamara, Chief, Special Services Division, FCC, to Arthur R. Still, (27 Nov.); online at http://www.arrl.org/FandES/field/regulations/rfi-legal/mcnamara.html

FCC (1993), Amendment of the Commission's Rules to Establish New Personal Communications Services, Second Report and Order, GEN Docket No. 90-314, 8 FCC Rcd 7700 (rel. Oct. 22).

FCC (1994), Implementation of Sections 3(n) and 332 of the Communications Act, Regulatory Treatment of Mobile Services, Third Report and Order, GN Docket No. 93-252, 9 FCC Rcd 7988 (rel. Sept. 23).

FCC (1994b), Letter from Ralph A. Haller, Chief, Private Radio Bureau, FCC, to Board of Zoning Appeals, Town of Hempstead, NY (25 Oct.); online at http://www.arrl.org/FandES/field/regulations/rfi-legal/haller.html

FCC (1995), Application for Review of Stephen Paul Dunifer, Berkeley, California, Memorandum Opinion and Order, FCC 95-333, 11 FCC Rcd 718 (rel. Aug. 2).

FCC (1995b), Amendment of Part 90 of the Commission's Rules to Facilitate Future Development of SMR Systems in the 800 MHz Frequency Band, First Report and Order, PR Docket No. 93-144, 11 FCC Rcd 1463 (rel. Dec. 15).

FCC (1995c), Amendment of the Commission's Rules Concerning the Inspection of Radio Installations on Large Cargo and Small Passenger Ships, Notice of Inquiry, CI Docket No. 95-55, 10 FCC Rcd 5424 (rel. May 16).

FCC (1995d), The FCC Votes To Restructure Its Compliance & Information Bureau, Report No. CI 95-16, 1995 WL 603456 (Oct. 13).

FCC (1996), Geographic Partitioning and Spectrum Disaggregation by Commercial Mobile Radio Services Licensees, Report and Order and Further Notice of Proposed Rulemaking, WT Docket NO. 96-148, 11 FCC 21831 (rel. Dec. 20).

FCC (1996b), Amendments of Parts 80 and 87 of the Commission's Rules to Permit Operation of Certain Domestic Ship and Aircraft Radio Stations Without Individual Licenses, Report and Order, WT Docket No. 96-82, 11 FCC Rcd 14849 (rel. Oct. 25, 1996).

FCC (1996c), Amendment of Part 80 of the Commission's Rules Regarding the Inspection of Great Lakes Agreement Ships, Report and Order, CI Docket 95-54, 11 FCC Rcd 18661 (rel. Apr. 26).

FCC (1996d), Voluntary Part 68 Reporting of Telephone RFI/Belltap Compliance," Public Notice (Oct. 9).

FCC (1996e), RFI- and Belltap-Immune Telephone Products, Public Notice (Nov. 14).

FCC (1996f), FCC Policy for Handling Complaints of Interference to Home Electronics Equipment, Public Notice (Apr. 5); online at http://www.fcc.gov/Bureaus/Compliance/News_Releases/nrci5009.txt

FCC (1997), Amendment of the Commission's Rules to Establish Part 27, the Wireless Communications Service ("WCS"), Report and Order, GN Docket No. 96-228, 12 FCC Rcd 10785 (rel. Feb. 19).

FCC (1998), Amendment of the Commission's Rules Concerning the Inspection of Radio Installations on Large Cargo and Small Passenger Ships, Report and Order, CI Docket No. 95-55, 13 FCC Rcd 13556 (rel. May 1).

FCC (1998b), Privatized Ship Radio Inspections Meet FCC Goals, Report No. CIB 98-23, 1998 WL 786457 (Nov. 13).

FCC (1998c), Ramsey L. Woodworth, Richard Rubin, Letter from Howard C. Davenport, Chief, Enforcement and Consumer Information Division, Wireless Telecommunications Bureau, 13 FCC Rcd 2486 (Jan. 30).

FCC (1999a), In the Matter of 1998 Biennial Regulatory Review – Conducted Emissions Limits Below 30 MHz for Equipment Regulated under Parts 15 and 18 of the Commission's Rules, Notice of Proposed Rule Making, ET Docket No. 98-80, (rel. Oct. 18, 1999), online at http://ftp.fcc.gov/Bureaus/Engineering_Technology/Notices/1999/fcc99296.txt

FCC (1999b), In the Matter of 1998 Biennial Regulatory Review – Amendment of Part 18 of the Commission's Rules to Update Regulations for RF Lighting Devices, First Report and Order, ET Docket No. 98-42 (rel. June 16), online at http://ftp.fcc.gov/Bureaus/Engineering_Technology/Orders/1999/fcc99135.pdf

FCC (2000), Amendment of Section 95.413 of the Commission's Rules Prohibiting Communications or Attempts to Communicate with Citizens Band Radio Service Stations More Than 250 Kilometers Away, Order, RM-9807 (rel. Aug. 21); online at http://hraunfoss.fcc.gov/edocs_public/attachmatch/DA-00-1907A1.pdf

FCC (2001), Definition Of Markets For Purposes Of The Cable Television Broadcast Signal Carriage Rules, Order on Reconsideration, CS Docket No. 95-178, 16 FCC Rcd 5022 (rel. Mar. 2).

FCC (2001b), Hughes Communications Inc., Application for Authority to Construct, Launch, and Operate a Ka-Band Satellite System in the Fixed-Satellite Service, DA 01-1686, 16 FCC Rcd 14,310 (rel. Aug. 3).

FCC (2001c), 1998 Biennial Regulatory Review – Amendment of Part 97 of the Commission's Amateur Radio Service Rules, Memorandum Opinion and Order, WT Docket No. 98-143; online at http://wireless.fcc.gov/services/amateur/releases/

FCC (2001d), Amendment of Section 95.413 of the Commission's Rules Prohibiting Communications or Attempts to Communicate with Citizens Band Radio Service Stations More Than 250 Kilometers Away, Order on Reconsideration, RM-9807 (rel. Aug. 1); online at http://hraunfoss.fcc.gov/edocs_public/attachmatch/DA-01-1831A1.pdf

FCC (2001e), Authorization and Use of Software Defined Radios, First Report and Order, ET Docket No. 00-47 (rel. Sept. 14); online at http://www.fcc.gov/Bureaus/Engineering_Technology/News_Releases/2001/nret0106.html

FCC (2001f), In re Application of VoiceStream Wireless Corp. and PowerTel Inc., Transferors, and Deutsche Telekom AG, Transferee, Memorandum Opinion and Order, IB Docket No. 00-187 (rel. Apr. 27); online at http://www.fcc.gov/Bureaus/International/Orders/2001/fcc01142.pdf

FCC (2002a), Fiscal Year 2003 Budget Estimates, Submitted to Congress, February, 2002; online at http://www.fcc.gov/Reports/fcc2003budget.html

FCC (2002b), "Telecommunications Industry Revenues 2000," Jim Lande and Kenneth Lyn, IAD CCB, Jan., online at http://www.fcc.gov/Bureaus/Common_Carrier/Reports/FCC-State_Link/IAD/telrev00.pdf

"Federal Control of Radio Broadcasting," *Yale Law Journal*, v. 39, pp. 247-56.

Federal Radio Commission [FRC] (1931*), Fifth Annual Report, Fiscal Year 1931* (Washington: GPO); online at http://www.fcc.gov/fcc-bin/assemble?docno=311207

FRC (1932), *Sixth Annual Report, Fiscal Year 1932* (Washington: GPO); online at http://www.fcc.gov/fcc-bin/assemble?docno=321205

Feyerabend, Paul (1975), *Against method: outline of an anarchistic theory of knowledge* (London: Humanities Press).

Field, Alexander J. (1992), "The Magnetic Telegraph, Price and Quantity Data, and the New Management of Capital," *Journal of Economic History*, vol. 52, no. 2 (June) pp. 401-13.

Field, Alexander J. (1998), "The Telegraphic Transmission of Financial Asset Prices and Orders to Trade: Implications for Economic Growth, Trading Volume, and Securities Market Regulation, in *Research in Economic History*, vol. 18 (JAI Press) pp. 145-84.

Fish, Stanley (1999), *The trouble with principle* (Cambridge, MA: Harvard University Press).

Fisher, Elizabeth (2001), “Unpacking the Toolbox: Or Why the Public/Private Divide is Important in EC Environmental Law,” Florida State University College of Law, Public Law and Legal Theory Working Paper No. 35 (Aug.); online at http://papers.ssrn.com/abstract=283295

Fradkin, Hillel (2002), “Parting of the Ways II: Jewish and Islamic Thought and 9/11,” Lecture delivered to the American Enterprise Institute; online at http://www.aei.org/past_event/conf020513.htm

Freeman, Jody (1999), “Private Parties, Public Functions and the New Administrative Law,” draft online at http://papers.ssrn.com/sol3/papers.cfm?abstract_id=165988

Futurepace Solutions (2001), “Response to UK Independent Radio Spectrum Management Review,” (19 Sept.); online at http://www.spectrumreview.radio.gov.uk/responses/futurepace19sept.pdf

Galbi, Douglas A. (2001a), “Communications Policy, Media Development, and Convergence,” online at http://www.galbithink.org and http://www.ssrn.com

Galbi, Douglas A. (2001b), “A New Account of Personalization and Effective Communication,” online at http://www.galbithink.org and http://www.ssrn.com

Gernsback, H. (1913), “Wireless and the Amateur: A Retrospect,” *Modern Electronics* (Feb.) pp. 1143-4.

Glaeser, Edward, Simon Johnson, and Andrei Shleifer (2001), “Coase versus the Coasians,” *Quarterly Journal of Economics*; online draft at http://post.economics.harvard.edu/faculty/shleifer/papers/qjecoasepaper.pdf

Glaeser, Edward L. and Andrei Shleifer (2001a), “Legal Origins,” Harvard Institute of Economic Research Discussion Paper No. 1920, online at http://papers.ssrn.com/paper.taf?abstract_id=267852

Glaeser, Edward L. and Andrei Shleifer (2001b), “The Rise of the Regulatory State,” online at http://papers.ssrn.com/sol3/papers.cfm?abstract_id=290287

Hagen, Jennifer (2001), “Can We Lose The Battle And Still Win The War?: The Fight Against Domestic Violence After The Death Of Title III Of The Violence Against Women Act,” *DePaul Law Review*, vol. 50 (Spring) pp. 919-91.

Hamilton, Scott J. (1994), “Allocation of Airspace as a Scarce National Resource,” *Transportation Law Journal*, vol. 22, pp. 251-90.

Harsch, Bradley A. (2001), "Finding a Sound Commerce Clause Doctrine: Time to Evaluate the Structural Necessity of Federal Legislation," *Seton Hall Law Review*, vol. 31, pp. 983-1041.

Hayne, Ian (1997), "Spectrum property rights and practical auction design: the Australian experience," Plenary Session 1, Invited paper 11, 1997 Industry Economics Conference, online at http://www.pc.gov.au/ic/research/confproc/iec1997/pap11.pdf

Hazlett, Thomas W. (2001a), "The Wireless Craze, The Unlimited Bandwidth Myth, The Spectrum Auction Faux Pas, and the Punchline to Ronald Coase's 'Big Joke': An Essay on Airware Allocation Policy," AEI-Brookings Joint Center for Regulatory Studies, Working Paper 01-01 (Jan.); online at http://www.aei.brookings.org/publications/abstract.asp?pID=117

Hazlett, Thomas W. (2001b), "The U.S. Digital TV Transition: Time to Toss the Negroponte Switch," AEI-Brookings Joint Center for Regulatory Studies, Working Paper 01-15 (Nov.); online at http://www.aei.brookings.org/publications/abstract.asp?pID=178

Hazlett, Thomas W. and Bruno E. Viani (2002), "Legislators v. Regulators: The Case of Low Power FM Radio," AEI-Brookings Joint Center for Regulatory Studies, Working Paper 02-1 (Feb.); online at http://www.aei.brookings.org/publications/working/working_02_01.pdf

Hirschman, Albert O. (1977), *The Passions and the Interests* (Princeton: Princeton Univ. Press).

Hirschman, Albert O. (1958), *The strategy of economic development* (New Haven, Yale University Press).

House Report (1910), *Radio Communication*, Mr. Greene, from the U.S. Committee on the Merchant Marine and Fisheries, Report to accompany H.R. 23595, 61'st Congress, 2n'd Session, House of Representatives, Report No. 924 (April 1).

Hundt, Reed (2000), *You Say You Want a Revolution: A Story of Information Age Politics* (New Haven and London: Yale University Press).

Ibarguen S., Giancarlo (2001), "Liberating the Radio Spectrum in Guatemala," paper presented at AEI, Nov. 9, 2001, forthcoming in *Telecommunications Policy*.

Isern, Josep and Maria Isabel Rios (2002), "Facing disconnection: Hard choices for Europe's telcos," *The McKinsey Quarterly*, No. 1; online at http://www.mckinseyquarterly.com/

IARU (2001), "ITU Adopts Recommendation on Amateur Qualifications," News Release, 23 Aug. 2001; online at http://www.iaru.org/rel010823.html

ITU (2001), *ITU Telecommunication Indicators Update*, April-May-June 2001, online at http://www.itu.int/ITU-D/ict/update/

ITU (2002a), "Key Global Telecom Indicators for the World Telecommunication Service Sector," online at http://www.itu.int/ITU-D/ict/statistics/at_glance/KeyTelecom99.html

ITU (2002b), "Basic Indicators, 2000", online at http://www.itu.int/ITU-D/ict/statistics/

ITU (2002), *ITU Telecommunication Indicators Update*, April-May-June 2000, online at http://www.itu.int/ITU-D/ict/update/

Jehiel, Philippe and Benny Moldovanu, "The European UMTS/IMT-2000 License Auctions," online at http://www.enpc.fr/ceras/jehiel/umts1.pdf

Johnson, Kenneth P. (1995), "Redefinition of the BEA Economic Areas, Survey of Current Business, (Feb.) pp. 75-81; online at http://www.bea.doc.gov/bea/ARTICLES/REGIONAL/PROJ/1995/0295rea.pdf

Jones, Adam (2000), "Gendercide and Genocide," Journal of Genocide Research, vol.2, no. 2 (June), pp. 185-211, online at http://www.gendercide.org/gendercide_and_genocide.html

Kahn, Frank J., ed. (1984), *Documents of American Broadcasting*, 4'th ed., (Englewood Cliffs, NJ: Prentice-Hall Inc.).

Katz, Eugenii (n.d.), "Leyden Jars," online at http://chem.ch.huji.ac.il/~eugeniik/instruments/archaic/leyden_jars.htm

Kennard, William E. (2000), "The FCC's New Enforcement Ethic," Remarks before the Competitive Carrier Summit 2000, Conference on Current U.S. Telecom Policy, Washington, DC (Jan. 19); online at http://www.fcc.gov/Speeches/Kennard/2000/spwek003.html

Killean, James J. (2000), "Der Grosse Lauschangriff: Germany Brings Home the War on Organized Crime," *Hastings International and Comparative Law Review*, vol. 23 (Winter) pp. 173-215.

King, Martin Luther, Jr. (1963), *Why We Can't Wait* (New York: NAL Penguin).

Kwerel, Evan R. and John R. Williams (1992), *Changing Channels: Voluntary Reallocation of UHF Television Spectrum*, FCC OPP Working Paper No. 27 (Nov.), NTIS PB93 114874.

Kwerel, Evan and John Williams (2001), "A Proposal for a Rapid Transition to Market Allocation of Spectrum," Paper presented at the AEI-Brookings Joint Center conference

"Practical Steps to Spectrum Markets" (Nov. 9); online at http://www.aei.brookings.org/events/011109/kwerel.pdf

Krattenmaker, Thomas G. and Lucas A. Powe, Jr. (1994), *Regulating Broadcast Programming* (Cambridge, MA; London: MIT Press).

Lee, Blewett (1925), "Power of Congress Over Radio Communication," *American Bar Association Journal*, v. 11, pp. 19-21.

Lemley, Mark (2001), "Rational Ignorance at the Patent Office," UC Berkeley School of Law Public Law and Legal Theory Working Paper No. 46 (Feb.); online at http://papers.ssrn.com/paper.taf?abstract_id=261400.

Lessig, Lawrence (1999), *Code and other laws of cyberspace* (New York: Basic Books).

Lessig, Lawrence (1998), "Understanding Federalism's Text," *George Washington Law Review* 66 (June-August) pp. 1218-37.

Lessig, Lawrence (1995), "Translating Federalism: United States v. Lopez, 1995 *Supreme Court Review*, pp. 125-215.

Lloyd, Marshall Davies (1998), "Polybius and the Founding Fathers: the separation of powers," online at http://www.sms.org/mdl-indx/polybius/polybius.htm

Lohr, Steve (1999), "Privacy On Internet Poses Legal Puzzle," *New York Times*, Apr. 19,1999; online at http://www.lcs.mit.edu/news/privacy.html

Lubrano, Annteresa (1997), *The telegraph: how technology innovation caused social change* (New York: Garland Pub.).

Lyotard, Jean-François (1984),*The postmodern condition: a report on knowledge* (*Condition postmoderne*), trans. by Geoff Bennington and Brian Massumi (Minneapolis: University of Minnesota Press).

MacIntyre , Alasdair (1988), *Whose justice? Which rationality?* (Notre Dame: University of Notre Dame Press).

Madison, James (1787), *Federalist No. 10*, online at http://memory.loc.gov/const/fed/fed_10.html

Marcus, Norman (1984), "Air Rights in New York City: TDR Zoning Lot Merger and Well-Consider Plan," *Brooklyn Law Review*, vol. 50 (Summer) pp. 867-911.

Maritain, Jacques (1928), *Three Reformers: Luther – Descartes -- Rousseau* (London: Sheed & Ward).

Merritt, Deborah Jones (1998), "The Third Translation of the Commerce Clause: Congressional Power to Regulate Social Problems," *George Washington Law Review*, vol. 66 (June-Aug.) pp. 1206-17.

Mitola, Joseph III (1996), *Software Radio - Cognitive Radio: Wireless Architectures for the 21st Century*; online at http://ourworld.compuserve.com/homepages/jmitola/

Morgenson, Gretchen (2001), "Telecom's Pied Piper: Whose Side Was He On," *The New York Times*, Nov. 18.

Morton, Robert A. (1909), "Wireless Interference," *Electrician and Mechanic*, April, pp. 422-427; online at http://www.ipass.net/~whitetho/1909ama.htm

Morton, Robert A. (1910), "The Amateur Wireless Operator," *The Outlook*, January 15, pp. 131-35; online at http://www.ipass.net/~whitetho/1910ama.htm

Mueller, Milton (2002), *Ruling the root: Internet governance and the taming of cyberspace* (Cambridge, Mass.: MIT Press); excerpt online at http://istweb.syr.edu/~mueller/

Muldrew, Craig (1998), *The economy of obligation: the culture of credit and social relations in early modern England* (NY: St. Martin's Press).

Mutschler, Linda (2002), "World View: Re-wiring the weak wireless industry," *Financial Times*, Feb. 17, 2002.

Nelson, Grant S. and Robert J. Pushaw (1999), "Rethinking the Commerce Clause: Applying First Principles to Uphold Federal Commercial Regulations But Preserve State Control Over Social Issues," *Iowa Law Review* vol. 85 (Oct.) pp. 1-172,

Neumann, A. Lin (2001), "The Great Firewall," Committee to Protect Journalists, Briefing, online at http://www.cpj.org/Briefings/2001/China_jan01/China_jan-1.html

"Note and Comment: The Radio and Interstate Commerce," 26 Mich. L. Rev. 919-21 (1928).

Oftel (2001), "The UK Telecommunications Industry: Market Information 2000/01," (Dec.); online at http://www.oftel.gov.uk/publications/market_info/2001/mia1201.pdf

Oftel (2002), "Market Information, Mobile Update", July to September 2001, published January 2002; online at http://www.oftel.gov.uk/publications/market_info/

Orange (2002), *Annual Report 2001*; online at http://www.orange.com/agm2002/2001%20annual%20report.pdf.

Ostrom, Elinor (1990), *Governing the commons: the evolution of institutions for collective action* (Cambridge [England]: Cambridge University Press).

Paglin, Max D. (1989), *A Legislative History of the Communications Act of 1934* (New York: Oxford University Press).

Paulu, Burton (1967), *Radio and Television Broadcasting on the European Continent* (Minneapolis, Univ. of Minnesota Press).

Peale, Norman Vincent (1956), *The Power of Positive Thinking* (New York, NY: Prentice-Hall).

Pierce Wells, Catharine (2001), "A Pragmatic Approach to Improving Tort Law," *Vanderbilt Law Review*, Vol. 54, No. 3 (Apr.) pp. 1447-65; online at http://law.vanderbilt.edu/lawreview/vol543/wells.pdf

Pitofsky, Robert (1979), "The Political Content of Antitrust," *University of Pennsylvania Law Review*, vol. 27, pp. 1051-75.

Polanyi, Karl (1944), *The Great Transformation* (New York: Farrar & Rinehart).

Powell, Michael K. (2001), Letter sent the leaders of the Senate and House Commerce and Appropriations Committees (May 4); online at http://www.fcc.gov/Bureaus/Common_Carrier/News_Releases/2001/nrcc0116.html

Productivity Commission (2002), Radiocommunications, Draft Report, AusInfo, Canberra; online at http://www.pc.gov.au/inquiry/radiocomms/draftreport/index.html

Pyle, Howard S. (1924), "Shake Hands with the 'R.I'", *Radio Broadcast* (Dec.) pp. 289-94; online at http://www.angelfire.com/nc/whitetho/shakeRI.htm

Radio Amateur News (1920), "The Autobiography of a Girl Amateur," *Radio Amateur News* (March) p. 490; online at http://www.angelfire.com/nc/whitetho/1920auto.htm

"The Radio and Interstate Commerce" (1928), *Michigan Law Review*, v. 26, pp. 919-21.

Radio Act (1912), "An Act to regulate radio communication," Public Law No. 264, Aug. 13, 1912, 62d Congress; online at http://www.ipass.net/~whitetho/1912act.htm and http://showcase.netins.net/web/akline/1912act.htm; also in Kahn (1984).

Radio Act (1927), "An Act for the regulation of radio communications, and for other purposes," Public Law No. 632, Feb. 23, 1927, 69'th Congress; online at http://showcase.netins.net/web/akline/1927act.htm; also in Kahn (1984).

Radio Division (1932), U.S. Dept. of Commerce, *Annual Report of the Director of the Radio Division to the Secretary of Commerce for the Fiscal Year Ended June 30, 1932* (Washington: GPO).

Radio Spectrum Management Group [RSMG] (1998), *Electromagnetic Noise Monitoring*, PIB 32, Ministry of Commerce, NZ (Wellington, NZ); online at http://www.med.govt.nz/rsm/publications/pibs/pib32.pdf

Radiocommunications Consultative Council [RCC] (1999), *Report of the Working Group on IMT-2000* (Dec.), online at http://auction.aca.gov.au/auction_results/2ghz_results_page/pdf/imtreport.pdf

Radio Regulations [RR] (1927), *International Radiotelegraph Convention and General Regulations, signed at the International Radiotelegraph Conference ending in Washington, Nov. 25, 1927* (Washington: GPO).

RR (1932), *International Telecommunication Convention, Madrid 1932, and annexed General Radio Regulations*, published in International Radiotelegraph Conference: Madrid 1932, U.S. Dept. of State Conference Series No. 15 (Washington: GPO).

RR (1947), *International Radio Regulations Annexed to the International Telecommunications Convention* (Atlantic City, 1947) (London: HMSO).

RR (1959), *Radio Regulations* (ITU: Geneva).

Rand McNally and Company (1992), *Commercial Atlas and Marketing Guide* (Chicago).

Rappaport, Jordan (1999), "Local Growth Empirics," Center for International Development Working Paper No. 23 (July), Harvard University, online at http://www.cid.harvard.edu/cidwp/023.htm

Rappaport, Jordan and Jeffrey Sachs (2001), "The U.S. as a Coastal Nation," Working Paper version July 1, 2001, online at http://www.kc.frb.org/publicat/reswkpap/rwp01-11.htm

Raymond, Eric S. (2000), "A Brief History of Hackerdom," online at http://tuxedo.org/~esr/writings/hacker-history/hacker-history.txt

Raymond, Eric S. (1997), "The Cathedral and the Bazaar," online at http://tuxedo.org/~esr/writings/cathedral-bazaar/

Raymond, Eric S. (2001), "How To Become a Hacker," online at http://tuxedo.org/~esr/faqs/hacker-howto.html

Raymond, Eric S. (2001b), Jargon File, version 4.3.1, 29 Jun 2001, online at http://www.tuxedo.org/~esr/jargon/index.html; book version available as *The New Hacker's Dictionary* (MIT Press).

Rennison, Callie Marie (2001), U.S. Bureau of Justice Statistics Special Report, Intimate Partner Violence, 1993-1990, NCJ 187635 (Oct.), online at http://www.ojp.usdoj.gov/bjs/pub/pdf/ipva99.pdf

Resnik,Judith (2001), "Categorical Federalism: Jurisdiction, Gender, and the Globe," *Yale Law Journal* 111 (Dec) pp. 619-80.

Rorty, Richard (1979), *Philosophy and the mirror of nature* (Princeton, NJ: Princeton University Press).

Rosen, Philip T. (1980), The Modern Stentors: Radio Broadcasting and the Federal Government, 1920-1934 (Westport, CN: Greenwood Press).

Rubin, Edward L. (1997), "The Fundamentality and Irrelevance of Federalism," *Georgia State Law Review* 13 (July) pp. 1009-.

Rubin, Edward L. and Malcolm Feeley (1994), "Federalism: Some Notes on a National Neurosis," *UCLA Law Review 41* (April) pp. 903-52.

Sachs, Jeffrey D. (2001), "Tropical Underdevelopment," NBER Working Paper 8119, online at http://www.nber.org/papers/w8119

Scalia, Antonin (1997), *A matter of interpretation* (Princeton, NJ: Princeton University Press).

Schmeckebier, Laurence F. (1932), *The Federal Radio Commission* (Washington: The Brookings Institution).

Scott, Jonathan (2000), *England's Troubles* (Cambridge, UK: Cambridge University Press).

Senior, T.B.A., D.L. Sengupta, and J.E. Ferris (1977), *TV and FM Interference by Windmills*, Final Report, 1 January 1976-21 December 1976, Energy Research and Development Administration, U.S. Dept. of Commerce, NTIS COO-2846-76-1.

Ship Act (1910), "An Act to require apparatus and operators for radio communication on certain ocean steamers" (approved June 24, 1910, amended July 23, 1912), online at http://www.ipass.net/~whitetho/1910act.htm

Silverstein, Shel (1974), *Where the Sidewalk Ends* (New York, N.Y.: HarperCollins).

Skinner, Quentin (1998), *Liberty Before Liberalism* (Cambridge: Cambridge Univ. Press).

Skomal, Edward N. (1978), *Man-Made Radio Noise* (New York: Van Nostrand Reinhold Co.).

Stallman, Richard (1998), "The GNU Project," online at http://www.gnu.org/gnu/the-gnu-project.html

Taugher, James Patrick (1928), "The Law of Radio Communication with Particular Reference to a Property Right in Radio Wavelength," *Marquette Law Review*, v. 12 n. 3 (April) pp. 179-91; no. 4 (June) pp. 299-317.

Tauranac, John (1995), *The Empire State Building: the making of a landmark* (NY: Scribner).

Taussig, Charles William (1922), *The book of radio* (New York, London: D. Appleton and Co.); excerpt online at http://www.ipass.net/~whitetho/1922ama2.htm

Taylor, Greg (2001), "The Commerce Clause – Commonwealth Comparisons," *Boston College International and Comparative Law Review*, 24 (Spring) pp. 235-51.

Technology Review (2001), "Unwiring the Web," Dec., pp. 22-23; online at http://www.personaltelcom.net/images/MIT2.jpg

Tomlinson, John D., *The International Control of Radiocommunications*, Thèse No. 41, Universitè de Genève (Genève: Journal de Genève, 1938).

Unger, Roberto Mangabeira (1987), *Politics, a work in constructive social theory; part 0, Social theory, its situation and its task; part 1, False necessity: anti-necessitarian social theory in the service of radical democracy* (Cambridge, New York: Cambridge University Press).

United States [US] (1927), International Radiotelegraph Conference of Washington, Draft of International Radio Convention, based on the proposals of the government of the United States of America for revisions of the International Radiotelegraph Convention of London of 1912 (Washington: GPO).

Washington Conference (1927), *International Radiotelegraph Convention, and General and Supplementary Regulations* (Washington: Government Printing Office).

Whitman, Dale A. (1999), "Digital Recording of Real Estate Conveyances," *John Marshall Law Review*, vol. 32 (Winter) pp. 227-68.

Whitman, James Q. (1996), "The Moral Menace of Roman Law and the Making of Commerce: Some Dutch Evidence," *Yale Law Review*, vol. 105 (May) pp. 1841-89.

Williams, John R. (1986), “Private Frequency Coordination in the Common Carrier Point-to-Point Microwave Service,” OPP Working Paper No. 21 (Sept.).

Williams, Sam (2002), *Free as in Freedom: Richard Stallman’s Crusade for Free Software* (O’Reilly); online version at http://www.oreilly.com/openbook/freedom/

Williams, Stephen F. (2002), “Radical Reform: Transitions to Liberal Democracy and the Rule of Law,” Bradley Lecture at the American Enterprise Institute, Jan. 7; online at http://www.aei.org/bradley/bl020107.htm

Williamson, Oliver (1985), *The economic institutions of capitalism: firms, markets, relational contracting* (New York: Free Press).

Zollmann, Carl (1927) “Radio Act of 1927,” *Marquette Law Review*, v. 11 (Apr.) pp. 121-7.

Zollmann, Carl (1930), *Cases on Air Law*, 1’st ed (St. Paul, MN: West Publishing).

## Appendix A
## Extended Notes to Tables

### Table 3

License totals for services other than broadcast services were tabulated from licenses with "active" status in FCC ULS data in April, 2002. Current ULS data are available at http://wireless.fcc.gov/cgi-bin/wtb-datadump.pl

Radio service codes were grouped according to headings given in the FCC list of radio services. See http://wireless.fcc.gov/uls/radioservices.html

In Table 3, the land (mobile) category combines:

- Cellular (CL)
- GMRS - General Mobile Radio Services (ZA)
- Land Mobile Commercial (LC)
- Land Mobile Private (LP)
- Paging (PG)
- Personal Communication Services - PCS (PC)

The land (fixed) category combines:

- Coast and Ground (CG)
- Microwave (MW)

The other categories in Table 3 map directly to associated headings in the radio services list.

Broadcast station totals are from FCC News Release, "Broadcast Station Totals as of September 30, 2001," (rel. Oct. 30, 2001); online at http://www.fcc.gov/Bureaus/Mass_Media/News_Releases/2001/nrmm0112.pdf

The "broadcast (FM radio)" category combines:

- FM commercial and educational (8275 stations)
- FM Translators and Boosters (3600 stations)

The "broadcast (television)" category combines:

- UHF and VHF commercial and educational TV stations (1686 stations)
- UHF and VHF translators (4762 stations)
- Class A TV stations (424 stations)
- Low Power TV stations (2212 stations)

The term of non-broadcast licenses was calculated as the average time period between the grant date and the expired date in the data record for the license (table HD in ULS). Records lacking entries for either of these dates were omitted in calculated the average license term.

47 U.S.C. 307(c)(1) states that a license for a broadcast station "shall be for a term of not to exceed 8 years." This figure was used for the term of broadcast licenses.

## Table 4

Monthly lists of FCC Field Office actions are available at http://www.fcc.gov/eb/rfo/actions.html

Field actions were categorized based on the second-level rule heading under which actions are listed. Notices of Violations under part 11 and part 17 headings comprise the emergency alert system and antenna rules categories, respectively. The cable system standards category is Notices under 47 C.F.R. § 76.605. The protocol and procedures category includes violations in these areas: station identification, station logs, inspection files and license information, and procedure in filing applications and responding to FCC notices. Free-to-air radio signal rules are all other notices. The two largest groups in this category are Notices under "47 C.F.R. § 1.903 – Authorization Required" (28 Notices) and "'47 C.F.R. § 90.403 – General Operating Requirements" (18 Notices).

Some stations are cited for multiple rule violations. The pattern of citations among multiple rule violations is similar to that for the primary headings.

## Table 5

U.S. states are a partition of U.S. counties. All geographies used thus far in FCC auctions are also partitions of counties. In Table 5, net bids for auctioned areas have been re-allocated, on a relative population basis, to the counties that make up the area. These county values were then summed across auctions for state totals. Table A1 shows the auctions included in the figures in Table 5, as well as other information related to the individual auctions. All offerings at auction of a nation-wide partition of spectrum, with auctioned areas about the size of states or smaller, are included in Table A1.

**Table A1**
**Auctions Included in Table 5**

| Auction | Auction Name | Geography | Geographic Areas | Licenses per Area | Net Bids | Notes |
|---|---|---|---|---|---|---|
| 4 | PCS A,B broadband | MTA | 51 | 2 | 8,149,470,541 | 1 |
| 5 | PCS C broadband | BTA | 493 | 1 | 10,069,802,341 | 2 |
| 6 | MMDS | BTA | 493 | 1 | 215,925,477 | |
| 7 | SMR | MTA | 51 | 20 | 204,143,678 | |
| 11 | PCS D,E,F | BTA | 493 | 3 | 2,516,367,759 | |
| 14 | WCS | MEA | 51 | 2 | 9,105,213 | 3 |
| 16 | SMR upper | EA | 175 | 3 | 95,380,410 | |
| 17 | LMDS | BTA | 493 | 2 | 578,467,404 | |
| 21 | LMS | EA | 175 | 3 | 3,431,404 | |
| 26 | Paging | MEA | 51 | 49 | 4,101,795 | |
| 30 | 39 Ghz | EA | 175 | 14 | 410,589,490 | |
| 33 | Upper 700 Guard Band | MEA | 51 | 2 | 519,839,250 | |
| 34 | SMR | EA | 175 | 6 | 318,730,110 | 4 |
| 36 | SMR lower | EA | 175 | 16 | 28,950,740 | |
| 40 | Paging lower | EA | 175 | 80 | 12,696,887 | 5 |
| 41 | PCS narrowband | MTA | 51 | 7 | 3,646,286 | 6 |
| 42 | Mult Address Systems | EA | 175 | 29 | 1,180,975 | |

Notes from last column of Table A1:

1. “A” block licenses for Washington, DC; New York; and LA/San Diego were award as pioneer preferences prior to the auction. Net bid values for the first two areas were imputed based on the A/B block net bid ratio for Boston-Providence. The net bid value for LA/San Diego was imputed based on the A/B net bid ratio for San Francisco.

2. Winning bidders defaulted on $874 million in net bids. The figure used is net bids, not excluding defaults.

3. Bids associated with 12 regional economic groups, a coarse partition, were not included.

4. Licenses offered in the 800 MHz Upper Band (861-865 MHz) were not included (in particular, a license for Honolulu).

5. Licenses unsold in Auction No. 26 (1514 upper band paging licenses) were not included. That allows, for particularly frequencies, the net figures used to represent a simultaneous geographic division of value.

6. Bids for eight nation-wide licenses were not included (too coarse of a partition).

Additional notes:

1. Areas defined for American Samoa, Northern Mariana Islands, and the Gulf of Mexico were excluded because they were not consistently auctioned, or population and area data were not available.

2. The Table is not strictly a list of states: it includes Washington, DC, and Puerto Rico, which are not U.S. states.

3. For the definition of auction geographies (MEA, EA, etc.), see the main text, Section III D.

4. For a summary of FCC auctions, see http://wireless.fcc.gov/auctions/summary.html#completed

Sources: For FCC auction data, see http://wireless.fcc.gov/auctions/data/roundresults.html
A county cross-reference database for the auction geographies is available at http://wireless.fcc.gov/auctions/data/maps.html#pops. That dataset also provides county population and area figures. The figures in Table 5 were calculated using the 1990 revised population figures.

**Table 8**

The figures for amateur radio users are from reports from national amateur radio societies to the International Amateur Radio Union. The figures are for "total operators" for the year 2000. See http://www.iaru.rog/statsum00.html

The figures for mobile telephony users and Internet users are from the ITU, year 2001 figures. See http://www.itu.int/ITU-D/ict/statistics/ For information about the quality of Internet access statistics, see Michael Mingers, "Counting the Net: Internet Access Indicators," ITU, online at http://www.isoc.org/isoc/conferences/inet/00/cdproceedings/8e/8e_1.htm

Population figures, which are for the year 2000, are from the World Bank's World Development Indicators database. See http://devdata.worldbank.org/data-query/

The sample consists of 164 countries and economies for which all the above data are available. For a full list of the individual countries and economies, along with income category, region, and users per thousand, see Appendix C.

## Appendix B
## A Desperate Case under the Commerce Clause: U.S. Courts' Role in Determining the Federal-State Balance in Radio Regulation

From about 1937 to 1994, the Commerce Clause of the U.S. Constitution was thought not to impose any effective limits on federal power.[1] During that period the Supreme Court turned aside every challenge it heard to federal authority under the Commerce Clause.[2] When the Constitution was written, you could not fly across the country in a matter of hours, you could not talk with a person thousands of miles away, nor could everyone all across the U.S. watch an event occurring in Washington, or on the moon. Our economy seems much more unified now. Federalism itself smells of slavery, succession, and dead bodies. Discussing federalism as a matter of constitutional law seems like an anachronism, a pointless and dangerous diversion, or a betrayal of our national unity.[3] Surely progress means defending the New Deal and upholding the supremacy of our national legislative process in areas classified as economic or social. Why should courts care about federalism? Why should courts care about the Commerce Clause, a most implausible aspect of federalism?

Federal radio regulation shows effects of not judiciously articulating federal power under the Commerce Clause. In the 1920s, most persons considered radio to be AM radio broadcasting. Many other uses of radio were known at the time, but they were much less popular. Fruitful development of radio broadcasting was widely thought to require a scheme of federal regulation. With remarkably little attention to statutory construction and judicial process, the courts allowed that consensus to become the basis for federal law covering all radio use. This federal-state balance has endured, despite relevant technological, social, political, and economic changes, through to the present. Thus an expedient and widely acclaimed solution to the regulatory crisis of the day has led to federal law covering all radio use.

This unprecedented legal development offers the opportunity to better understand the meaning of the Commerce Clause. The natural sense of the word "commerce" was much different in the eighteenth century than it is today. Commerce then meant intercourse, ongoing, deeply enmeshing relationships among persons. While much has changed since the eighteenth century, general patterns of human behavior indicate that most persons still seek to realize themselves more fully through relationships with others. Such relationships are not necessarily alternatives to the relationships that define the national polity.[1] Human relationships concern, for the most part, mundane human activities. The Commerce Clause, understood in its eighteenth century meaning, provides constitutional protection against legal developments that might too easily suppress broad categories of relationships.[2] The Commerce Clause should have provided constitutional protection against federal suppression of radio use supporting distinctive local relationships but no substantial interstate relation. Understanding the Commerce Clause in this way leads to a better interpretation of important Commerce Clause cases. Understanding the Commerce Clause in this way also helps to articulate a better federal-state balance in radio regulation.

[1] The Commerce Clause is a name commonly used for a clause in Article I, Sec. 8, of the U.S. Constitution. Article I., Sec. 8, enumerates powers of the federal government. These powers include the power "To regulate commerce with foreign nations, and among the several states, and with the Indian tribes". This statement of federal power is called the Commerce Clause.

[2] Nelson and Pushaw (1999) pp. 79-86.

[3] Rubin and Feeley (1994) describes federalism as a "national neurosis," while Rubin (1997) argues that federalism is enjoyable to discuss because it has long been and remains irrelevant to any real issues in the U.S. That is not the case in Australia and Canada. In those federations, constitutional clauses similar to the Commerce Clause of the U.S. Constitution have real, contemporary significance. See Taylor (2001).

### B-I. Case History of the Federal-State Balance in Radio Regulation

Three district court decisions in the late 1920s seem to have been formally influential in interpreting the Commerce Clause with respect to radio regulation. In *Whitehurst v. Grimes* (decided Sept. 17, 1927), an amateur radio operator brought suit in a district court in Kentucky. The amateur radio operator sought to void a municipal ordinance imposing a license tax on radio operation. In the third and concluding paragraph of its decision, the Court declared:

> *Radio communications are all interstate. This is so, though they may be intended only for intrastate transmission; and interstate transmission of such communications may be seriously affected by communications intended only for intrastate transmission. Such communications admit of and require a uniform system of regulation and control throughout the United States, and Congress has covered the field by appropriate legislation. It follows that the ordinance is void, as a regulation of commerce.* [3]

The breadth of this decision, along with its economy of analysis and forthright statement of policy necessity, points to contemporary scholarly consensus regarding radio regulation.[4] This decision, in turn, made this consensus more significant by giving it legal sanction.

In *White v. Federal Radio Commission* (decided Oct. 16, 1928), a district court in Illinois addressed even more concisely the federal-state balance in radio regulation. The plaintiffs sought to enjoin enforcement of a Federal Radio Commission (FRC) order.

---

[1] Cf. Rubin (1997), which focus on human affiliations as a matter of political identity. Williams (2002) emphasizes that rule of law does not mean "a society in which most or lots of issues are resolved by 'law,' i.e. rules created and enforced by the state, or by agencies using the power of the state." He points out that forming voluntary organizations is important to liberal democracy because such actions provide practice in working out relationships. It is not necessary that the relationships themselves have intrinsic political significance, nor that they are alternatives to a particular type of political affiliation.

[2] Tocqueville (1835) argues that if persons in a democracy "did not acquire the practice of associating with each other in ordinary life, civilization itself would perish." Without the counterbalancing effect of growth in civic associations across a range of desires and objects, Tocqueville felt that more democratic societies would tend more toward torpor in sentiments and ideas. See ibid, vol. 2, section 2, Chapter V.

[3] 21 F.2d 787.

[4] See Section II.B of the main paper.

One of the motions for action was that the Radio Act of 1927 is unconstitutional. The court denied this motion with a single sentence:

> *The regulation of radio communication is a valid exercise of the power of Congress under the commerce clause.*[1]

Subsequent cases cited *White* as establishing that Congress can regulate all radio communications under the Commerce Clause.

The judge that decided *White* provided a more extensive opinion in *United States v. American Bond & Mortgage Co.* (decided Mar. 1, 1929). The U.S. Attorney General sought injunction against the defendant broadcasting without a federal license in Illinois, about twenty-five miles south of Chicago. The defendant attacked the constitutionality of the Radio Act of 1927 and questioned "the scope of the act as a proper regulation of interstate and foreign commerce."

The Court's judgment in *American Bond & Mortgage* takes different forms. Anticipating the Communications Amendments Act of 1982, the Court described the Radio Act of 1927 as asserting government control "over all channels of radio transmissions" rather than "over all channels of interstate and foreign radio transmissions."[2] In addition, the decision provided data relating the power of a radio station to the radius of service, differentiated by service quality. The terrain, weather conditions, and quality of receiving equipment were not specified, and the data source was not given. The Court presented this data as establishing necessary conditions:

> *It is also apparent that the nuisance area of a station using power of 50 watts (the lowest for which a station is licensed) will almost certainly pass beyond the borders of the state in which the broadcasting station is located. This may cause interference within the state with radio waves coming into the state from other states; with broadcasting stations within the state whose service area extends beyond the state; with broadcasting stations in other state which are broadcasting across other state lines. The engineering showing demonstrates that unified public control is essential to secure to the owners of broadcasting stations an assured channel, which they may use without interference, and to secure to the listening public satisfactory reception of the programs broadcast on those channels.*[3]

Given that the FRC refused to license broadcasting stations with less than 50 watts power, the significance of this power level to the validity of federal licensing of all radio communications is not clear. But analysis of the proper scope of federal regulation of radio communication wasn't, after all, really necessary for the Court:

> *It does not seem to be open to question that radio transmission and reception among the states are interstate commerce. … A device in one state produces energy which reaches every part, however small, of the space affected by its power.…It is intercourse, and that intercourse is commerce. …*

---

[1] 29 F.2d 113, 114.

[2] 31 F.2d 448 at 451. For discussion of the Communications Amendment Act of 1982, see Section II.B of the main paper.

[3] Ibid at 453. The "radius of nuisance area" for a 50 watts station was listed as 300 miles.

> *The provisions of the act prescribed the only method by which order could be brought out of chaos and this form of interstate commerce saved from destruction.*[1]

This decision has been regularly cited as support for the proposition that all radio communications can be regulated under the Commerce Clause. The Court noted that emphasis in analyzing radio regulation should be "on the receiving public, whose interest it is the duty of the Government, *parens patriae*, to protect."[2]

Beginning in 1929, the DC Circuit issued *dicta* that, although often inconsistent and indirect, played an important role in providing judicial support for federal regulation of all radio communications. Under the Radio Act of 1927, this Court handled all appeals of FRC decisions. Appeals typically concerned license denials. An appeal of a FRC license denial would not seem to present for judgment the constitutionality of federal licensing. Nonetheless, in addressing appeals of FRC decisions, the Court repeatedly made statements and references addressing the constitutionality of federal licensing.

> *General Electric Co. vs. Federal Radio Commission* (decided Feb. 25, 1929):
> *This act [Radio Act of 1927], which is yet in force, forbids all radio broadcasting in this country, except under and in accordance with a license granted under the provisions of the act. ... under the commerce clause of the Constitution (article 1, Sec. 8, cl. 3), Congress has the power to provide for the reasonable regulation of the use and operation of radio stations in this country.... Without such national regulation of radio, a condition of chaos in the air would follow...Davis, Law of Radio, p. 71; Zollman, Law of the Air, pp. 102, 103.*[3]

> *Technical Radio Lab. v. Federal Radio Commission* (decided Nov. 4, 1929):
> *...Congress has power to regulate interstate commerce, and radio communication in general falls within this classification. Whitehurst v. Grimes (D.C.) 21 F.(2d) 787; 35 Op. Attys. Gen. 126; White v. Federal Radio Commission (D.C.) 29 F.(2d) 113; United States v. Am. Bond & Mtg. Co. (D.C.) 31 F.(2d) 448; Davis, Law of Radio Communication, p. 29.*[4]

> *City of New York v. Federal Radio Commission* (decided Nov. 4, 1929):
> *In our opinion the interstate broadcasting of radio communications is a species of interstate commerce, and as such is subject to federal regulations. Whitehurst v. Grimes (D.C.) 21 F.(2d) 787; United States v. American Bond & Mortgage Co. (D.C.) 31 F.(2d) 448; General Electric Co. v. Federal Radio Commission, 58 App.D.C. 386, 31 F.(2d) 630; Davis, Law of Radio, 71; Zollman, Law of the Air, 102, 103.*[5]

> *Chicago Fed. of Labor v. Federal Radio Commission* (decided May 5, 1930):

---

[1] Ibid at 454 (first three sentences quoted are from the same paragraph), 456.
[2] Ibid at 455.
[3] 58 App.D.C. 386, 31 F.2d 630.
[4] 59 App.D.C. 125, 36 F.2d 111.
[5] 59 App.D.C. 129, 36 F.2d 115.

*This statement [FRC cannot arbitrarily withdraw licenses] does not imply any derogation of the controlling rule that all broadcasting privileges are held subject to the reasonable regulatory powers of the United States…*[1]

*KFKB Broadcasting Ass'n v. Federal Radio Commission* (decided Feb. 2, 1931): *We have held that the business of broadcasting, being a species of interstate commerce, is subject to the reasonable regulation of Congress. Technical Radio Lab. v. Fed. Radio Comm., 59 App.D.D. 125, 36 F. (2d) 111, 66 A.LR. 1355; City of New York v. Fed. Radio Comm., 59 App.D.C. 333, 41 F. (2d) 422.*[2]

*Trinity Methodist Church v. Federal Radio Commission* (decided Nov. 28, 1932): *We have already held that radio communication, in the sense contemplated by the act, constituted interstate commerce, KFKB Broadcasting Ass'n v. Federal Radio Commission, supra; General Elec. Co. v. Federal Radio Commission, 58 App. D.C. 386, 31 F.(2d) 630, and in this respect we are supported by many decisions of the Supreme Court, Pensacola Telegraph Co. v. Western Union Tel Co., 96 U.S. 1,9, 24 L.Ed. 708; International Text-Book Co. v. Pigg, 217 U.S. 91, 106, 107, 30 S. Ct. 481, 54 L Ed. 678, 27 L.R.A. (N.S.) 493, 18 Ann. Cas. 1103; Western Union Teleg. Co. V. Pendleton, 122 U.S. 347, 356, 7 S.Ct. 1126, 30 L.Ed. 1187.*[3]

These statements supported federal regulation of all radio communications in a variety of formally interesting ways. They also present a largely unexplored constitutional "reasonableness" constraint on at least this area of federal legislation. But most importantly, these statements show that, from 1929 to 1932, the DC Circuit struggled to establish a solid legal foundation for federal regulation of all radio communications.

Supreme Court *dicta* aided this project. In *Federal Radio Commission v. Nelson Bros. Bond & Mortgage Co.* (decided May 8, 1933), the Court considered whether the FRC could grant a licensing to a radio broadcaster in one state, and, to avoid interference, terminate licenses to two radio broadcasters in another state. The Court declared:

*No question is presented as to the power of the Congress, in its regulation of interstate commerce, to regulate radio communications. No state lines divide the radio waves, and national regulation is not only appropriate but essential to the efficient use of radio facilities.*[4]

The first sentence above is literally true. The following sentence, however, points to the Court's position: Federal regulation of all radio communications is unquestionably legal under the Commerce Clause. The Court cited three of the above DC Circuit decisions in support of this position. But the Court's statement suggests that this position seemed not to have required legal justification then.

[1] 59 App.D.C. 333, 41 F.2d 422.
[2] 60 App.D.C. 79, 47 F.2d 670.
[3] 61 App.D.C. 311, 62 F.2d 850.
[4] 289 U.S. 266, 279, 53 S.Ct. 627, 633-4.

The most recent Supreme Court *dicta* that bear any formal resemblance to a ruling came in *Fisher's Blend Station v. Tax Commission of State of Washington* (decided Mar. 30, 1936). In this case, the licensee of two stations, one broadcasting across eleven states, another broadcasting across the U.S., sought to enjoin the Tax Commission from collecting tax on its stations. The arguments concerned whether factors other than the range of broadcasting implied that the station operator was engaged in intrastate commerce. The court rejected the arguments and voided the tax as a tax on interstate commerce. The Court also declared:

> *By its very nature broadcasting transcends state lines and is national in its scope and importance – characteristics which bring it within the purpose and protection, and subject it to the control, of the commerce clause. See Federal Radio Commission v. Nelson Bond & Mortgage Co., 289 U.S. 266, 279, 53 S.Ct. 627, 77 L.Ed. 1166.*[1]

This statement clearly pertains to AM radio broadcasting. That it has more general legal significance for the federal-state balance of power in regulating radio use seems questionable.

Subsequent Supreme Court cases simply describe the radio and communications statutes as establishing federal control over all radio use. Compare the texts of those statutes, which are quoted and analyzed in Part II.B of the attached main paper, to the Court's descriptions of them. In *Federal Communications Commission v. Pottsville Broadcasting Co.* (1940), the Court declared:

> *By this Act [Radio Act of 1927] Congress, in order to protect the national interest involved in the new and far-reaching science of broadcasting, formulated a unified and comprehensive regulatory system for the industry.*[2]

*National Broadcasting Co. v. United States* (1943) pushed a similar description back to the Radio Act of 1912:

> *The Statute [Radio Act of 1912] forbade the operation of radio apparatus without a license from the Secretary of Commerce and Labor;*[3]

By 1963, *Head v. New Mexico Board of Examiners in Optometry*, the description became an abbreviated statutory reference placed in a footnote:

> *It is to be noted that this case in no way involves the Commission's jurisdiction over technical matters such as a frequency allocation, over which federal control is clearly exclusive. 47 U.S.C. s 301.*[4]

None of the three quotes above are literally correct. The jurisdictional qualifications written in the federal communications statutes have not merely been rendered without effect. They have been obliterated – one can read the text of the statutes and literally not see them.

Court decisions through to the present do not appear to recognize the problems with precedent in this area. The status of federal control of all radio use is perceived to have been authoritatively settled. In 1984, *Federal Communications Commission v. League of Women Voters of California*, all the Supreme Court had to say was this:

> *First, we have long recognized that Congress, acting pursuant to the Commerce Clause, has power to regulate the use of this scarce and valuable national resource [radio spectrum].*[1]

[1] 297 U.S. 650, 655, 56 S.Ct. 608, 609.
[2] 309 U.S. 134, 137, 60 S.Ct. 437, 439.
[3] 319 U.S. 190, 210, 63 S.Ct. 997, 1007.
[4] 374 U.S. 424, 430, 83 S.Ct. 1759, 1763, n. 6.

The cited case history most directly relevant is *National Broadcasting Co. v. FCC* and *Federal Radio Comm'n v. Nelson Bros. Bond & Mortgage Co*. See the analysis of these cases above. Lower courts have generally followed the Supreme Court's lead. For example, a Circuit Court decision in 1994 declared:

> *As the Supreme Court recognizes, the FCC's jurisdiction "over technical matters" associated with the transmission of radio signals "is clearly exclusive." Head v. New Mexico Board of Examiners in Optometry, 374 U.S. 424, 430 n. 6, 10 L.Ed.2d 983 (1963);*[2]

Whether bad precedent can be converted into good precedent through time and repetition seems to be the relevant legal question here.

To see existing precedent in action, consider an FCC Order that addressed the constitutionality of an unlicensed low power radio station. The station was broadcasting with a power of about 10 watts from an apartment in Berkeley, CA in 1993. According to a determination based on established FCC technical regulations, a 10 watt station with a 100 meter antenna has a service radius of 5.9 kilometers. The Order noted that the Communications Act (as amended in 1982) requires a license for intrastate radio transmissions. To support the constitutionality of requiring a federal license, the Order cited *National Broadcasting Co. v United States* and *FCC v. League of Women Voters of California*. The Order noted:

> *There is no support whatever in these or any other Supreme Court decisions for [the station operator's] view that the licensing requirement of Section 301 of the Communications Act cannot be enforced unless there is a specific showing that the unlicensed radio transmissions caused or may cause harmful interference.*[3]

The FCC saw no need to consider statutory and legal history or the evolution of radio technology and applications. What appeared to be a clear, general legal rule simply decided the case.

One can find, however, analysis that dared to show more sensitivity to particular, contemporary facts and that exercised more analytical discipline. In 1927, an important treatise on radio law noted:

> *Perhaps the applicable rule can be expressed by saying that if the station is in fact an isolated unit, keeping its emanations within the state boundaries, separate and apart from the general mass of interstate communications, it remains free from Federal control.*[4]

This statement was clearly meant to have no practical effect. It succeeded. Court rulings cited the treatise to support the position that all radio communication is necessarily interstate. While most courts saw no need to maintain space for deliberation about effects and jurisdiction, a district court in 1931 impressively reasoned:

> *The plaintiff contends that all radio communication is necessarily interstate, and in the present state of the art, this appears to be correct. However, it is not inconceivable that radio communication may in the future be so perfected that it may be confined strictly intrastate; but we do not consider it necessary to make any ruling upon that point now.*[1]

---

[1] 468 U.S. 364, 376, 104 S.Ct. 3106, 3115.
[2] Broyde v. Gotham Tower, 13 F.3d 994, 997.
[3] FCC (1995) at 727.
[4] Davis (1927) p. 29.

Other parts of the opinion contradict the above statement.[2] The court seemed unable to resist making statements supporting the prevailing view that all radio communications are categorically interstate commerce.

The most careful, fact-based analysis of jurisdiction is in *United States v. Gregg*. A district court in Texas decided *Gregg* on Jan. 10, 1934.[3] The case concerned a 2-4 watt radio station in Houston, called "Voice of Labor, Inc," which had a service radius of approximately 30 miles. The Court found that the station did not ordinarily communicate with receivers in other states or on vessels at sea; it did not ordinarily have effects in other states; it did not ordinarily interfere with radio communications from Texas to other states; and it did not interfere with communications among states other than Texas. The Court noted, however, that other stations in Miami, Fla., and Shreveport, La. broadcast on the frequency that "The Voice of Labor" used. "The Voice of Labor"'s broadcasts interfered with reception in Houston of those interstate broadcasts. The Court thus found that "The Voice of Labor" was prohibited under the Radio Act of 1927 from operating without a license.

The Court then addressed the constitutionality of the portion of the Radio Act of 1927 that forbid "The Voice of Labor" from broadcasting without a license. The Court reviewed in detail a number of cases regarding regulation of interstate commerce. After three and half pages of such review, the Court curiously added the following paragraph:

> *See, also, cases involving various phases of radio broadcasting: American Bond & Mortgage Co. v United States (C.C.A.) 52 F.(2d)319; Station WBT v. Poulnot (D.C.) 46 F.(2d) 671; Whitehurst v. Grimes (D.C.) 21 F.(2d) 787; United States v. American Bond & Mfg. Co. (D.C.) 31 F.(2d) 448; General Electric Co. v. Federal Radio Commission, 58 App.D.C. 386, 31 F.(2d) 630, 631; White v. Federal Radio Com. (D.C.) 29 F.(2d) 113.*[4]

---

[1] *Station WBT v. Poulnot*, 46 F.2d 671, 675 (1931).

[2] "All radio communication, anywhere in the United States, travels actually or potentially across state lines, and even if certain radio electric wave energy, through an accident or otherwise, should lose its force before crossing the state line, yet it potentially interferes with other radio communication passing interstate. Congress has assumed control of all communications by radio, acting through the Federal Radio Commission, which activity and continuously supervises all such communication." Ibid at 672. "There can be no doubt that communications by radio constitutes interstate commerce." Ibid at 675.

[3] 5 F.Supp. 848. On March 4, 1934, the Supreme Court decided *Nebbia v New York*, 291 U.S. 502, 54 S.Ct. 505. This decision essentially eliminated the existing distinction between public and private businesses and effectively expanded the range of the Commerce Clause. See Cushman (2000) pp. 1132-7.

[4] 5 F. Supp. 848, 857.

These cases may have gone unanalyzed because their value as precedent was not related to specific written rulings in them. The Court followed that paragraph immediately with its ruling:

> *In light of these cases and many others that may be cited, I have no difficulty in concluding that the Congress may lawfully, as has been done in this act, require the licensing and regulation of intrastate radio broadcasting stations where, as here, the operation thereof interferes with interstate commerce.*

Note that this ruling is narrow: it relates to radio broadcasting of the type discussed in the case. The Court, deciding in equity, then quickly found the Radio Act of 1927 reasonable:

> *That it is reasonable will be seen by reflecting that a sufficient number of unlicensed and unregulated intrastate radio broadcasting stations, such as is defendants', broadcasting on different frequencies in each community, could and would not only interfere with, but destroy, all interstate radio broadcasting.*

This statement contrasts sharply with the style of reasoning and findings in the rest of the opinion. It reasons from an extreme hypothetical far removed from any actual findings.

Precedent for federal control of all radio use, though widely regarded as rock-solid, at best settles the matter legally just for AM radio broadcasting. *Stare decisis* is an important legal principle. But in considering "technical" matters related to "radio" frequencies, courts might apply *stare decisis* much more carefully and much more narrowly. Areas of human activity that once might have been considered technical radio matters are now central public concerns.[1] Radio regulation, to promote the public interest, needs judicious legal review.

### B-II. Misunderstanding Commerce in the Real World

In legal deliberations in the 1920s, AM radio broadcasting was abstracted into radio communications, and radio waves were analogized to instruments of commerce such as railroads and the telegraph.[2] Thus a natural aspect of the physical world – electromagnetic spectrum – was implicitly conceptualized as an instrument. The sense that the physical world is not an instrument but part of creation, and very good, was lost. Dominion over the physical world, in the sense of good stewardship of a household, can easily be mistaken for merely ruling over. Keeping the physical world, in the sense of keeping a garden, can be easily mistaken for keeping a purchase. Both dominion and gardening evoke relationships. AM radio broadcasting, as historically incarnated, can be characterized in terms of relationships among persons. But all activities associated with a particular physical aspect of the world, e.g. activities using (radio) spectrum, cannot be articulated in this way. Because of an unappreciated transformation of its subject, radio regulation developed a totalizing approach radically different from regulation for other aspects of the physical world.

---

[1] That at least is the argument of the main paper.

[2] Davis (1927) pp. 29-31, Zollmann (1927) pp. 102-3, "Note and Comment" (1928) p. 920, Taugher (1928) pp. 313-4.

To see how easily this transformation occurred, consider an important, non-judicial opinion. Early in 1926, the Commerce Department apparently was pondering the legality of its implementation of the Radio Act of 1912. On June 4, 1926, the Secretary of Commerce requested an opinion from the Attorney General on this question:

> *Does the 1912 Act require broadcasting stations to obtain licenses, and is the operation of such a station without a license an offense under the Act?*[1]

On July 8, 1926, the Attorney General replied:,

> *There is no doubt whatever that radio communication is a proper subject for Federal regulation under the commerce clause of the Constitution. Pensacola Telegraph Company v. Western Union Telegraph Company, 96 U.S. 1, 9, 24 L. Ed. 708; 24 Op. Attys. Gen. 100. And it may be noticed in passing that even purely intrastate transmission of radio waves may fall within the scope of Federal power when it disturbs the air in such a manner as to interfere with interstate communication, a situation recognized and provided for in the Act.*[2]

From the question to the answer there is a shift from (AM radio) broadcasting to radio communication. This shift in formal categories might have been of little real significance in 1926. At that time, for the vast majority of persons, radio communications practically meant such broadcasting. But the shift had deep, subsequent significance for radio regulation.

Federal radio regulation became law with a different constitutional status than federal regulation of other aspects of the physical world. Consider land use. State and local governments are involved in regulating land use through mechanisms such as property law, zoning statutes, nuisance laws, and the use of land in providing public services (garbage collection, schools, etc.). Moreover, land use directly relates to state and local taxation. Economic studies of the sort discussed in *United States v. Lopez* and *United States v. Morrison* could easily be produced to document that land use affects interstate commerce.[3] However, in *Jones v. United States*, a unanimous Supreme Court found that the statutory language "…property used in interstate or foreign commerce or in any activity affecting interstate or foreign commerce" does not cover an owner-occupied residence not used for any commercial purpose.[4] The Court made careful and well-justified formal distinctions between "used in any activity affecting commerce" and "affecting commerce," and between statutory and constitutional questions. Does a realistic appraisal of the meaning of *Jones* imply limits on federal regulation of land use under the Commerce Clause? Are there constitutional constraints on federal regulation of land use other than the constitutional rules of the national political process? Such questions would be vigorously deliberated if anyone really thought that the answers were not obvious.

Consider also federal regulation of water use. *Gibbons v. Ogden* ruled that the Commerce Clause comprehended the power to regulate navigation. Navigation, a particular water use feasible in some bodies of water, was considered in *Gibbons* as an aspect of trade.[1] Most other water uses have been primarily of local concern. Physiographic, climatic, economic, and historical factors create significant differences in water use across the U.S.[2] States and persons hold title to most water on the land and tidal areas of the U.S.[3] Water use law has primarily developed within states and localities. As a Supreme Court decision that reviewed the federal-state balance in water law noted:

> *The history of the relationship between the Federal Government and the States in the reclamation of the arid lands of the Western States is both long and involved, but through it runs the consistent thread of purposeful and continued deference to state water law by Congress.*[4]

---

[1] 35 Op. Attys. Gen. 126, reprinted in Zollmann (1930) pp. 292-8. Citation from p. 292.
[2] Ibid. p. 293.
[3] 514 U.S. 549, 563-5, 619-25, 631-44, 115 S.Ct. 1624, 1632-3, 1659-61,1665-71 (1995) (*Lopez*) and 529 U.S. 598, 628-37, 120 S.Ct. 1740, 1760-4 (2000) (*Morrison*).
[4] 529 U.S. 848, 120 S.Ct. 1904 (2000).

Whether Congress could regulate all water use under the Commerce Clause is a question that has been unnecessary to discuss. Even to contemplate it seems impolite.

With respect to specific activities, a federal statute might rationally be interpreted to have regulated water use to the fullest extent possible under the Commerce Clause. Section 404(a) of the Clean Water Act regulates discharging dredged or fill material into "navigable waters." "Navigable waters" are defined under the Act as "the waters of the United States, including the territorial seas."[5] In *The Solid Waste Agency of Northern Cook County v. United States Corps of Engineers*, the petitioner sought to use an abandoned sand and gravel mining site to dispose baled, non-hazardous solid waste. The site included "a scattering of permanent and seasonal ponds of varying size (from under one-tenth of an acre to several acres) and depth (from several inches to several feet)."[6] The Supreme Court found that Congress had not made a clear statement indicating that the Act covers such sites.[7] The Court noted "…the States' traditional and primary power over land and water use," cited precedent indicating a preference for avoiding constitutional questions, and read the Act as not applying to the site.

A vigorous dissent argued that the Act clearly addresses its regulations to all "waters over which federal authority may properly be asserted."[8] The dissent analyzed this federal authority under the Commerce Clause and concluded that Congress has the power to prohibit "filling any part of the 31 acres of ponds" on the site.[1] Even dissenting Supreme Court justices, by virtue of their position in the U.S. judiciary, deserve to have their formal statements categorized as rational.[2] While the correct legal interpretation may still be debated among scholars, the Clean Water Act surely can be rationally interpreted to regulate a particular water use to the fullest extent possible under the Commerce Clause.

---

[1] 22 U.S. 1, 190-4 (1824).
[2] *California v. United States*, 438 U.S. 645, 648, 98 S.CT. 2985, 2987 (1978).
[3] Title to coastal tidelands and the beds of navigable rivers belong to states as part of the constituting legal framework for states. The beds of non-navigable rivers and other bodies of water are owned under the same general legal framework as for land ownership. Others may have rights to the use of such water under legal doctrines such as the prior appropriation doctrine. Acts such as the Desert Land Act of 1877 (43 USC §321) made water on federal lands subject to state laws of water appropriation.
[4] *California v. United States*, 438 U.S. 645, 653, 98 S.CT. 2985, 2990 (1978).
[5] See 33 U.S.C. § 1344(a) and § 1362(7).
[6] 531 U.S. 159, 163, 121 S.Ct. 675, 678 (2001). Acting under the Clean Water Act, the United States Army Corps of Engineers blocked this use of the site.
[7] In fact the majority stated, "We find § 404(a) [regulations interpreting the scope of waters covered by the Clean Water Act] to be clear, but even were we to agree with respondents, we would not extend Chevron deference here." Ibid at 172. That statement means that the subsequent statements concerning the Constitutional questions in this case are *dicta*.
[8] Ibid at 182.

The point is that the Clean Water Act identifies a particular water use. As the dissent pointed out:

> *The activity being regulated in this case (and by the Corps § 404 regulations in general) is the discharge of fill material into water. The Corps did not assert jurisdiction over petitioner's land simply because the waters were "used as habitat by migratory birds." It asserted jurisdiction because petitioner planned to* discharge fill *[emphasis in original] into waters "used as habitat by migratory birds."*[3]

More generally, the dissent noted that the Clean Water Act proclaimed the goal of ending water pollution by 1985.[4] Legislative history of the Act includes a statement that the "main purpose" of the Act is "to establish a comprehensive long-range policy for the elimination of water pollution."[5] Water pollution is a much narrower class of activities than water use. Moreover, many types of water pollution effect relationships of harm substantially between persons within a state and persons outside the state.[6] Such relationships clearly come under the meaning of interstate commerce.

The development of aviation regulation also shows the importance of understanding the Commerce Clause in relation to specific activities. Law of the air in the 1920s meant law concerning aviation and law concerning radio.[7] With the development of aviation and radio broadcasting, the physical characteristics of air gained new significance. Just as persons struggled to communicate the meaning of radio spectrum ("the ether"), a judge in 1934 similarly struggled to define airspace:

> *What is the sky? Who can tell where it begins or define its meaning in terms of the law? When can it be said that a plane is above the sky or below it?*[8]

Many subsequent cases struggled with such issues, and law in this area remains unclear to this day.[9] Federal law of the air has, however, remained focused on aviation.

---

[1] Ibid at 192.

[2] Whether members of Congress merit in U.S. law this same type of respect is still an open question in some areas of law. Categorizing statements as rational by virtue of their relation to particular institutions, times, and manners should not be done indiscriminately. Certainly the writings of academics and government bureaucrats deserve no such respect.

[3] Ibid at 193,

[4] Ibid at 175. 33 USC § 1251(a)(7) states: "it is the national goal that the discharge of pollutants into navigable waters be eliminated by 1985."

[5] Ibid at 179, quoting Senate Report No. 92-414, p. 95.

[6] Not all such uses do so. Consider a farmer who lets pigs drink from a pond on his land. Some Jewish and Muslim scholars might argue that the pond is polluted by this activity. Yet if no other persons have the right or the need to use the water from the pond, surely that pollution does not effect a relationship between the farmer and anyone else.

[7] See Zollmann (1927).

[8] Thrasher v. City of Atlanta 173 S.E. 817, 825 (1934).

[9] For a review, see Cahoon (1990).

The Empire State Building illustrates the extent of state and local regulation of air use. Technical analysis of the flight patterns of airplanes, such as a biplane with a motor rated at 90 horsepower and a monoplane with motor rated at 220 horsepower, established 500 feet of altitude as the lower boundary of airspace under federal aviation regulation in the late 1920s.[1] Courts, beginning in 1930 through to the present, have given this early technical judgment enduring importance in aviation law.[2] That legal boundary did not, however, affect early skyscraper construction. A former governor of New York led construction of the Empire State Building in 1930 in a race between New York and Chicago to claim the world's tallest building.[3] When completed in 1931, the Empire State Building reached 1250 feet into the Manhattan skyline. The building included a final 200 feet section intended as a dirigible mooring mast. The significance of this building to aviation was tragically emphasized in 1945, when a B-25 bomber in dense fog crashed into the 79'th floor.[4] Such an accident surely provided reason and opportunity to establish comprehensive, unified federal regulation of the airways.

Despite the significance of tall objects and airports to aviation, federal regulations carefully assert only an advisory role in regulating real estate development. Federal Aviation Administration (FAA) rules require that it be given notification of plans for a construction extending more than 200 feet above ground level.[5] The FAA analyzes the implications of the construction for air navigation and categorizes the construction as a hazard or not a hazard. The FAA does not assert authority to prohibit constructions that it deems hazardous to air navigation.[6] The FAA also requires notification for construction of an airport.[7] Upon such notification, the FAA does an aeronautical study of the proposal, and then states objections, if any, and in some cases conditions for ameliorating objections. The FAA does not claim authority to prohibit construction in any case.[8] Thus FAA regulations show considerable respect for state and local governance of activities significantly affecting the use of the airways, i.e. the use of air for aviation.

A federal law regulating all uses of the air or all uses of the sky would raise serious constitutional issues under the Commerce Clause. The Air Commerce Act of 1926, like the Radio Act of 1927, distinguished between interstate and intrastate activities.[9] In the 1930s, states passed statutes regulating intrastate aviation.[1] Over time, however, federal law came to cover both interstate and intrastate aviation. This development of federal law seems to have sound legal and practical bases. But despite many public concerns relating to aviation, airspace law was not totally federalized. A complex and vitally important law governing air rights relating to real estate developed at the state and local level.[2] State and local law regarding airport zoning and associated nuisances also developed.[3] These developments indicate awareness that the Commerce Clause places limits on federal regulation of air use.

---

[1] See discussion in *Smith v. New England Aircraft*, 270 Mass. 511 515-20, 170 N.E. 385, 387-9 (1930) and *Swetland v. Curtiss Airports Corp*, 41 F.2d 929, 938-42 (1930). Over congested parts of cities, towns, or settlements, the minimum height of flight was declared to be 1000 feet.

[2] See Cahoon (1990), esp. Sections IX and X.

[3] Other buildings in New York at the time reached 808, 792, 700, 625, and 612 feet into the sky. See *Smith v. New England Aircraft Co*., 270 Mass. 511, 522, 170 N.E. 385, 389 (1930).

[4] This accident killed 14 persons and injured about 25 others. For details, see Tauranac (1995) pp. 317-331.

[5] FAA Notice of Construction or Alternation of Objects Affecting Navigable Airspace, 14 CFR 77.13 (2002).

[6] Hamilton (1994) p. 266.

[7] FAA Notice of Construction, Alteration, Activation, and Deactivation of Airports, 14 CFR 157.3 (2002).

[8] Hamilton (1994) p. 270.

[9] The distinction between interstate and intrastate aviation was a major issue in deliberations about the Air Commerce Act of 1926. See Downs (2001).

Electromagnetic spectrum is similar to land, water, and air. Electromagnetic spectrum, like land, water, and air, is a general aspect of the physical world. Electromagnetic spectrum, like land, water, and air, is not itself an instrumentality of commerce. New technological developments are rapidly expanding the range and variety of human activities that use radio. That one federal law would assert authority under the Commerce Clause to govern interstate highways and backyard sandboxes (land use), or water pollution and swimming (water use), or aviation and speech (air use), seems so remote and gratuitously provocative as to be not worth deliberating. But current federal law that governs "disturbing" radio spectrum is that type of law.[4]

## B-III. Re-Articulating Commerce

To identify and maintain boundaries for radio regulation under the Commerce Clause, one must recover the natural sense of the word "commerce" in eighteenth century thought. "Commerce" meant intercourse.[5] The word "commerce" has been used in English only since the sixteenth century, when it came into use to supplement "merchandise," meaning the buying, selling or bartering of moveable goods, or the moveable goods themselves.[6] With the rapid growth of the English economy in the second half of the sixteenth century and the development of a Dutch empire in world trade, the activity of merchandizing became a systematic network of ongoing relationships.[7] Ongoing, deeply enmeshing relationships is what "commerce" added to the terms "merchandise," "traffic," and "trade."[8]

---

[1] See *Swetland* and *Smith*, supra.
[2] For a review, see, for example, Marcus (1984).
[3] For discussion, see Hamilton (1994).
[4] On disturbing radio spectrum, see above discussion of 35 Op. Attys. Gen. 126 (1926).
[5] Other scholars have briefly noted and dismissed this meaning of commerce with respect to the Commerce Clause. See Barnett (2001) pp. 128-9; Nelson and Pushaw (1999) p. 19, 41; Lessig (1995) p. 141 n. 37.
[6] OED (1989), heading note for "commerce".
[7] In a fascinating account of commercial morality in seventeenth century Dutch society, Whitman (1996) emphasizes the hospitality that Dutch Calvinist moralists in the seventeenth century extended to values found in Roman commercial law. In England, where Roman law and Protestant theology had much less significance, the way "commerce" was understood helped to distinguish it from "unbrotherly" or "uncommunal" activities.
[8] Muldrew (1998) describes the period 1550-1580 in England as a period of rapidly increasing internal trade and consumption -- the most intensely concentrated period of economic growth before the late eighteenth century. Muldrew explains that trade then was deeply connected to an intricate system of personal relations. Credit meant not just money but social standing and character. Given that the word

Key works in the English language illustrate this meaning. Consider Christian scripture. In the King James Version of the Bible, a translation finished in England in 1611, the words "merchandise," "trade," and "traffick" occur repeatedly, but "commerce" never occurs.[1] Words used in a sense most closely approaching the eighteenth century meaning of commerce are in Matthew 22:5 ("But they made light of it, and went their ways, one to his farm, another to his merchandise") and in John 2:16 ("make not my Father's house a house of merchandise.") In the Vulgate, St. Jerome's Latin translation of Christian scripture in 400 CE, these phrases are rendered with forms of "negotiation," meaning wholesale or banking business, or more generally, business or traffic.[2] Twentieth-century translations use "business" for "merchandise" in Matthew 22:5, and "into a market," "place of business," or "house of trade" for "house of merchandise" in John 2:16.[3] These meanings were too narrow to invoke the use of the word "commerce" in early seventeenth century English.

Shakespeare shows that the natural sense of commerce then covered the full sense of human relationships. The word "commerce" occurs four times in Shakespeare's oeuvre.[4] Malvolio, a courtier deluded that has become a great political and romantic favorite of the Countess Olivia, is described by Olivia's maid at the end of a playful dialogue as being engaged "in some commerce with my lady."[5] That is an insinuation of intercourse in many human dimensions. Regarding Archilles' intimate relationship with a prominent woman in the enemy city of Troy, Ulysses says to him: "All the commerce that you have had with Troy / As perfectly is ours as yours, my lord."[6] The significance to the state of this relationship – this commerce – means that it could not be kept hidden. To Agamemnon, the leader of the Greek camp, Ulysses urges the importance of respecting order in relationships, and he contrasts "peaceful commerce from dividable shores" with bounded waters flooding the globe, sons striking fathers, and power, will, and appetite perverting justice.[7] Finally, in a scene underscoring disordered relationships, Ophelia, returning love letters to a Hamlet deeply troubled by the events in his family and in the

---

"commerce" came into the English language about this time, these developments shed light on the meaning of that word.

[1] The King James Version, as well as a number of other translations, can be searched at The Bible Gateway, online at http://www.biblegateway.org. "Commerce" was derived from the Latin "commercium". This term, and related forms, does not occur in the Vulgate. An online search of the Vulgate is available at ARTFL Project, online at http://www.lib.uchicago.edu/efts/ARTFL/public/bibles/vulgate.search.html

[2] See Vulgate text, ibid. For an online Latin dictionary, see the Perseus Digital Library, online at http://www.perseus.tufts.edu/cgi-bin/resolveform?lang=Latin. The early Greek texts use, in Matthew 22:5 and John 2:16 respectively, "emporia" and "emporion". These terms mean trade or a place of trade. See interlinear Greek, Strong's Concordance, and Vine's Expository Dictionary linked to verses online at Blue Letter Bible, http://www.blueletterbible.org/index.html. A translation in Rheims in 1582 used "merchandise" and "traffick" in the same respective passages. See Douay-Rheims translation at The Unbound Bible, online at http://unbound.biola.edu/.

[3] See various Bible translations at links mentioned in footnotes above.

[4] For online searches of Shakespeare's oeuvre, see the Rhyme Zone, http://www.rhymezone.com/shakespeare/ and Matty Farrow's search, http://www.it.usyd.edu.au/~matty/Shakespeare/test.html

[5] *Twelfth Night*, Act 3, Scene 4.

[6] *Troilus and Cressida*, Act 3, Scene 3.

[7] Ibid, Act 1, Scene 3.

royal court, asks Hamlet: "Could beauty, my lord, have better commerce than with honesty?"[1] Hamlet responds by speculating on the transforming power of relationships between virtues and of relationships between persons. Thus commerce means intercourse so intimate and comprehensive as to be transformative.

For eighteenth century intellectuals and public figures, the natural and obvious meaning of commerce comprised a broad sense of relationship. The 1776 Declaration of Independence declared the "United Colonies" to have "…full Power to levy War, conclude Peace, contract Alliances, establish Commerce, and to do all other Acts and Things which Independent States may of right do." Consider the four distinguished acts and things: "War," "Peace," "Alliances," and "Commerce." The selection and order of these terms indicates that commerce, balanced against "alliances," is a serious, informal, general, peaceful relationship not merely limited to trade.[2]

Montesquieu discussed *doux commerce* – sweetness and gentleness that commerce engendered in relations among persons. This state of relations he set in contrast to hostility and violence.[3] A Scottish historian in 1769 declared:

> *Commerce tends to wear off those prejudices which maintain distinction and animosity between nations. It softens and polishes the manners of men.*[4]

Concern about commerce meant concern about relationships among persons. Thus James Madison argued that the purpose of the Commerce Clause is to avoid practices that would "nourish unceasing animosities" and that might "terminate in serious interruptions of public tranquillity."[5]

Alexander Hamilton recognized contemporary views on the transformative power of commerce and of republican government:

> *The genius of republics (say they) is pacific; the spirit of commerce has a tendency to soften the manners of men, and to extinguish those inflammable humors which have so often kindled into wars. Commercial republics, like ours, will never be disposed to waste themselves in ruinous contentions with each other. They will be governed by mutual interest, and will cultivate a spirit of mutual amity and concord.*[6]

But he ridiculed such a view as abstract speculation contrary to real experience:

> *Are not popular assemblies frequently subject to the impulses of rage, resentment, jealousy, avarice, and of other irregular and violent propensities? Is it not well known that their determinations are often governed by a few individuals in whom they place confidence, and are, of course, liable to be tinctured by the passions and views of those individuals? Has commerce hitherto done anything more than change the objects of war? Is not the love of wealth as domineering and enterprising a passion as that of power or glory? Have there not been as many*

---

[1] *Hamlet, Prince of Denmark*, Act 3, Scene 1.
[2] Note as well that the expressed causes for the Declaration contain only one reference to trade.
[3] Hirschman (1977) p. 60.
[4] Ibid., p. 61, quoting William Robertson, *View of the Progress of Society in Europe* (1769).
[5] Federalist No. 42, online at http://memory.loc.gov/const/fed/fed_42.html
[6] Federalist No. 6, online at http://memory.loc.gov/const/fed/fed_06.html

*wars founded on commercial motives since that has become the prevailing system of nations, as were before occasioned by the cupidity of territory or dominion? Has not the spirit of commerce, in many instances, administered new incentives to the appetite, both for the one and for the other? Let experience, the least fallible guide of human opinions, be appealed to for an answer to these inquiries.*[1]

Hamilton argued that the structure of commercial governance affected the nature of relationships among citizens of the United States:

*The interfering and unneighborly regulations of some States, contrary to the true spirit of Union, have, in different instances, given just cause of umbrage and complaint to others, and it is to be feared that examples of this nature, if not restrained by a national control, would be multiplied and extended till they became not less serious sources of animosity and discord than injurious impediments to the intercourse between the different parts of the Confederacy. ... we may reasonably expect, from the gradual conflicts of State regulations, that the citizens of each would at length come to be considered and treated by the others in no better light than that of foreigners and aliens.*[2]

Thus, to Hamilton, commerce itself does not constitute harmonious human relationships. Under different structures of governance, the qualities of commerce can be very different.

*Gibbons v. Ogden*, a Supreme Court case decided in 1824, inadvertently helped to create space for a much different legal perspective on commerce. The primary ruling of the Court declared:

*Commerce, undoubtedly, is traffic, but it is something more: it is intercourse. It describes commercial intercourse between nations, and parts of nations, in all its branches, and is regulated by prescribing rules for carrying on that intercourse....*[3]

The opinion clearly identifies "commerce" as "intercourse," meaning more than "traffic" or "trade." Yet the opinion also uses the phrases "no sort of trade," "commercial intercourse," and "trading intercourse." These phrases may have contributed to misunderstanding the eighteenth century meaning of commerce as being only a particular dimension of intercourse, and thus less than intercourse.[4] The concurring opinion shows meaning moving in that direction:

---

[1] Ibid.

[2] Federalist No. 22. online at http://memory.loc.gov/const/fed/fed_22.html

[3] 22 U.S. 1, 189-90 (1824).

[4] Marshall seems to have attached the adjectives "commercial" and "trading" to "intercourse" only for emphasis in the context of a specific description. "It has, we believe, been universally admitted, that these words ['with foreign nations, and among the several States, and with the Indian tribes.'] comprehend every species of commercial intercourse between the United States and foreign nations. No sort of trade can be carried on between this country and any other, to which this power does not extend." Ibid at 193-4. "What is commerce 'among' them [states]; and how is it to be conducted? Can a trading expedition between two adjoining States, commence and terminate outside each? And if the trading intercourse be between two States remote from each other, must it not commence in one, terminate in the other, and probably pass through a third? Commerce among the States must, of necessity, be commerce with the States." Ibid. at 196.

*Commerce, in its simplest signification, means an exchange of goods; but in the advancement of society, labour, transportation, intelligence, care, and various mediums of exchange, become commodities, and enter into commerce…*[1]

This is an early glimpse of what Marx saw in the British Industrial Revolution. Commerce, and economics more generally, was evolving to focus on concepts and commodities abstracted from persons.

Cases addressing the telegraph contributed to this change in the meaning of commerce. In *Pensacola Telegraph v. Western Union* (decided in 1877), the Supreme Court declared:

*Since the case of Gibbons v. Ogden (9 Wheat. 1), it has never been doubted that commercial intercourse is an element of commerce which comes within the power of Congress. ... it is not only the right, but the duty, of Congress to see to it that intercourse among the States and the transmission of intelligence are not obstructed or unnecessarily encumbered by State legislation.*[2]

One might consider the transmission of intelligence to be intrinsically related to human intercourse. But *Pensacola*'s reference point was not the primary ruling in *Gibbons* but the concurrence. *Pensacola* presented intelligence as a commodity – something "transmitted" like a commercial good.

Four years later, *Telegraph v. Texas* showed this understanding in analogy:

*A telegraph company occupies the same relation to commerce as a carrier of messages, that a railroad company does as a carrier of goods. Both companies are instruments of commerce, and their business is commerce itself.*[3]

The wonder of the telegraph was that it separated communication from physical relation and movement of objects:

*But the telegraph transports nothing visible and tangible; it carries only ideas, wishes, orders, and intelligence.*[4]

In these decisions and others, the Court found telegraphic communication across state lines to be interstate commerce. Rather than understanding communication as inter-subjectivity extended in space, courts, following *Pensacola* and reasoning consistent with the intellectual and practical circumstances of their times, presented communications as the transmission of intelligence analogized to a commodity.

Yet by the late nineteenth century, the telegraph had substantial and well-recognized affects on relationships spanning the nation. Telegraph lines expanded with the railroads, enabling considerable capital savings and thus enabling more persons to be brought more quickly within the scope of this national transportation network.[5] Logistical control through telegraphic communication prompted a new national division of labor in the slaughtering and meat packing industries, a national trade in fresh fruits and vegetables, and the growth of mail order houses and mass merchandising.[1] The telegraph also transformed commodities and securities trading. By 1874, Western Union Telegraph transmitted near-real-time transaction data from the New York Stock Exchange to 116 other cities.[2] By that time Western Union was the first national monopoly, pointing toward a new, important national political issue. Working with the U.S. Navy, Western Union also transmitted nationally time synchronization pulses that standardized time across the nation.[3] Moreover, arising directly from the opportunities that the telegraph presented and closely tied to Western Union, the Associated Press had become "one of the most powerful centripetal forces shaping American society."[4] By the late 1880s, Associated Press wire dispatches made up 80-100% of news copy in mid-western small town dailies.[5]

---

[1] Ibid at 229-30.

[2] 96 U.S. 1, 9 (1877).

[3] 105 U.S. 460 (1881).

[4] *Western Union Tel. Co. v. Pendleton*, 122 U.S. 347, 356, 7 S.Ct. 1126, 1127 (1887).

[5] The telegraph provided control that allowed rail lines to be single-tracked. In 1890, the U.S. rail system was 73% single-tracked. The additional cost of double-tracking the system would have been about one billion dollars in 1890. For comparison, the total book value for construction of all railroads and for the equipment on them was about $8 billion at that time. See Field (1992) pp. 406-9.

These realities should be recognized when reading the *Pensacola* decision of 1877. That decision merely briefly noted that the telegraph had "changed the habits of business" and had fostered communications among government officials spread across the country and the world.[6] Certainly the telegraph had done much, much more than that. The telegraph greatly transformed relationships among persons across the nation. Yet the heart of the *Pensacola* decision was to judge intelligence to be a commodity. Judging the scope of commerce came to be about identifying commodities and estimating effects on commodity streams.[7] The loss of the eighteenth century understanding of commerce in

---

[1] Huge stockyards and slaughtering operations developed in Chicago and other major mid-Western cities. Ibid. pp. 410-2.

[2] Field (1998) p. 167, which also notes:

> *Physical proximity to the exchange was no longer a prerequisite for a trader whose strategies depended on knowing what was happening on the floor, reacting to it, transmitting trading instructions, and having them executed and confirmed within very short time spans.*

These were important changes in relationships of trade. More generally, see ibid. pp. 163-9.

[3] Lubrano (1997) pp. 118-20.

[4] Blondheim (1994) p. 7.

[5] Blondheim's impressively detailed history of news wire services noted:

> *The Associated Press replaced the great party organs and popular metropolitan newspapers as the primary source of national news. Unlike predecessors such as the Washington National Intelligencer or the New York Weekly Tribune, the wire service was not a local institution. Wire service reports were a compilation of dispatches, by hundreds of anonymous local agents, printed verbatim in all the country's major newspapers simultaneously. And thus they were not perceived as representing any local institution or interest.*

Ibid. p. 195.

[6] Compare the Court's description to this newspaper description more than 30 years earlier:

> *Washington is as near to us now as our up-town wards. We can almost hear through the Telegraph, members of Congress as they speak. The country now will be excited from its different capitals. Man will immediately respond to man. An excitement will thus be general, and cease to be local. Whether good or ill is to come from all this, we cannot foresee, but that the Telegraph is to exert an important bearing over our political and social interest, no man can doubt.*

Ibid. p. 191, quoting the New York Express from 1846. Developments from the time that account was written to the time the Court issued its opinion confirmed what the journalist regarded as indubitable.

[7] To get a sense for the extent to which analysis of commerce got unmoored from the eighteenth century understanding of commerce, consider this analysis in a 1928 law review article examining the law of radio communication (Taugher (1928) p. 314):

*Pensacola* now seems natural and unremarkable. Commerce as an economic reality is now firmly entrenched in a grand, deterministic narrative of the rise of a single national market.[1] A better reading of *Pensacola* looks backwards and finds a different story. In the silences of *Pensacola* – the obvious, unnoted reality that railroads and telegraphs dramatically changed human relations across the country – one can recognize the resources for much more vigorous and fruitful deliberations about the Commerce Clause.

### B-IV. Spurring Fruitful Deliberation

Considering commerce in its eighteenth century meaning can help to revitalize deliberations about the Commerce Clause. One might well consider many past Commerce Clause cases to be correctly decided, or at least to be prudently preserved. But decisions are written to be discussed. And discussions, like barn-raisings, require good tools, in addition to what might be called rational calculations. The eighteenth century meaning of commerce, etched in still-accessible artifacts, is one such tool.

With this tool, one can better understand influential precedent. For example, the *Shreveport* decision:

> *...does not ground federal power to regulate rates for intrastate carriage solely in their effect on commerce. [That decision] repeatedly emphasizes that it is the intrastate operations of "agencies" or "instruments" of interstate commerce that*

---

> *[quoting Pensacola] "...The powers thus granted to Congress are not confined to the instrumentalities of commerce known or in use when the Constitution was adopted, but they keep pace with the progress of the country, and adapt themselves to the new developments of time and circumstances." It would seem, therefore, that the analogy between the radio wave as an instrumentality of interstate commerce and the telegraph is a much closer one than that between the telegraph and the railroad, as the electro-magnetic energy which is the basis of communication in the case of both radio and telegraphy by wire has already received judicial cognizance as a means of commercial intercourse.*

The telegraph's use of electromagnetic energy to transmit messages has little to do with the intercourse related to the telegraph. Locomotives' use of iron railroad tracks likewise has little to do with the intercourse associated with railroads.

[1] A more critical approach to this narrative is needed. What does it mean to say that the U.S. (or the European Union) is a "single market"? The contextual significance of such references are usually promotion of a particular set of contemporary economic policies. In attempting to give this use of "market" additional historical significance, Lessig (1995) p. 138 suggests:

> *There is no simply way to describe this difference in the extent of integration, and no handy way to quantify it. But I don't believe that we need data to make the point that I want to make here: That integration in the sense I suggest has increased; that more operates in a national rather than local market; and that this change in the extent of the market properly has consequences for the scope of federal and state power.*

Consider one long-run economic trend that Lessig neglects: the growth of (monetized) services in the economy. Transactions for services tend to be more individualized, more localized, and involve more relationship-specific investments than transactions for goods. That the "market" for health care now is much more national than the "market" for corn was in the mid-nineteenth century is not at all clear. The same seems to be the case for the economy as a whole. On trends in shared symbolic experience and cultural homogeneity, see Galbi (2001b).

> *are being regulated. That is, [it] focuses not just on the activity regulated, but also on the interstate nature of the entity regulated.*[1]

An entity in the sense used above can be understood as a network of human relations. The scope of these relations – commerce in the eighteenth century meaning of the word – matters in determining what falls within the boundaries of the Commerce Clause.

The eighteenth century meaning of commerce can also spur needed discussion of the many Commerce Clause cases categorized as non-economic. Consider *Heart of Atlanta Motel v. United States* (1964), *Katzenbach v. McClung* (1964), *Daniel v. Paul* (1969), and *United States v. Morrison* (2000).[2] The first three found that the Commerce Clause provides authority to enjoin racial discrimination in certain motels, restaurants, and snack bars in the 1960s. The latter found that the Commerce Clause does not provide authority to enact civil remedies against gender-motivated violence in the 1990s. These decisions imply a judgment directly related to the eighteenth century notion of commerce. Through them the Court judged that racial discrimination in the 1960s was a national problem in commerce – intercourse – to an extent that gender-motivated violence in the 1990s was not.

That judgment might seem controversial or questionable, but it has a rational basis. In an ideal speech situation, which characterizes a type of rationality, most persons might agree with the Court's judgment. Alternatively, based on lived experience, actual discussions with real persons, and moral intuition, most persons might agree with the Court's judgment. A mass of statistics and citations might also, on a purely formal level, be impressive enough to convince some. Others might seek a clear, disciplined, and coherent analysis of statistics, preferably calculated using data from sources collecting data for broad purposes, in a methodical way, and with full documentation of definitions and measurement techniques. Today there is no general agreement about what a rational basis for a view formally entails. If there ever were such agreement, it certainly broke down more than a century ago. A dominant thrust of philosophical and literary thought over the past decades has been to show the tenuous and historically contingent character of the basis or foundation for reasoning and knowledge.[3] Nonetheless, the Supreme Court is the head of a judicial process that deserves considerable respect. Its decisions, in a serious sense, define an important type of rationality. Therefore, that racial discrimination in the 1960s was a national problem in commerce to an extent that gender-motivated violence in the 1990s was not is a statement with a rational basis.

The Supreme Court's rationality has significance not just because it has legal force, but also because the Court provides written decisions that can stimulate fruitful discussions. Such discussions do not naturally emerge via the workings of some invisible hand or tongue. Good discussion, like every other aspect of human existence, requires resources and confronts constraints. To see how the eighteenth century meaning of commerce can support more disciplined reading of statutes, prompt recognition of previously unrecognized but relevant facts, eliminate some utterability constraints, and stimulate imaginative discussion, consider in more detail *Morrison*.[1]

---

[1] Cushman (2000) p. 1130.

[2] *Heart of Atlanta Motel v. United States* 379 U.S. 241, 85 S.Ct. 348 (1964), *Katzenbach v. McClung* 379 U.S. 294, 85 S.Ct. 377 (1964), *Daniel v. Paul* 395 U.S. 298, 89 S.Ct. 1697 (1969), and *United States v. Morrison,* 529 U.S. 598, 120 S.Ct. 1740 (2000).

[3] See, e.g. Feyerabend (1975), Rorty (1979), Lyotard (1984), Derrida (1988), MacIntryre (1988), and Fish (1999).

*Morrison* concerns gender and the Violence Against Women Act.[2] What does "gender" mean in the Violence Against Women Act? Gender there, as in much of the gender literature, seems to mean concern only about women. However, interpreting "gender" to mean "women" probably would be an inappropriately cramped reading of the Violence Against Women Act. "Gender" might better be read in the Act as a simple substitute for "sex" (among those who consider "sex" to be impolite, undesirable, or unfashionable), or as a word emphasizing the social and political construction of sexual categories, particular from a post-modern perspective that does not take the intrinsic differences between males and females to be significant.[3] A hypothetical (reasonable) Member of Congress in the year 1994 probably would have had the courage to use "sex" in a statute and probably would not have felt compelled by decorum to choose the more ambiguous substitute "gender." Moreover, a (hypothetical) reasonable Member of Congress in the year 1994 or in the year 2000 probably would have wanted the statute to apply to a man, hostile to the idea of women voting, who assaulted, for that reason, a man voting dressed as a woman. Thus, in judicial interpretation of the Act, gender should be understood to refer to the social and political construction of sexual categories.

Now consider the Violence Against Women Act in relation to the Commerce Clause. The Violence Against Women Act addresses "a crime of violence committed because of gender or on the basis of gender, and due, at least in part, to an animus based on the victim's gender."[4] Criminal violence, without further qualification, has long been recognized as a matter of intrastate commerce. One might analyze whether that is still appropriate under current circumstances by considering whether the relationships associated with criminal violence have a substantial interstate component.[5] But criminal violence does not describe well the relationship that the Act governs. The relationship that the Act governs is better described as gender hostility to such an extent that it is expressed in criminal violence.

Attention to appropriately characterizing the relationships of concern matters for analysis.

---

[1] For an analysis of utterability constraints, see Lessig (1998). For an additional attempt to address utterability constraints in a scholarly context, see Galbi (2001b), Section III.

[2] PL 103-322, September 13, 1994, 108 Stat 1796, Sec. 40001.

[3] Forms that instruct the respondent to identify "gender" as male or female are a fine example of the first meaning of "gender."

[4] 42 U.S.C. § 13981(d)(1).

[5] As part of such inquiry, one might ask whether the significance of violence in intercourse considered from a national perspective differs significantly from the significance considered in states as separate entities. Considering Commerce Clause cases from a "commerce is intercourse" approach provides better connection to existing precedent, more focus on case-specific facts, and a more tightly defined framework for analysis than the "structural necessity" approach proposed in Harsch (2001). The "commerce is intercourse" approach has similar advantages relative to the suggestion in Merritt (1998) p. 1210, that, under the Commerce Clause, "Congress should have the power to regulate whenever the states cannot do the job effectively."

A lower court made specific findings that the alleged crime in *Morrison* expressed gender animus. But such evidence does not go to the constitutionality of the Act under the Commerce Clause. Cited statistics from the "mountain of data" put forward to support Commerce Clause authority consisted only of evidence that criminal violence significantly affects women.[1] Such violence is repugnant and an insult to human dignity. Thus data describing it are an affective showing. These data do not, however, provide useful material for discussing Commerce Clause authority.

Other aspects of the formal use of data in *Morrison* have similar weaknesses. The principal dissent cites statistics concerning the effects of violence against women on "the supply and demand for goods in interstate commerce," as well as on "reductions in the [size of] the work force."[2] These data seem to reflect "silly questions rather than meaningful ones" and "tedious and somewhat arcane attempts to show how various social problems affect the economy."[3] The statistical material in *Morrison* does not enrich deliberation about the Court's judgment that racial discrimination in the 1960s was a national problem in commerce to a greater extent than gender-motivated violence in the 1990s.

With a better understanding of commerce, the Court could have easily found statistics and pushed forward analysis that would have stimulated meaningful deliberation. Consider, for example, that in 1999, 12,785 males and 4,104 females died from assault (homicide).[4] Gender unquestionably has something to do with the fact that more than three times as many males were killed. One would explore whether gender animus, in particular, notions of male disposability and social norms suggesting that killing a male is less horrible than killing a female, are part of the explanation for this disparity. With respect for the Commerce Clause, one would also consider whether these norms reflect a systematic national pattern of human relationships that is not encompassed by relationships in the states taken separately.[5] Such facts and such analysis do not appear to have been seriously considered in the intellectually and rhetorically mechanistic deliberations about the Violence Against Women Act, or in the Court's analysis of its constitutionality under the Commerce Clause. Doing justice in a case is not a matter of displaying a mass of statistics, or even analyzing a couple.[1] But facts can be useful in identifying truth. The eighteenth century meaning of commerce can help to inform the selection and analysis of relevant facts.

[1] 529 U.S. 598, 628-36, 120 S.Ct. 1740, 1760-4, esp. fn. 2.
[2] Ibid. at 636.
[3] Merritt (1998) p. 1213. These descriptions were made in the context of discussion of Commerce Clause cases in general. They appear to me to be directly relevant to some aspects of judicial reasoning in the principal dissents in *Morrison* and in *Lopez*.
[4] See National Center for Health Statistics (NCHS), "Deaths: Final Data for 1999" available online at http://www.cdc.gov/nchs/about/major/dvs/mortdata.htm NCHS is a broad source of statistics with considerable relevant expertise. Its website provides comprehensive descriptions of the data and documentation of statistical methods used. Compare these statistics and its citation to a statement in the principal dissent in *Morrison*: "Supply and demand for goods in interstate commerce will also be affected by the deaths of 2,000 to 4,000 women annually at the hands of domestic abusers, see S.Rep. No. 101-1545, at 36…" *Morrison*, supra, 120 S.Ct. 1740, 1764. One should wonder at the plausibility and meaning of the implication that roughly 50-100% of females that die from assault are women who die "at the hands of domestic abusers." FBI homicides reports for 1999 indicate that 1,218 females were killed, inside or outside the home, by intimate partners, defined as "current or former spouses, boyfriends or girlfriends." See Rennison (2001) p. 2. One should interrogate the symbolic domination that creates the statistics cited in *Morrison*, and effaces those cited *infra*.
[5] Being sent far from home for long periods to risk death in wars is a culturally salient form of disposition historically associated with maleness in the U.S. While the U.S. now has a sex-integrated military, the Military Selective Service Act currently requires only males to register. The Supreme Court has upheld such discrimination (see *Rostker v. Goldberg*, 435 U.S. 57, 101 S.Ct. 2646 (1981); both the majority

This understanding of commerce also provides better intuition about boundaries. A focus on economic effects exacerbates the un-Donne challenge:

> *No man is an island, entire of itself; every man is a piece of the continent, a part of the main. If a clod be washed away by the sea, Europe is the less, as well as if a promontory were, as well as if a manor of thy friend's or thine own were: any man's death diminishes me, because I am involved in mankind, and therefore never send to know for whom the bell tolls; it tolls for thee.*[2]

These words are easily recognized as poetry, which most persons in the U.S. today would regard as ignorable. But viewed through the discourse of twentieth century economics, they present a formidable challenge:

> *"I really know of no place," [Supreme Court Justice] Jackson confessed, "where we can bound the doctrine of competition as expounded in the Shreveport, the Wrightwood, and the Wickard cases. I suppose that soy beans compete with wheat, and buckwheat competes with soy beans, and a man who spends his money for corn liquor affects the interstate commerce in corn because he withdraws that much purchasing power from that market." As so the jig was up. "If we were to*

---

opinion and the two sharp dissents lack any serious consideration of men as male, or of the social construction of maleness). That a large share of military jobs currently require physical characteristics highly disproportionately found among males seems doubtful. Women currently serve in military positions very closely related to combat. The Military Selective Service Act might be understood as current national law supporting norms of male disposability. Disposal of persons in the criminal justice system also fosters stereotypes of male disposability. In 1997, 1.23% of males ages 20 and over were in federal and state prisons under a sentence longer than 1 year. The comparable figure for women is 0.07%. See Bureau of Justice Statistics, Correctional Populations in the United States, 1997, Table 1.8, p. 4 (online at http://www.ojp.usdoj.gov/bjs/abstract/cpus97.htm) and U.S. Census Bureau, Current Population Reports P25-1130, Table 2, p. 46 (online at http://www.census.gov/prod/1/pop/p25-1130/). This sex disparity, which concerns some of the most disadvantaged persons in the U.S. and has attracted little political concern, is typically legitimated via an essentialist stereotype of males that re-enforces their disposability. Violence directed at males by virtue of their sex is recorded in some of the earliest written human records (see Exodus 1:16, 22) and across a variety of cultures (see Jones (2000), who notes that "gendercide, at least when it targets males, has attracted virtually no attention at the level of scholarship or public policy." (p. 186)). Domestic violence against males has received virtually no public policy attention, perhaps because of archaic and overbroad stereotypes that presume women not capable of the full range of human emotions and not capable of devising tools to attack men who on average are physically larger and stronger. A highly credible recent study found that among heterosexual partners women were slightly more likely than men to "use one or more acts of physical aggression and to use such acts more frequently," and that 38% of those physically injured by a partner were men. *See* Archer (2000) p. 651. The narrow focus of the literature on gender and the Violence Against Women Act might indicate that violence toward males continues at a relatively high level because of gender animus that has become a truly national problem. Cf. Hagen (2001), Resnik (2001) and Bonner (2002).

[1] For those interested in considering additional statistics, I suggest starting with those in the footnote above.

[2] Donne, John (1623), "For whom the Bell Tolls," online at http://www.incompetech.com/authors/donne/bell.html

*be brutally frank," he wrote, "I suspect what we would say is that in any case where Congress thinks there is an effect on interstate commerce, the Court will accept that judgment. All of the efforts to set up formulae to confine the commerce power have failed. When we admit it is an economic matter, we pretty nearly admit that it is not a matter which courts may judge."*[1]

With more emphasis on intercourse, the situation may be more appealing. While everyone may in some sense be related to everyone else, one intuitively recognizes that the relationship between spouses is of different quality than the relationship between foreigners.[2] No arcane and tedious calculations of economic effects are necessary. More generally, compared to such calculations, evidence and judgments about the boundaries of ongoing, systematic relationships are likely to produce a more coherent, stable, understandable, and satisfying body of law.[3]

Current case law tends to interpret the "substantially affects" test to describe an absolute measure of a single, all-encompassing quantity – dollars. In *Lopez*, the Supreme Court "identified three broad categories of activity that Congress may regulate under its commerce power."[4] The final broad category was described as follows: "Congress' commerce authority includes the power to regulate those activities that have a substantial relation to interstate commerce i.e. those activities that substantially affect interstate commerce..."[5] Such language does not prescribe a particular method of analysis. The requirement for "substantial relation" is entirely consistent with an eighteenth-century understanding of "commerce." Nonetheless, toting up the "effects" of an activity in dollar terms predominates in asserting the constitutionality of a federal statute under the Commerce Clause. This may be related to the intellectual aura of cost-benefit analysis. But even under this spell, law in this area struggles to establish legal discipline.[6]

Extent of relationship provides a better metric for disciplined analysis of relevant facts. Common discourse, as well as scholarship outside the discipline of economics, often identifies as values sex and power, in addition to money. But all three of these values appear to be, at least in academic discourse, reducible to any one. Real relationships, in contrast, are case-specific and not reducible.[1] The definition of the relationships of concern itself often provides a natural, direct measure for judging whether there is a substantial interstate component. Thus water pollution, understood in terms of ecological relationships, has a substantial interstate component. The allocation of water among persons living along a stream within a state does not. Growing wheat for home consumption in the U.S. in the year 2002 might concern only relationships within households, while in a time and place where agriculture was key to national economic stability, the relevant set of relationships might be much different.[2]

---

1 Cushman (2000) p. 1145-6, quoting letter from Justice Jackson to his friend Sherman Minton.

2 Strong relationships of similar type are not necessarily conflicting. A person may give herself in love fully to another person. Yet doing so might complement, rather than conflict with, that person loving God with all her heart, and with all her soul, and with all her might. See Ruth 1:16. *U.S. Term Limits v. Thornton*, 514 U.S. 779, 115 S.Ct. 1842 (1995), presents a challenging case concerning the nature of political relationships. See, in particular, the intellectually impressive dissenting opinion.

3 Merritt (1998) suggests that, under the Commerce Clause, "Congress should have the power to regulate whenever the states cannot do the job effectively…." (p. 1210). Harsch (2001) similarly argues that courts "ought to ask whether a solution at the federal level is necessary, because the states and the federal government are politically structured in such as way as to preclude the states from dealing with the problem at hand." (p. 985). While these approaches make sense for policy inquiry, practical legal needs seem rather different. These approaches move far from the text of the Commerce Clause and provide little general structure for positive inquiry into facts. What type of analysis would these approaches imply for considering federal regulation of all radio use? What value would precedent offer under that kind of analysis?

4 514 U.S. 549, *557, 115 S.CT. 1624, 1629.

5 Internal citations omitted.

6 Among other problems, there is little basis for defining a relevant threshold value for cost-benefit analysis. See Lessig (1995) p. 208.

A relational understanding of commerce has additional advantages with respect to jurisdictional elements. Suppose Congress passes a law making it a federal crime to kill a child where the child wears any article of clothing home-sewn or otherwise affecting interstate commerce. Would such a law mass constitutional mustard? The significance of murdering clothed children to intercourse would seem to be about how threats to children, and parental fears of such threats, affect children's opportunities to participate in social interaction. Thus the clothes that a child wears would not matter in judging whether such a law comes under the scope of the Commerce Clause. In contrast, a jurisdictional element such as participation in activities sponsored by national organizations, and a finding that threats to children participating in such activities are in fact greater, would be relevant to Commerce Clause analysis. Analysis would thus recognize real circumstances and make meaningful distinctions.[3]

**B-V. Doing Justice in Communications**

If, in 1934, the *Gregg* case had been judged in terms of the relational understanding of commerce natural in the eighteenth century, the ruling may well have been different. "The Voice of Labor" described itself as follows:

> *That the purpose of the station is to furnish a medium, free of charge, whereby the laboring people of Houston, Texas, may be kept in daily touch with all matters of public interest affecting the interests of working people.*
> *That each day there is broadcast over the station addresses and talks by men prominent in labor circles. That these talks counsel and advise laboring people to wholeheartedly support the President of the United States in his effort to restore normalcy.*

---

[1] Consider fatherhood. Legally, fatherhood in the U.S. today is defined in terms of a DNA test for paternity, i.e. particular effects of a particular sex act. Moreover, the consequences of this sex act can be effectively nothing more than a legal obligation to pay money, i.e. child support. Despite this legal regime, many persons still believe that fatherhood cannot be reduced to sex and money.

[2] The literature known as new institutional economics provides considerable insight into relationships. See Williamson (1985). Commerce, understood as intercourse – ongoing, systematic relationships – is associated with relationship-specific investment (getting to know persons). Thus the mere functioning of a national commodity market does not imply anything about such relationships. Certain understandings of markets place markets in direct opposition to human relationships, e.g. Polanyi (1944). On the other hand, a common mistake is to consider markets too abstractly. Understanding the relationships relevant to a particular law requires a case-specific enquiry into real circumstances.

[3] Brown (2001) reviews the debate about federal criminal law and emphasizes the importance of analysis of jurisdictional elements.

*...That the station's facilities are daily used, free of charge, for conducting religious services. Any denomination is welcome, absolutely without cost, to conduct religious services, etc. over said station.*
*An interdenominational 'Go to Church' hour is conducted, free of charge, over this station.*
*...The facilities of the station are offered, free of charge, for any emergency use, and for all civic and charitable uses.*[1]

"The Voice of Labor" seems predominately oriented toward developing local community relationships. Moreover, "The Voice of Labor" asserted:

*...that said station is furnishing a valuable public service to this community, free of charge, which is far more beneficial, convenient and helpful to this community than would be an occasional fragmentary broadcast which might under extraordinary conditions occasionally be heard in this locality from a 'local' licensed station in Shreveport, Louisiana or other out-of-state station broadcasting on the same frequency as is 'The Voice of Labor.'*[2]

*...that 'The Voice of Labor' station does not impair, does not destroy and does not interfere with any interstate or foreign radio communication or transmission and that even though 'The Voice of Labor' station was not on the air that the people of Houston and of Harris County, Texas, would not be served with any satisfactory, reasonable or consistent radio entertainment by any licensed station broadcasting on a frequency of 1310 kilocycles during the hours of the day when 'The Voice of Labor' broadcasts on such frequency.*

*...That during the hours when 'The Voice of Labor' station is broadcasting the Federal Radio Commission and the Plaintiff are making no use of said 1310 kilocycle channel within the area served by 'The Voice of Labor.'*

*...In the event at a future date the Federal Radio Commission licenses some other station at Houston, Texas, to broadcast interstate or foreign radio communications and the broadcast of 'The Voice of Labor' illegally interferes with such licensed station these Defendants will, upon notice of such illegal interference by 'The Voice of Labor' station, immediately discontinue such interference.*[3]

U.S. communications regulation views localism and universal service as key public policy objectives. The ruling in *Gregg* fostered neither of these objectives. From a legal perspective, in considering whether federal suppression of "The Voice of Labor" is constitutional under the Commerce Clause, the ruling did not consider whether stations like "The Voice of Labor" were actually engaged in interstate commerce. The facts of the case indicate that "The Voice of Labor" employees, listeners, and programmers did not have a substantial interstate relation relevant to the case. On that basis and with appreciation for the natural sense of commerce as understood at the time the Constitution was written, the Court might have let "The Voice of Labor" continue broadcasting to the community about Houston, Texas.

[1] 5 F. Supp. 848, 853 (1934) fn. 1.
[2] Ibid.
[3] Above three sentences are from ibid. at 858, fn. 2.

One can imagine today equally compelling cases that would not be brought because the prospects for justice now appear so dim. Suppose an inner city school in Atlanta wants to use a device utilizing short-range radio emissions to screen students for guns. The device has been successful deployed in another country. While it uses radio spectrum categorically forbidden for such devices in the US, there is no evidence that it causes harmful interference to any other radio use. Suppose the Georgia Public Service Commission, unwilling to accept the FCC's refusal to license the device, authorized the school to use the device. Why should courts be unable to deliberate about an FCC enforcement action against this (federally) unauthorized radio use?

Here's another possibility. Suppose that a poor, retired cowboy in Idaho lives twenty-five miles, by deeply rutted dirt roads, from the nearest oncological specialist. He needs a weekly visual examination to confirm the shrinkage of a tumor. The most cost-effective wireless video conferencing system under the geographic and meteorological conditions in Idaho uses radio spectrum allocated in the U.S. to broadcast television services. The Idaho Department of Health and Welfare and the Idaho Public Utilities Commission, in light of pressing rural healthcare needs, jointly declare that such systems can be used in Idaho so long as they do not cause harmful and irremediable interference to any other radio users. Why should courts be unable to deliberate about an FCC enforcement action against this (federally) unauthorized radio use?

For a more mundane but very contentious concern, consider mobile phone suppression technology. Maintaining a certain aural atmosphere is an important concern in opera theatres, movie houses, courtrooms, and other places. In such places, ringing mobile phones ringing are a disturbing and interfering sound. Radio technology now available allows the organization that operates a space to suppress electronically mobile phone communications. Mobile phone suppression technology is illegal in most developed countries, but permitted in Israel and Japan. As of mid-2001, Hong Kong and Canada were considering legalizing mobile phone suppression.[1] There is some evidence of low-level debate about this issue in the U.S.[2] One might imagine state or local politics engaging with the issue and experimenting with different policies. Why should the only forum for considering this issue in the U.S. be the Federal Communications Commission?[3]

The significance to radio communications of inarticulate Commerce Clause jurisprudence is growing. Wireless technology is developing rapidly, with costs plummeting and the range of applications expanding dramatically. Such technology is being adopted quickly in many countries around the world. Moreover, many countries are likely to permit short-to-medium-range radio services in spectrum bands where such services are not permitted in the U.S. Geographic, demographic, and economic conditions vary significantly across U.S. states. Federal radio regulation will not be able to respond completely, with no contested decisions, to the wide range of radio communications needs and opportunities that will be presented to U.S. state governments. Yet a simplistic, widely held, and scarcely deliberated view of the Commerce Clause's meaning for U.S. radio regulation might prevent courts from ever seriously considering desperate pleas for justice.[1]

---

[1] See "Canadians debate use of cell-phone 'jammers'", USA Today, April, 23, 2001; online at http://www.usatoday.com/life/cyber/tech/review/2001-04-23-cell-phone-jam.htm

[2] See the online "Talkback" section at the end of the article, "Silencing Cell Phones," techtv, Aug. 16, 2001; online at http://www.techtv.com/news/culture/story/0,24195,3342655,00.htm

[3] Thus far public discussion of this issue at the FCC seems to consist only of posting a web page indicating that the Communications Act of 1934, as amended, does not permit the use of transmitters to prevent or jam the operation of wireless devices. See http://wireless.fcc.gov/services/cellular/operations/blockingjamming.html

---

[1] Mercy and truth can meet over radio regulation. The Superior Court of Pennsylvania recently stated:

> *Empathy for others woes and travails may evoke sympathy, but not jurisdiction. Rather, the decision of whether a court is seized with the ability to exercise subject matter jurisdiction must be predicated upon a careful statutory and constitutional foundation.*

See Fetterman v. Green, 455 Pa.Super.639,646,689 JA.2d 228, 293 (1997). If courts follow that statement, this paper should help offer hope for easing some woes and travails.

# Appendix C
# Communications Capabilities around the World

**Table C1**
**Countries, Categories, and Data**
**(used in Tables 10 and 11)**

| | | | | | Users Per Thousand Persons | | |
|---|---|---|---|---|---|---|---|
| Country or Economy | Class | Region | Pop. (ths.) | GNI PPP (ths.$) | Amateur | Internet | Mobile |
| Albania | 3 | 9 | 3411 | 3600 | 0.0047 | 2.9 | 102.6 |
| Algeria | 3 | 2 | 30400 | 5040 | 0.0020 | 2.0 | 3.3 |
| Andorra | 1 | 1 | 67 | | 1.4030 | 104.5 | 350.7 |
| Angola | 4 | 2 | 13100 | 1180 | 0.0004 | 4.6 | 6.6 |
| Antigua | 1 | 4 | 68 | 10000 | 0.2941 | 73.5 | 367.6 |
| Argentina | 2 | 6 | 37000 | 12050 | 0.4565 | 81.1 | 188.5 |
| Aruba | 1 | 4 | 101 | | 0.7723 | 237.6 | 524.8 |
| Australia | 1 | 8 | 19200 | 24970 | 0.7983 | 375.0 | 581.7 |
| Austria | 1 | 1 | 8110 | 26330 | 0.7662 | 320.6 | 809.6 |
| Bahamas | 1 | 4 | 303 | 16400 | 0.0330 | 55.8 | 200.0 |
| Bahrain | 2 | 3 | 691 | | 0.0767 | 202.9 | 433.6 |
| Bangladesh | 4 | 7 | 131000 | 1590 | 0.0002 | 1.1 | 4.0 |
| Barbados | 2 | 4 | 267 | 15020 | 0.5618 | 37.5 | 106.7 |
| Belarus | 3 | 9 | 10000 | 7550 | 0.1359 | 42.2 | 13.8 |
| Belgium | 1 | 1 | 10300 | 27470 | 0.5141 | 279.7 | 746.6 |
| Belize | 2 | 5 | 240 | 5240 | 0.2083 | 75.0 | 117.5 |
| Benin | 4 | 2 | 6272 | 980 | 0.0008 | 4.0 | 19.9 |
| Bermuda | 1 | 4 | 63 | | 1.4286 | 396.8 | 211.1 |
| Bolivia | 3 | 6 | 8329 | 2360 | 0.1663 | 14.4 | 89.3 |
| Bosnia | 3 | 9 | 3977 | | 0.4420 | 11.3 | 58.7 |
| Botswana | 2 | 2 | 1602 | 7170 | 0.0112 | 15.6 | 173.5 |
| Brazil | 2 | 6 | 170000 | 7300 | 0.1885 | 47.1 | 169.1 |
| Bulgaria | 3 | 9 | 8167 | 5560 | 0.6061 | 74.1 | 189.8 |
| Burkina Faso | 4 | 2 | 11300 | 970 | 0.0005 | 1.9 | 6.6 |
| Burundi | 4 | 2 | 6807 | 580 | 0.0001 | 0.9 | 2.9 |
| Cambodia | 4 | 7 | 12000 | 1440 | 0.0004 | 0.8 | 18.6 |
| Cameroon | 4 | 2 | 14900 | 1590 | 0.0013 | 3.0 | 20.8 |
| Canada | 1 | 5 | 30800 | 27170 | 1.4294 | 438.3 | 322.2 |
| Cape Verde | 3 | 2 | 441 | 4760 | 0.0091 | 27.2 | 71.4 |
| Central Africa | 4 | 2 | 3717 | 1160 | 0.0054 | 0.5 | 3.0 |
| Chad | 4 | 2 | 7694 | 870 | 0.0013 | 0.5 | 2.9 |
| Chile | 2 | 6 | 15200 | 9100 | 0.4641 | 204.1 | 346.8 |
| China | 3 | 7 | 1260000 | 3920 | 0.0006 | 26.7 | 114.9 |
| Chinese Taipei | 1 | 7 | 22200 | 23140 | 3.0942 | 340.1 | 974.5 |
| Colombia | 3 | 6 | 42300 | 6060 | 0.1537 | 27.3 | 74.7 |
| Costa Rica | 2 | 5 | 3811 | 7980 | 0.2459 | 100.8 | 81.7 |
| Cote d'Ivoire | 4 | 2 | 16000 | 1500 | 0.0313 | 4.4 | 45.5 |

**Table C1**
**Countries, Categories, and Data**
**(used in Tables 10 and 11)**

| Country or Economy | Class | Region | Pop. (ths.) | GNI PPP (ths.$) | Users Per Thousand Persons Amateur | Internet | Mobile |
|---|---|---|---|---|---|---|---|
| Croatia | 2 | 9 | 4380 | 7960 | 0.3219 | 57.1 | 400.7 |
| Cuba | 3 | 4 | 11200 | | 0.1670 | 10.7 | 0.7 |
| Cyprus | 1 | 1 | 757 | 20780 | 0.7635 | 198.2 | 415.3 |
| Czech Rep. | 2 | 9 | 10300 | 13780 | 0.6880 | 135.9 | 657.2 |
| Denmark | 1 | 1 | 5336 | 27250 | 1.8853 | 449.8 | 741.0 |
| Djibouti | 3 | 2 | 632 | | 0.0095 | 5.2 | 4.7 |
| Dominica | 2 | 4 | 73 | | 0.9589 | 82.2 | 16.4 |
| Dominicana | 3 | 4 | 8373 | 5710 | 0.2811 | 22.2 | 128.2 |
| DPR Congo | 4 | 2 | 50900 | | 0.0001 | 0.1 | 2.9 |
| Ecuador | 3 | 6 | 12600 | 2910 | 0.1592 | 26.0 | 68.2 |
| Egypt | 3 | 2 | 64000 | 3670 | 0.0007 | 9.4 | 43.7 |
| El Salvador | 3 | 5 | 6276 | 4410 | 0.1120 | 8.0 | 127.5 |
| Eq. Guinea | 3 | 2 | 457 | 5600 | 0.0109 | 2.0 | 32.8 |
| Estonia | 2 | 9 | 1369 | 9340 | 0.7305 | 313.9 | 475.7 |
| Ethiopia | 4 | 2 | 64300 | 660 | 0.0004 | 0.4 | 0.4 |
| Fiji | 3 | 8 | 812 | 4480 | 0.0185 | 18.5 | 93.6 |
| Finland | 1 | 1 | 5177 | 24570 | 1.1397 | 431.8 | 781.1 |
| Fr. Polynesia | 1 | 8 | 235 | 23340 | 0.7319 | 68.1 | 285.1 |
| France | 1 | 1 | 58900 | 24420 | 0.3141 | 265.8 | 609.9 |
| FYR Macedonia | 3 | 9 | 2031 | 5020 | 0.3053 | 34.5 | 109.9 |
| Gabon | 2 | 2 | 1230 | 5360 | 0.0285 | 12.2 | 97.6 |
| Gambia | 4 | 2 | 1303 | 1620 | 0.0146 | 13.8 | 33.0 |
| Germany | 1 | 1 | 82200 | 24920 | 0.9692 | 365.0 | 684.2 |
| Ghana | 4 | 2 | 19300 | 1910 | 0.0016 | 2.1 | 10.0 |
| Greece | 1 | 1 | 10600 | 16860 | 0.2785 | 132.1 | 751.1 |
| Grenada | 2 | 4 | 98 | 6960 | 1.7143 | 53.1 | 65.3 |
| Guatemala | 3 | 5 | 11400 | 3770 | 0.0339 | 17.5 | 99.5 |
| Guinea | 4 | 2 | 7415 | 1930 | 0.0007 | 2.0 | 7.5 |
| Guyana | 3 | 6 | 761 | 3670 | 0.0657 | 124.8 | 51.9 |
| Haiti | 4 | 4 | 7959 | 1470 | 0.0126 | 3.8 | 11.5 |
| Honduras | 3 | 5 | 6417 | 2400 | 0.0464 | 6.2 | 37.0 |
| Hong Kong | 1 | 7 | 6797 | 25590 | 0.2219 | 456.1 | 838.9 |
| Hungary | 2 | 9 | 10000 | 11990 | 0.9000 | 148.0 | 496.8 |
| Iceland | 1 | 1 | 281 | 28710 | 0.4982 | 694.0 | 837.7 |
| India | 4 | 7 | 1020000 | 2340 | 0.0105 | 6.9 | 5.6 |
| Indonesia | 4 | 7 | 210000 | 2830 | 0.1325 | 19.0 | 25.3 |
| Iran | 3 | 3 | 63700 | 5910 | 0.0002 | 6.3 | 23.3 |
| Ireland | 1 | 1 | 3794 | 25520 | 0.4370 | 235.9 | 738.0 |
| Israel | 1 | 3 | 6233 | 19330 | 0.1965 | 240.6 | 843.9 |
| Italy | 1 | 1 | 57700 | 23470 | 0.5199 | 277.3 | 844.0 |
| Jamaica | 3 | 4 | 2633 | 3440 | 0.0319 | 38.0 | 265.9 |
| Japan | 1 | 7 | 127000 | 27080 | 0.1020 | 455.9 | 573.2 |
| Jordan | 3 | 3 | 4887 | 3950 | 0.0272 | 43.4 | 152.6 |
| Kenya | 4 | 2 | 30100 | 1010 | 0.0012 | 16.6 | 16.6 |

**Table C1**
**Countries, Categories, and Data**
**(used in Tables 10 and 11)**

| Country or Economy | Class | Region | Pop. (ths.) | GNI PPP (ths.$) | Users Per Thousand Persons Amateur | Internet | Mobile |
|---|---|---|---|---|---|---|---|
| Kiribati | 3 | 8 | 91 | | 0.2205 | 22.1 | 4.4 |
| Kuwait | 1 | 3 | 1984 | 18690 | 0.0635 | 100.8 | 246.5 |
| Latvia | 3 | 9 | 2372 | 7070 | 0.0801 | 71.7 | 276.9 |
| Lebanon | 2 | 3 | 4328 | 4550 | 0.0474 | 69.3 | 171.7 |
| Lesotho | 4 | 2 | 2035 | 2590 | 0.0064 | 2.5 | 16.2 |
| Lithuania | 3 | 9 | 3695 | 6980 | 0.2192 | 67.7 | 252.2 |
| Luxembourg | 1 | 1 | 438 | 45470 | 1.1975 | 228.1 | 986.3 |
| Madagascar | 4 | 2 | 15500 | 820 | 0.0013 | 2.3 | 9.5 |
| Malawi | 4 | 2 | 10300 | 600 | 0.0019 | 1.9 | 5.4 |
| Malaysia | 2 | 7 | 23300 | 8330 | 0.0164 | 244.6 | 305.9 |
| Maldives | 3 | 7 | 276 | 4240 | 0.0181 | 36.2 | 66.7 |
| Mali | 4 | 2 | 10800 | 780 | 0.0021 | 2.8 | 4.2 |
| Malta | 2 | 1 | 390 | 16530 | 1.2051 | 253.8 | 355.9 |
| Marshall Is. | 3 | 8 | 52 | | 0.1923 | 17.3 | 9.6 |
| Mauritania | 4 | 2 | 2665 | 1630 | 0.0075 | 2.6 | 2.7 |
| Mauritius | 2 | 2 | 1186 | 9940 | 0.0346 | 133.2 | 252.9 |
| Mexico | 2 | 5 | 98000 | 8790 | 0.0720 | 35.7 | 205.5 |
| Moldova | 4 | 9 | 4282 | 2230 | 0.0448 | 14.0 | 49.0 |
| Mongolia | 4 | 7 | 2398 | 1760 | 0.0079 | 16.7 | 81.3 |
| Morocco | 3 | 2 | 28700 | 3450 | 0.0136 | 13.9 | 166.3 |
| Mozambique | 4 | 2 | 17700 | 800 | 0.0028 | 0.8 | 9.6 |
| Namibia | 3 | 2 | 1757 | 6410 | 0.0501 | 25.6 | 56.9 |
| Nepal | 4 | 7 | 23000 | 1370 | 0.0004 | 2.6 | 0.8 |
| Netherlands | 1 | 1 | 15900 | 25850 | 0.9138 | 333.3 | 748.4 |
| New Caledonia | 1 | 8 | 213 | 21820 | 0.6112 | 112.8 | 234.6 |
| New Zealand | 1 | 8 | 3831 | 18530 | 1.4263 | 285.0 | 630.9 |
| Nicaragua | 4 | 5 | 5071 | 2080 | 0.1112 | 9.9 | 30.8 |
| Nigeria | 4 | 2 | 127000 | 800 | 0.0012 | 1.6 | 2.6 |
| Norway | 1 | 1 | 4491 | 29630 | 1.1806 | 601.2 | 832.1 |
| Oman | 2 | 3 | 2395 | | 0.0392 | 50.1 | 135.5 |
| Pakistan | 4 | 7 | 138000 | 1860 | 0.0016 | 3.6 | 5.8 |
| Panama | 2 | 5 | 2856 | 5680 | 0.7087 | 31.5 | 210.1 |
| Paraguay | 3 | 6 | 5496 | 4450 | 0.9252 | 10.9 | 209.2 |
| Peru | 3 | 6 | 25700 | 4660 | 0.1198 | 116.7 | 60.1 |
| Philippines | 3 | 7 | 75600 | 4220 | 0.0529 | 26.5 | 139.8 |
| Poland | 2 | 9 | 38700 | 9000 | 0.4134 | 98.2 | 259.7 |
| Portugal | 1 | 1 | 10000 | 16990 | 0.4200 | 360.0 | 797.8 |
| Qatar | 1 | 3 | 585 | | 0.0359 | 68.4 | 305.7 |
| RO Congo | 4 | 2 | 3018 | 570 | 0.0017 | 0.2 | 49.7 |
| RO Korea | 2 | 7 | 47300 | 17300 | 2.9810 | 515.4 | 614.1 |
| Romania | 3 | 9 | 22400 | 6360 | 0.1563 | 44.6 | 172.3 |
| Russia | 3 | 9 | 146000 | 8010 | 0.2603 | 29.5 | 38.1 |
| Rwanda | 4 | 2 | 8508 | 930 | 0.0024 | 2.4 | 7.6 |
| Senegal | 4 | 2 | 9530 | 1480 | 0.0043 | 10.5 | 41.0 |

**Table C1**
**Countries, Categories, and Data**
**(used in Tables 10 and 11)**

| Country or Economy | Class | Region | Pop. (ths.) | GNI PPP (ths.$) | Users Per Thousand Persons: Amateur | Internet | Mobile |
|---|---|---|---|---|---|---|---|
| Seychelles | 2 | 2 | 81 | | 0.1231 | 110.8 | 542.9 |
| Sierra Leone | 4 | 2 | 5031 | 480 | 0.0072 | 1.4 | 5.3 |
| Singapore | 1 | 7 | 4018 | 24910 | 0.0236 | 373.3 | 711.5 |
| Slovakia | 2 | 9 | 5402 | 11040 | 0.1722 | 120.3 | 397.5 |
| Slovenia | 1 | 9 | 1988 | 17310 | 3.2696 | 301.8 | 762.4 |
| Solomon Is. | 4 | 8 | 447 | 1710 | 0.0134 | 4.5 | 2.2 |
| South Africa | 2 | 2 | 42800 | 9160 | 0.1402 | 71.7 | 214.9 |
| Spain | 1 | 1 | 39500 | 19260 | 1.4861 | 187.0 | 670.7 |
| Sri Lanka | 3 | 7 | 19400 | 3460 | 0.0103 | 7.7 | 37.1 |
| St. Kitts | 2 | 4 | 41 | 10960 | 0.9756 | 48.8 | 29.3 |
| St. Vincent | 3 | 4 | 115 | 5210 | 0.2609 | 30.4 | 20.9 |
| Sudan | 4 | 2 | 31100 | 1520 | 0.0002 | 1.8 | 3.4 |
| Suriname | 3 | 6 | 417 | 3480 | 0.2230 | 34.8 | 201.7 |
| Swaziland | 3 | 2 | 1045 | 4600 | 0.0536 | 13.4 | 63.2 |
| Sweden | 1 | 1 | 8869 | 23970 | 1.2196 | 518.7 | 774.3 |
| Switzerland | 1 | 1 | 7180 | 30450 | 0.7660 | 406.3 | 727.9 |
| Syria | 3 | 3 | 16200 | 3340 | 0.0006 | 3.7 | 12.3 |
| Tajikistan | 4 | 9 | 6170 | 1090 | 0.0078 | 0.5 | 0.3 |
| Tanzania | 4 | 2 | 33700 | 520 | 0.0051 | 8.9 | 12.7 |
| Thailand | 3 | 7 | 60700 | 6320 | 2.3269 | 58.3 | 124.4 |
| Togo | 4 | 2 | 4527 | 1410 | 0.0022 | 11.0 | 21.0 |
| Tonga | 3 | 8 | 100 | | 0.1497 | 10.0 | 1.0 |
| Trinidad | 2 | 4 | 1301 | 8220 | 0.3420 | 92.2 | 173.3 |
| Tunisia | 3 | 2 | 9564 | 6070 | 0.0021 | 41.8 | 40.7 |
| Turkey | 2 | 1 | 65300 | 7030 | 0.0184 | 38.3 | 306.3 |
| Turkmenistan | 4 | 9 | 5199 | 3800 | 0.0037 | 1.5 | 1.8 |
| UAE | 1 | 3 | 2905 | | 0.0034 | 309.8 | 657.2 |
| Uganda | 4 | 2 | 22200 | 1210 | 0.0009 | 2.7 | 14.5 |
| UK | 1 | 1 | 59700 | 23550 | 0.9787 | 402.0 | 787.7 |
| Ukraine | 4 | 9 | 49500 | 3700 | 0.3488 | 12.1 | 44.9 |
| Uruguay | 2 | 6 | 3337 | 8880 | 1.1927 | 119.9 | 155.8 |
| USA | 1 | 5 | 282000 | 34100 | 2.4109 | 506.5 | 450.4 |
| Vanuatu | 3 | 8 | 197 | 2960 | 0.1269 | 27.9 | 1.5 |
| Venezuela | 2 | 6 | 24200 | 5740 | 0.4380 | 53.7 | 268.2 |
| Vietnam | 4 | 7 | 78500 | 2000 | 0.0003 | 5.1 | 15.9 |
| Western Samoa | 3 | 8 | 170 | 5050 | 0.0882 | 17.6 | 17.6 |
| Yugoslavia | 3 | 9 | 10600 | | 0.0334 | 56.6 | 188.5 |
| Zambia | 4 | 2 | 10100 | 750 | 0.0045 | 2.5 | 9.7 |
| Zimbabwe | 4 | 2 | 12600 | 2550 | 0.0076 | 7.9 | 26.1 |

Sources: See notes for Appendix A, notes for Table 8.

Income Class Codes

| | |
|---|---|
| 1 | High Income |
| 2 | Upper Middle Income |
| 3 | Lower Middle Income |
| 4 | Low Income |

Region Codes

| | |
|---|---|
| 1 | Western Europe |
| 2 | Africa |
| 3 | Middle East |
| 4 | Caribbean |
| 5 | North and Central America |
| 6 | South America |
| 7 | Asia |
| 8 | Oceania |
| 9 | Central and Eastern Europe |